\documentclass{class/llncs}

\usepackage{epsfig}
\usepackage{fancyhdr}
\usepackage{plain}

\usepackage{footmisc}
\usepackage[FIGBOTCAP,TABBOTCAP,center,nooneline]{subfigure}
\usepackage[dvips]{rotating}
\usepackage[noend]{algorithmic}
\usepackage[normalem]{ulem}
\usepackage{amscd}
\usepackage{amsfonts}
\usepackage{amsmath}
\usepackage{booktabs}
\usepackage{color}
\usepackage{comment}
\usepackage{datetime}
\usepackage{epsfig}
\usepackage{graphicx}
\usepackage{pdfpages}
\usepackage{latexsym}
\usepackage{marginnote}
\usepackage{mathtools}
\usepackage{multirow}
\usepackage{paralist}
\usepackage{placeins}
\usepackage{rotate} 
\usepackage{tikz}
\usepackage{times} 
\usepackage{url}
\usepackage{varwidth}
\usepackage{verbatim} 
\usepackage{wrapfig}
\usepackage{xcolor,colortbl}
\usepackage{transparent}
\usepackage{xspace}
\usepackage[formats]{listings}
\usepackage{courier}
\usepackage{relsize}
\usepackage{float}
\usepackage{afterpage}
\usepackage{tabularx}
\usepackage[linesnumbered,ruled,vlined]{algorithm2e}
\usepackage[framemethod=tikz]{mdframed}
\usepackage{cleveref}
\usepackage{etoolbox}
\usepackage{array}
\usepackage{eqparbox}
\usepackage[T1]{fontenc}
\usepackage[utf8]{inputenc}
\usepackage{bibentry}
\usepackage{nicefrac}

\newcommand{\longversion}{true}
\newcommand{\withproofs}{true}

\newcommand{\blue}[1]{\textcolor{blue}{#1}}

\newcommand{\green}[1]{\textcolor{darkgreen}{#1}}

\definecolor {snow}                {rgb}{1.00,0.98,0.98}
\definecolor {ghostwhite}          {rgb}{0.97,0.97,1.00}
\definecolor {whitesmoke}          {rgb}{0.96,0.96,0.96}
\definecolor {gainsboro}           {rgb}{0.86,0.86,0.86}
\definecolor {floralwhite}         {rgb}{1.00,0.98,0.94}
\definecolor {oldlace}             {rgb}{0.99,0.96,0.90}
\definecolor {linen}               {rgb}{0.98,0.94,0.90}
\definecolor {antiquewhite}        {rgb}{0.98,0.92,0.84}
\definecolor {papayawhip}          {rgb}{1.00,0.94,0.84}
\definecolor {blanchedalmond}      {rgb}{1.00,0.92,0.80}
\definecolor {bisque}              {rgb}{1.00,0.89,0.77}
\definecolor {peachpuff}           {rgb}{1.00,0.85,0.73}
\definecolor {navajowhite}         {rgb}{1.00,0.87,0.68}
\definecolor {moccasin}            {rgb}{1.00,0.89,0.71}
\definecolor {cornsilk}            {rgb}{1.00,0.97,0.86}
\definecolor {ivory}               {rgb}{1.00,1.00,0.94}
\definecolor {lemonchiffon}        {rgb}{1.00,0.98,0.80}
\definecolor {seashell}            {rgb}{1.00,0.96,0.93}
\definecolor {honeydew}            {rgb}{0.94,1.00,0.94}
\definecolor {mintcream}           {rgb}{0.96,1.00,0.98}
\definecolor {azure}               {rgb}{0.94,1.00,1.00}
\definecolor {aliceblue}           {rgb}{0.94,0.97,1.00}
\definecolor {lavender}            {rgb}{0.90,0.90,0.98}
\definecolor {lavenderblush}       {rgb}{1.00,0.94,0.96}
\definecolor {mistyrose}           {rgb}{1.00,0.89,0.88}
\definecolor {white}               {rgb}{1.00,1.00,1.00}
\definecolor {black}               {rgb}{0.00,0.00,0.00}
\definecolor {darkslategray}       {rgb}{0.18,0.31,0.31}
\definecolor {dimgray}             {rgb}{0.41,0.41,0.41}
\definecolor {slategray}           {rgb}{0.44,0.50,0.56}
\definecolor {lightslategray}      {rgb}{0.47,0.53,0.60}
\definecolor {gray}                {rgb}{0.75,0.75,0.75}
\definecolor {lightgrey}           {rgb}{0.83,0.83,0.83}
\definecolor {midnightblue}        {rgb}{0.10,0.10,0.44}
\definecolor {navy}                {rgb}{0.00,0.00,0.50}
\definecolor {cornflowerblue}      {rgb}{0.39,0.58,0.93}
\definecolor {darkslateblue}       {rgb}{0.28,0.24,0.55}
\definecolor {slateblue}           {rgb}{0.42,0.35,0.80}
\definecolor {mediumslateblue}     {rgb}{0.48,0.41,0.93}
\definecolor {lightslateblue}      {rgb}{0.52,0.44,1.00}
\definecolor {mediumblue}          {rgb}{0.00,0.00,0.80}
\definecolor {royalblue}           {rgb}{0.25,0.41,0.88}
\definecolor {blue}                {rgb}{0.00,0.00,1.00}
\definecolor {dodgerblue}          {rgb}{0.12,0.56,1.00}
\definecolor {deepskyblue}         {rgb}{0.00,0.75,1.00}
\definecolor {skyblue}             {rgb}{0.53,0.81,0.92}
\definecolor {lightskyblue}        {rgb}{0.53,0.81,0.98}
\definecolor {steelblue}           {rgb}{0.27,0.51,0.71}
\definecolor {lightsteelblue}      {rgb}{0.69,0.77,0.87}
\definecolor {lightblue}           {rgb}{0.68,0.85,0.90}
\definecolor {powderblue}          {rgb}{0.69,0.88,0.90}
\definecolor {paleturquoise}       {rgb}{0.69,0.93,0.93}
\definecolor {darkturquoise}       {rgb}{0.00,0.81,0.82}
\definecolor {mediumturquoise}     {rgb}{0.28,0.82,0.80}
\definecolor {turquoise}           {rgb}{0.25,0.88,0.82}
\definecolor {cyan}                {rgb}{0.00,1.00,1.00}
\definecolor {lightcyan}           {rgb}{0.88,1.00,1.00}
\definecolor {cadetblue}           {rgb}{0.37,0.62,0.63}
\definecolor {mediumaquamarine}    {rgb}{0.40,0.80,0.67}
\definecolor {aquamarine}          {rgb}{0.50,1.00,0.83}
\definecolor {darkgreen}           {rgb}{0.00,0.39,0.00}
\definecolor {darkolivegreen}      {rgb}{0.33,0.42,0.18}
\definecolor {darkseagreen}        {rgb}{0.56,0.74,0.56}
\definecolor {seagreen}            {rgb}{0.18,0.55,0.34}
\definecolor {mediumseagreen}      {rgb}{0.24,0.70,0.44}
\definecolor {lightseagreen}       {rgb}{0.13,0.70,0.67}
\definecolor {palegreen}           {rgb}{0.60,0.98,0.60}
\definecolor {springgreen}         {rgb}{0.00,1.00,0.50}
\definecolor {lawngreen}           {rgb}{0.49,0.99,0.00}
\definecolor {green}               {rgb}{0.00,1.00,0.00}
\definecolor {chartreuse}          {rgb}{0.50,1.00,0.00}
\definecolor {mediumspringgreen}   {rgb}{0.00,0.98,0.60}
\definecolor {greenyellow}         {rgb}{0.68,1.00,0.18}
\definecolor {limegreen}           {rgb}{0.20,0.80,0.20}
\definecolor {yellowgreen}         {rgb}{0.60,0.80,0.20}
\definecolor {forestgreen}         {rgb}{0.13,0.55,0.13}
\definecolor {olivedrab}           {rgb}{0.42,0.56,0.14}
\definecolor {darkkhaki}           {rgb}{0.74,0.72,0.42}
\definecolor {khaki}               {rgb}{0.94,0.90,0.55}
\definecolor {palegoldenrod}       {rgb}{0.93,0.91,0.67}
\definecolor {lightgoldenrodyellow} {rgb}{0.98,0.98,0.82}
\definecolor {lightyellow}         {rgb}{1.00,1.00,0.88}
\definecolor {yellow}              {rgb}{1.00,1.00,0.00}
\definecolor {gold}                {rgb}{1.00,0.84,0.00}
\definecolor {lightgoldenrod}      {rgb}{0.93,0.87,0.51}
\definecolor {goldenrod}           {rgb}{0.85,0.65,0.13}
\definecolor {darkgoldenrod}       {rgb}{0.72,0.53,0.04}
\definecolor {rosybrown}           {rgb}{0.74,0.56,0.56}
\definecolor {indianred}           {rgb}{0.80,0.36,0.36}
\definecolor {saddlebrown}         {rgb}{0.55,0.27,0.07}
\definecolor {sienna}              {rgb}{0.63,0.32,0.18}
\definecolor {peru}                {rgb}{0.80,0.52,0.25}
\definecolor {burlywood}           {rgb}{0.87,0.72,0.53}
\definecolor {beige}               {rgb}{0.96,0.96,0.86}
\definecolor {wheat}               {rgb}{0.96,0.87,0.70}
\definecolor {sandybrown}          {rgb}{0.96,0.64,0.38}
\definecolor {tan}                 {rgb}{0.82,0.71,0.55}
\definecolor {chocolate}           {rgb}{0.82,0.41,0.12}
\definecolor {firebrick}           {rgb}{0.70,0.13,0.13}
\definecolor {brown}               {rgb}{0.65,0.16,0.16}
\definecolor {darksalmon}          {rgb}{0.91,0.59,0.48}
\definecolor {salmon}              {rgb}{0.98,0.50,0.45}
\definecolor {lightsalmon}         {rgb}{1.00,0.63,0.48}
\definecolor {orange}              {rgb}{1.00,0.65,0.00}
\definecolor {darkorange}          {rgb}{1.00,0.55,0.00}
\definecolor {coral}               {rgb}{1.00,0.50,0.31}
\definecolor {lightcoral}          {rgb}{0.94,0.50,0.50}
\definecolor {tomato}              {rgb}{1.00,0.39,0.28}
\definecolor {orangered}           {rgb}{1.00,0.27,0.00}
\definecolor {red}                 {rgb}{1.00,0.00,0.00}
\definecolor {hotpink}             {rgb}{1.00,0.41,0.71}
\definecolor {deeppink}            {rgb}{1.00,0.08,0.58}
\definecolor {pink}                {rgb}{1.00,0.75,0.80}
\definecolor {lightpink}           {rgb}{1.00,0.71,0.76}
\definecolor {palevioletred}       {rgb}{0.86,0.44,0.58}
\definecolor {maroon}              {rgb}{0.69,0.19,0.38}
\definecolor {mediumvioletred}     {rgb}{0.78,0.08,0.52}
\definecolor {violetred}           {rgb}{0.82,0.13,0.56}
\definecolor {magenta}             {rgb}{1.00,0.00,1.00}
\definecolor {violet}              {rgb}{0.93,0.51,0.93}
\definecolor {plum}                {rgb}{0.87,0.63,0.87}
\definecolor {orchid}              {rgb}{0.85,0.44,0.84}
\definecolor {mediumorchid}        {rgb}{0.73,0.33,0.83}
\definecolor {darkorchid}          {rgb}{0.60,0.20,0.80}
\definecolor {darkviolet}          {rgb}{0.58,0.00,0.83}
\definecolor {blueviolet}          {rgb}{0.54,0.17,0.89}
\definecolor {purple}              {rgb}{0.63,0.13,0.94}
\definecolor {mediumpurple}        {rgb}{0.58,0.44,0.86}
\definecolor {thistle}             {rgb}{0.85,0.75,0.85}
\definecolor {snow2}               {rgb}{0.93,0.91,0.91}
\definecolor {snow3}               {rgb}{0.80,0.79,0.79}
\definecolor {snow4}               {rgb}{0.55,0.54,0.54}
\definecolor {seashell2}           {rgb}{0.93,0.90,0.87}
\definecolor {seashell3}           {rgb}{0.80,0.77,0.75}
\definecolor {seashell4}           {rgb}{0.55,0.53,0.51}
\definecolor {antiquewhite1}       {rgb}{1.00,0.94,0.86}
\definecolor {antiquewhite2}       {rgb}{0.93,0.87,0.80}
\definecolor {antiquewhite3}       {rgb}{0.80,0.75,0.69}
\definecolor {antiquewhite4}       {rgb}{0.55,0.51,0.47}
\definecolor {bisque2}             {rgb}{0.93,0.84,0.72}
\definecolor {bisque3}             {rgb}{0.80,0.72,0.62}
\definecolor {bisque4}             {rgb}{0.55,0.49,0.42}
\definecolor {peachpuff2}          {rgb}{0.93,0.80,0.68}
\definecolor {peachpuff3}          {rgb}{0.80,0.69,0.58}
\definecolor {peachpuff4}          {rgb}{0.55,0.47,0.40}
\definecolor {navajowhite2}        {rgb}{0.93,0.81,0.63}
\definecolor {navajowhite3}        {rgb}{0.80,0.70,0.55}
\definecolor {navajowhite4}        {rgb}{0.55,0.47,0.37}
\definecolor {lemonchiffon2}       {rgb}{0.93,0.91,0.75}
\definecolor {lemonchiffon3}       {rgb}{0.80,0.79,0.65}
\definecolor {lemonchiffon4}       {rgb}{0.55,0.54,0.44}
\definecolor {cornsilk2}           {rgb}{0.93,0.91,0.80}
\definecolor {cornsilk3}           {rgb}{0.80,0.78,0.69}
\definecolor {cornsilk4}           {rgb}{0.55,0.53,0.47}
\definecolor {ivory2}              {rgb}{0.93,0.93,0.88}
\definecolor {ivory3}              {rgb}{0.80,0.80,0.76}
\definecolor {ivory4}              {rgb}{0.55,0.55,0.51}
\definecolor {honeydew2}           {rgb}{0.88,0.93,0.88}
\definecolor {honeydew3}           {rgb}{0.76,0.80,0.76}
\definecolor {honeydew4}           {rgb}{0.51,0.55,0.51}
\definecolor {lavenderblush2}      {rgb}{0.93,0.88,0.90}
\definecolor {lavenderblush3}      {rgb}{0.80,0.76,0.77}
\definecolor {lavenderblush4}      {rgb}{0.55,0.51,0.53}
\definecolor {mistyrose2}          {rgb}{0.93,0.84,0.82}
\definecolor {mistyrose3}          {rgb}{0.80,0.72,0.71}
\definecolor {mistyrose4}          {rgb}{0.55,0.49,0.48}
\definecolor {azure2}              {rgb}{0.88,0.93,0.93}
\definecolor {azure3}              {rgb}{0.76,0.80,0.80}
\definecolor {azure4}              {rgb}{0.51,0.55,0.55}
\definecolor {slateblue1}          {rgb}{0.51,0.44,1.00}
\definecolor {slateblue2}          {rgb}{0.48,0.40,0.93}
\definecolor {slateblue3}          {rgb}{0.41,0.35,0.80}
\definecolor {slateblue4}          {rgb}{0.28,0.24,0.55}
\definecolor {royalblue1}          {rgb}{0.28,0.46,1.00}
\definecolor {royalblue2}          {rgb}{0.26,0.43,0.93}
\definecolor {royalblue3}          {rgb}{0.23,0.37,0.80}
\definecolor {royalblue4}          {rgb}{0.15,0.25,0.55}
\definecolor {blue2}               {rgb}{0.00,0.00,0.93}
\definecolor {blue4}               {rgb}{0.00,0.00,0.55}
\definecolor {dodgerblue2}         {rgb}{0.11,0.53,0.93}
\definecolor {dodgerblue3}         {rgb}{0.09,0.45,0.80}
\definecolor {dodgerblue4}         {rgb}{0.06,0.31,0.55}
\definecolor {steelblue1}          {rgb}{0.39,0.72,1.00}
\definecolor {steelblue2}          {rgb}{0.36,0.67,0.93}
\definecolor {steelblue3}          {rgb}{0.31,0.58,0.80}
\definecolor {steelblue4}          {rgb}{0.21,0.39,0.55}
\definecolor {deepskyblue2}        {rgb}{0.00,0.70,0.93}
\definecolor {deepskyblue3}        {rgb}{0.00,0.60,0.80}
\definecolor {deepskyblue4}        {rgb}{0.00,0.41,0.55}
\definecolor {skyblue1}            {rgb}{0.53,0.81,1.00}
\definecolor {skyblue2}            {rgb}{0.49,0.75,0.93}
\definecolor {skyblue3}            {rgb}{0.42,0.65,0.80}
\definecolor {skyblue4}            {rgb}{0.29,0.44,0.55}
\definecolor {lightskyblue1}       {rgb}{0.69,0.89,1.00}
\definecolor {lightskyblue2}       {rgb}{0.64,0.83,0.93}
\definecolor {lightskyblue3}       {rgb}{0.55,0.71,0.80}
\definecolor {lightskyblue4}       {rgb}{0.38,0.48,0.55}
\definecolor {slategray1}          {rgb}{0.78,0.89,1.00}
\definecolor {slategray2}          {rgb}{0.73,0.83,0.93}
\definecolor {slategray3}          {rgb}{0.62,0.71,0.80}
\definecolor {slategray4}          {rgb}{0.42,0.48,0.55}
\definecolor {lightsteelblue1}     {rgb}{0.79,0.88,1.00}
\definecolor {lightsteelblue2}     {rgb}{0.74,0.82,0.93}
\definecolor {lightsteelblue3}     {rgb}{0.64,0.71,0.80}
\definecolor {lightsteelblue4}     {rgb}{0.43,0.48,0.55}
\definecolor {lightblue1}          {rgb}{0.75,0.94,1.00}
\definecolor {lightblue2}          {rgb}{0.70,0.87,0.93}
\definecolor {lightblue3}          {rgb}{0.60,0.75,0.80}
\definecolor {lightblue4}          {rgb}{0.41,0.51,0.55}
\definecolor {lightcyan2}          {rgb}{0.82,0.93,0.93}
\definecolor {lightcyan3}          {rgb}{0.71,0.80,0.80}
\definecolor {lightcyan4}          {rgb}{0.48,0.55,0.55}
\definecolor {paleturquoise1}      {rgb}{0.73,1.00,1.00}
\definecolor {paleturquoise2}      {rgb}{0.68,0.93,0.93}
\definecolor {paleturquoise3}      {rgb}{0.59,0.80,0.80}
\definecolor {paleturquoise4}      {rgb}{0.40,0.55,0.55}
\definecolor {cadetblue1}          {rgb}{0.60,0.96,1.00}
\definecolor {cadetblue2}          {rgb}{0.56,0.90,0.93}
\definecolor {cadetblue3}          {rgb}{0.48,0.77,0.80}
\definecolor {cadetblue4}          {rgb}{0.33,0.53,0.55}
\definecolor {turquoise1}          {rgb}{0.00,0.96,1.00}
\definecolor {turquoise2}          {rgb}{0.00,0.90,0.93}
\definecolor {turquoise3}          {rgb}{0.00,0.77,0.80}
\definecolor {turquoise4}          {rgb}{0.00,0.53,0.55}
\definecolor {cyan2}               {rgb}{0.00,0.93,0.93}
\definecolor {cyan3}               {rgb}{0.00,0.80,0.80}
\definecolor {cyan4}               {rgb}{0.00,0.55,0.55}
\definecolor {darkslategray1}      {rgb}{0.59,1.00,1.00}
\definecolor {darkslategray2}      {rgb}{0.55,0.93,0.93}
\definecolor {darkslategray3}      {rgb}{0.47,0.80,0.80}
\definecolor {darkslategray4}      {rgb}{0.32,0.55,0.55}
\definecolor {aquamarine2}         {rgb}{0.46,0.93,0.78}
\definecolor {aquamarine4}         {rgb}{0.27,0.55,0.45}
\definecolor {darkseagreen1}       {rgb}{0.76,1.00,0.76}
\definecolor {darkseagreen2}       {rgb}{0.71,0.93,0.71}
\definecolor {darkseagreen3}       {rgb}{0.61,0.80,0.61}
\definecolor {darkseagreen4}       {rgb}{0.41,0.55,0.41}
\definecolor {seagreen1}           {rgb}{0.33,1.00,0.62}
\definecolor {seagreen2}           {rgb}{0.31,0.93,0.58}
\definecolor {seagreen3}           {rgb}{0.26,0.80,0.50}
\definecolor {palegreen1}          {rgb}{0.60,1.00,0.60}
\definecolor {palegreen2}          {rgb}{0.56,0.93,0.56}
\definecolor {palegreen3}          {rgb}{0.49,0.80,0.49}
\definecolor {palegreen4}          {rgb}{0.33,0.55,0.33}
\definecolor {springgreen2}        {rgb}{0.00,0.93,0.46}
\definecolor {springgreen3}        {rgb}{0.00,0.80,0.40}
\definecolor {springgreen4}        {rgb}{0.00,0.55,0.27}
\definecolor {green2}              {rgb}{0.00,0.93,0.00}
\definecolor {green3}              {rgb}{0.00,0.80,0.00}
\definecolor {green4}              {rgb}{0.00,0.55,0.00}
\definecolor {chartreuse2}         {rgb}{0.46,0.93,0.00}
\definecolor {chartreuse3}         {rgb}{0.40,0.80,0.00}
\definecolor {chartreuse4}         {rgb}{0.27,0.55,0.00}
\definecolor {olivedrab1}          {rgb}{0.75,1.00,0.24}
\definecolor {olivedrab2}          {rgb}{0.70,0.93,0.23}
\definecolor {olivedrab4}          {rgb}{0.41,0.55,0.13}
\definecolor {darkolivegreen1}     {rgb}{0.79,1.00,0.44}
\definecolor {darkolivegreen2}     {rgb}{0.74,0.93,0.41}
\definecolor {darkolivegreen3}     {rgb}{0.64,0.80,0.35}
\definecolor {darkolivegreen4}     {rgb}{0.43,0.55,0.24}
\definecolor {khaki1}              {rgb}{1.00,0.96,0.56}
\definecolor {khaki2}              {rgb}{0.93,0.90,0.52}
\definecolor {khaki3}              {rgb}{0.80,0.78,0.45}
\definecolor {khaki4}              {rgb}{0.55,0.53,0.31}
\definecolor {lightgoldenrod1}     {rgb}{1.00,0.93,0.55}
\definecolor {lightgoldenrod2}     {rgb}{0.93,0.86,0.51}
\definecolor {lightgoldenrod3}     {rgb}{0.80,0.75,0.44}
\definecolor {lightgoldenrod4}     {rgb}{0.55,0.51,0.30}
\definecolor {lightyellow2}        {rgb}{0.93,0.93,0.82}
\definecolor {lightyellow3}        {rgb}{0.80,0.80,0.71}
\definecolor {lightyellow4}        {rgb}{0.55,0.55,0.48}
\definecolor {yellow2}             {rgb}{0.93,0.93,0.00}
\definecolor {yellow3}             {rgb}{0.80,0.80,0.00}
\definecolor {yellow4}             {rgb}{0.55,0.55,0.00}
\definecolor {gold2}               {rgb}{0.93,0.79,0.00}
\definecolor {gold3}               {rgb}{0.80,0.68,0.00}
\definecolor {gold4}               {rgb}{0.55,0.46,0.00}
\definecolor {goldenrod1}          {rgb}{1.00,0.76,0.15}
\definecolor {goldenrod2}          {rgb}{0.93,0.71,0.13}
\definecolor {goldenrod3}          {rgb}{0.80,0.61,0.11}
\definecolor {goldenrod4}          {rgb}{0.55,0.41,0.08}
\definecolor {darkgoldenrod1}      {rgb}{1.00,0.73,0.06}
\definecolor {darkgoldenrod2}      {rgb}{0.93,0.68,0.05}
\definecolor {darkgoldenrod3}      {rgb}{0.80,0.58,0.05}
\definecolor {darkgoldenrod4}      {rgb}{0.55,0.40,0.03}
\definecolor {rosybrown1}          {rgb}{1.00,0.76,0.76}
\definecolor {rosybrown2}          {rgb}{0.93,0.71,0.71}
\definecolor {rosybrown3}          {rgb}{0.80,0.61,0.61}
\definecolor {rosybrown4}          {rgb}{0.55,0.41,0.41}
\definecolor {indianred1}          {rgb}{1.00,0.42,0.42}
\definecolor {indianred2}          {rgb}{0.93,0.39,0.39}
\definecolor {indianred3}          {rgb}{0.80,0.33,0.33}
\definecolor {indianred4}          {rgb}{0.55,0.23,0.23}
\definecolor {sienna1}             {rgb}{1.00,0.51,0.28}
\definecolor {sienna2}             {rgb}{0.93,0.47,0.26}
\definecolor {sienna3}             {rgb}{0.80,0.41,0.22}
\definecolor {sienna4}             {rgb}{0.55,0.28,0.15}
\definecolor {burlywood1}          {rgb}{1.00,0.83,0.61}
\definecolor {burlywood2}          {rgb}{0.93,0.77,0.57}
\definecolor {burlywood3}          {rgb}{0.80,0.67,0.49}
\definecolor {burlywood4}          {rgb}{0.55,0.45,0.33}
\definecolor {wheat1}              {rgb}{1.00,0.91,0.73}
\definecolor {wheat2}              {rgb}{0.93,0.85,0.68}
\definecolor {wheat3}              {rgb}{0.80,0.73,0.59}
\definecolor {wheat4}              {rgb}{0.55,0.49,0.40}
\definecolor {tan1}                {rgb}{1.00,0.65,0.31}
\definecolor {tan2}                {rgb}{0.93,0.60,0.29}
\definecolor {tan4}                {rgb}{0.55,0.35,0.17}
\definecolor {chocolate1}          {rgb}{1.00,0.50,0.14}
\definecolor {chocolate2}          {rgb}{0.93,0.46,0.13}
\definecolor {chocolate3}          {rgb}{0.80,0.40,0.11}
\definecolor {firebrick1}          {rgb}{1.00,0.19,0.19}
\definecolor {firebrick2}          {rgb}{0.93,0.17,0.17}
\definecolor {firebrick3}          {rgb}{0.80,0.15,0.15}
\definecolor {firebrick4}          {rgb}{0.55,0.10,0.10}
\definecolor {brown1}              {rgb}{1.00,0.25,0.25}
\definecolor {brown2}              {rgb}{0.93,0.23,0.23}
\definecolor {brown3}              {rgb}{0.80,0.20,0.20}
\definecolor {brown4}              {rgb}{0.55,0.14,0.14}
\definecolor {salmon1}             {rgb}{1.00,0.55,0.41}
\definecolor {salmon2}             {rgb}{0.93,0.51,0.38}
\definecolor {salmon3}             {rgb}{0.80,0.44,0.33}
\definecolor {salmon4}             {rgb}{0.55,0.30,0.22}
\definecolor {lightsalmon2}        {rgb}{0.93,0.58,0.45}
\definecolor {lightsalmon3}        {rgb}{0.80,0.51,0.38}
\definecolor {lightsalmon4}        {rgb}{0.55,0.34,0.26}
\definecolor {orange2}             {rgb}{0.93,0.60,0.00}
\definecolor {orange3}             {rgb}{0.80,0.52,0.00}
\definecolor {orange4}             {rgb}{0.55,0.35,0.00}
\definecolor {darkorange1}         {rgb}{1.00,0.50,0.00}
\definecolor {darkorange2}         {rgb}{0.93,0.46,0.00}
\definecolor {darkorange3}         {rgb}{0.80,0.40,0.00}
\definecolor {darkorange4}         {rgb}{0.55,0.27,0.00}
\definecolor {coral1}              {rgb}{1.00,0.45,0.34}
\definecolor {coral2}              {rgb}{0.93,0.42,0.31}
\definecolor {coral3}              {rgb}{0.80,0.36,0.27}
\definecolor {coral4}              {rgb}{0.55,0.24,0.18}
\definecolor {tomato2}             {rgb}{0.93,0.36,0.26}
\definecolor {tomato3}             {rgb}{0.80,0.31,0.22}
\definecolor {tomato4}             {rgb}{0.55,0.21,0.15}
\definecolor {orangered2}          {rgb}{0.93,0.25,0.00}
\definecolor {orangered3}          {rgb}{0.80,0.22,0.00}
\definecolor {orangered4}          {rgb}{0.55,0.15,0.00}
\definecolor {red2}                {rgb}{0.93,0.00,0.00}
\definecolor {red3}                {rgb}{0.80,0.00,0.00}
\definecolor {red4}                {rgb}{0.55,0.00,0.00}
\definecolor {deeppink2}           {rgb}{0.93,0.07,0.54}
\definecolor {deeppink3}           {rgb}{0.80,0.06,0.46}
\definecolor {deeppink4}           {rgb}{0.55,0.04,0.31}
\definecolor {hotpink1}            {rgb}{1.00,0.43,0.71}
\definecolor {hotpink2}            {rgb}{0.93,0.42,0.65}
\definecolor {hotpink3}            {rgb}{0.80,0.38,0.56}
\definecolor {hotpink4}            {rgb}{0.55,0.23,0.38}
\definecolor {pink1}               {rgb}{1.00,0.71,0.77}
\definecolor {pink2}               {rgb}{0.93,0.66,0.72}
\definecolor {pink3}               {rgb}{0.80,0.57,0.62}
\definecolor {pink4}               {rgb}{0.55,0.39,0.42}
\definecolor {lightpink1}          {rgb}{1.00,0.68,0.73}
\definecolor {lightpink2}          {rgb}{0.93,0.64,0.68}
\definecolor {lightpink3}          {rgb}{0.80,0.55,0.58}
\definecolor {lightpink4}          {rgb}{0.55,0.37,0.40}
\definecolor {palevioletred1}      {rgb}{1.00,0.51,0.67}
\definecolor {palevioletred2}      {rgb}{0.93,0.47,0.62}
\definecolor {palevioletred3}      {rgb}{0.80,0.41,0.54}
\definecolor {palevioletred4}      {rgb}{0.55,0.28,0.36}
\definecolor {maroon1}             {rgb}{1.00,0.20,0.70}
\definecolor {maroon2}             {rgb}{0.93,0.19,0.65}
\definecolor {maroon3}             {rgb}{0.80,0.16,0.56}
\definecolor {maroon4}             {rgb}{0.55,0.11,0.38}
\definecolor {violetred1}          {rgb}{1.00,0.24,0.59}
\definecolor {violetred2}          {rgb}{0.93,0.23,0.55}
\definecolor {violetred3}          {rgb}{0.80,0.20,0.47}
\definecolor {violetred4}          {rgb}{0.55,0.13,0.32}
\definecolor {magenta2}            {rgb}{0.93,0.00,0.93}
\definecolor {magenta3}            {rgb}{0.80,0.00,0.80}
\definecolor {magenta4}            {rgb}{0.55,0.00,0.55}
\definecolor {orchid1}             {rgb}{1.00,0.51,0.98}
\definecolor {orchid2}             {rgb}{0.93,0.48,0.91}
\definecolor {orchid3}             {rgb}{0.80,0.41,0.79}
\definecolor {orchid4}             {rgb}{0.55,0.28,0.54}
\definecolor {plum1}               {rgb}{1.00,0.73,1.00}
\definecolor {plum2}               {rgb}{0.93,0.68,0.93}
\definecolor {plum3}               {rgb}{0.80,0.59,0.80}
\definecolor {plum4}               {rgb}{0.55,0.40,0.55}
\definecolor {mediumorchid1}       {rgb}{0.88,0.40,1.00}
\definecolor {mediumorchid2}       {rgb}{0.82,0.37,0.93}
\definecolor {mediumorchid3}       {rgb}{0.71,0.32,0.80}
\definecolor {mediumorchid4}       {rgb}{0.48,0.22,0.55}
\definecolor {darkorchid1}         {rgb}{0.75,0.24,1.00}
\definecolor {darkorchid2}         {rgb}{0.70,0.23,0.93}
\definecolor {darkorchid3}         {rgb}{0.60,0.20,0.80}
\definecolor {darkorchid4}         {rgb}{0.41,0.13,0.55}
\definecolor {purple1}             {rgb}{0.61,0.19,1.00}
\definecolor {purple2}             {rgb}{0.57,0.17,0.93}
\definecolor {purple3}             {rgb}{0.49,0.15,0.80}
\definecolor {purple4}             {rgb}{0.33,0.10,0.55}
\definecolor {mediumpurple1}       {rgb}{0.67,0.51,1.00}
\definecolor {mediumpurple2}       {rgb}{0.62,0.47,0.93}
\definecolor {mediumpurple3}       {rgb}{0.54,0.41,0.80}
\definecolor {mediumpurple4}       {rgb}{0.36,0.28,0.55}
\definecolor {thistle1}            {rgb}{1.00,0.88,1.00}
\definecolor {thistle2}            {rgb}{0.93,0.82,0.93}
\definecolor {thistle3}            {rgb}{0.80,0.71,0.80}
\definecolor {thistle4}            {rgb}{0.55,0.48,0.55}
\definecolor {gray1}               {rgb}{0.01,0.01,0.01}
\definecolor {gray2}               {rgb}{0.02,0.02,0.02}
\definecolor {gray3}               {rgb}{0.03,0.03,0.03}
\definecolor {gray4}               {rgb}{0.04,0.04,0.04}
\definecolor {gray5}               {rgb}{0.05,0.05,0.05}
\definecolor {gray6}               {rgb}{0.06,0.06,0.06}
\definecolor {gray7}               {rgb}{0.07,0.07,0.07}
\definecolor {gray8}               {rgb}{0.08,0.08,0.08}
\definecolor {gray9}               {rgb}{0.09,0.09,0.09}
\definecolor {gray10}              {rgb}{0.10,0.10,0.10}
\definecolor {gray11}              {rgb}{0.11,0.11,0.11}
\definecolor {gray12}              {rgb}{0.12,0.12,0.12}
\definecolor {gray13}              {rgb}{0.13,0.13,0.13}
\definecolor {gray14}              {rgb}{0.14,0.14,0.14}
\definecolor {gray15}              {rgb}{0.15,0.15,0.15}
\definecolor {gray16}              {rgb}{0.16,0.16,0.16}
\definecolor {gray17}              {rgb}{0.17,0.17,0.17}
\definecolor {gray18}              {rgb}{0.18,0.18,0.18}
\definecolor {gray19}              {rgb}{0.19,0.19,0.19}
\definecolor {gray20}              {rgb}{0.20,0.20,0.20}
\definecolor {gray21}              {rgb}{0.21,0.21,0.21}
\definecolor {gray22}              {rgb}{0.22,0.22,0.22}
\definecolor {gray23}              {rgb}{0.23,0.23,0.23}
\definecolor {gray24}              {rgb}{0.24,0.24,0.24}
\definecolor {gray25}              {rgb}{0.25,0.25,0.25}
\definecolor {gray26}              {rgb}{0.26,0.26,0.26}
\definecolor {gray27}              {rgb}{0.27,0.27,0.27}
\definecolor {gray28}              {rgb}{0.28,0.28,0.28}
\definecolor {gray29}              {rgb}{0.29,0.29,0.29}
\definecolor {gray30}              {rgb}{0.30,0.30,0.30}
\definecolor {gray31}              {rgb}{0.31,0.31,0.31}
\definecolor {gray32}              {rgb}{0.32,0.32,0.32}
\definecolor {gray33}              {rgb}{0.33,0.33,0.33}
\definecolor {gray34}              {rgb}{0.34,0.34,0.34}
\definecolor {gray35}              {rgb}{0.35,0.35,0.35}
\definecolor {gray36}              {rgb}{0.36,0.36,0.36}
\definecolor {gray37}              {rgb}{0.37,0.37,0.37}
\definecolor {gray38}              {rgb}{0.38,0.38,0.38}
\definecolor {gray39}              {rgb}{0.39,0.39,0.39}
\definecolor {gray40}              {rgb}{0.40,0.40,0.40}
\definecolor {gray42}              {rgb}{0.42,0.42,0.42}
\definecolor {gray43}              {rgb}{0.43,0.43,0.43}
\definecolor {gray44}              {rgb}{0.44,0.44,0.44}
\definecolor {gray45}              {rgb}{0.45,0.45,0.45}
\definecolor {gray46}              {rgb}{0.46,0.46,0.46}
\definecolor {gray47}              {rgb}{0.47,0.47,0.47}
\definecolor {gray48}              {rgb}{0.48,0.48,0.48}
\definecolor {gray49}              {rgb}{0.49,0.49,0.49}
\definecolor {gray50}              {rgb}{0.50,0.50,0.50}
\definecolor {gray51}              {rgb}{0.51,0.51,0.51}
\definecolor {gray52}              {rgb}{0.52,0.52,0.52}
\definecolor {gray53}              {rgb}{0.53,0.53,0.53}
\definecolor {gray54}              {rgb}{0.54,0.54,0.54}
\definecolor {gray55}              {rgb}{0.55,0.55,0.55}
\definecolor {gray56}              {rgb}{0.56,0.56,0.56}
\definecolor {gray57}              {rgb}{0.57,0.57,0.57}
\definecolor {gray58}              {rgb}{0.58,0.58,0.58}
\definecolor {gray59}              {rgb}{0.59,0.59,0.59}
\definecolor {gray60}              {rgb}{0.60,0.60,0.60}
\definecolor {gray61}              {rgb}{0.61,0.61,0.61}
\definecolor {gray62}              {rgb}{0.62,0.62,0.62}
\definecolor {gray63}              {rgb}{0.63,0.63,0.63}
\definecolor {gray64}              {rgb}{0.64,0.64,0.64}
\definecolor {gray65}              {rgb}{0.65,0.65,0.65}
\definecolor {gray66}              {rgb}{0.66,0.66,0.66}
\definecolor {gray67}              {rgb}{0.67,0.67,0.67}
\definecolor {gray68}              {rgb}{0.68,0.68,0.68}
\definecolor {gray69}              {rgb}{0.69,0.69,0.69}
\definecolor {gray70}              {rgb}{0.70,0.70,0.70}
\definecolor {gray71}              {rgb}{0.71,0.71,0.71}
\definecolor {gray72}              {rgb}{0.72,0.72,0.72}
\definecolor {gray73}              {rgb}{0.73,0.73,0.73}
\definecolor {gray74}              {rgb}{0.74,0.74,0.74}
\definecolor {gray75}              {rgb}{0.75,0.75,0.75}
\definecolor {gray76}              {rgb}{0.76,0.76,0.76}
\definecolor {gray77}              {rgb}{0.77,0.77,0.77}
\definecolor {gray78}              {rgb}{0.78,0.78,0.78}
\definecolor {gray79}              {rgb}{0.79,0.79,0.79}
\definecolor {gray80}              {rgb}{0.80,0.80,0.80}
\definecolor {gray81}              {rgb}{0.81,0.81,0.81}
\definecolor {gray82}              {rgb}{0.82,0.82,0.82}
\definecolor {gray83}              {rgb}{0.83,0.83,0.83}
\definecolor {gray84}              {rgb}{0.84,0.84,0.84}
\definecolor {gray85}              {rgb}{0.85,0.85,0.85}
\definecolor {gray86}              {rgb}{0.86,0.86,0.86}
\definecolor {gray87}              {rgb}{0.87,0.87,0.87}
\definecolor {gray88}              {rgb}{0.88,0.88,0.88}
\definecolor {gray89}              {rgb}{0.89,0.89,0.89}
\definecolor {gray90}              {rgb}{0.90,0.90,0.90}
\definecolor {gray91}              {rgb}{0.91,0.91,0.91}
\definecolor {gray92}              {rgb}{0.92,0.92,0.92}
\definecolor {gray93}              {rgb}{0.93,0.93,0.93}
\definecolor {gray94}              {rgb}{0.94,0.94,0.94}
\definecolor {gray95}              {rgb}{0.95,0.95,0.95}
\definecolor {gray97}              {rgb}{0.97,0.97,0.97}
\definecolor {gray98}              {rgb}{0.98,0.98,0.98}
\definecolor {gray99}              {rgb}{0.99,0.99,0.99}
\definecolor {darkgrey}            {rgb}{0.66,0.66,0.66}

\newcommand{\resp}[1]{[resp. #1]}

\newcommand{\TODO}[1]{{}}

\newcommand{\ignore}[1]{}
\newcommand{\todo}[1] {}

\newcommand{\marg}[1]{\marginpar{\ \\{\em #1\/}}}
\renewcommand{\marg}[1]{\marginpar{\ \\\textcolor{blue}{\ {\sf #1\/}}}}

 \newcommand{\ignoreinshort}[1]{}
 \newcommand{\ignoreinlong}[1]{{#1}}

\def\makenewenumerate#1#2{%
\newcounter{cnt#1}
\newenvironment{#1}%
{\begin{list}{\makebox[0pt][r]{#2}}%
{\setlength{\itemsep}{0pt}%
 \setlength{\parsep}{.2em}%
 \setlength{\leftmargin}{1.5em}%
 \setlength{\labelwidth}{.4em}%
 \usecounter{cnt#1}}}
{\end{list}}}

\makenewenumerate{myenumerate}{\arabic{cntmyenumerate}.}

\makenewenumerate{renumerate}{\rm(\roman{cntrenumerate})}
\makenewenumerate{renumerateprime}{\rm(\roman{cntrenumerateprime}$'$)}
\makenewenumerate{renumeratesecond}{\rm(\roman{cntrenumeratesecond}$''$)}

\makenewenumerate{aenumerate}{({\it\alph{cntaenumerate}})}
\makenewenumerate{aenumerateprime}{({\it\alph{cntaenumerateprime}$'$})}
\makenewenumerate{aenumeratesecond}{({\it\alph{cntaenumeratesecond}$''$})}

\def\newplaintheorem#1#2{%
\newtheorem{#1plain}{#2}[section]%
\newenvironment{#1}{\begin{#1plain}\rm }{\end{#1plain}}}

\newplaintheorem{notation}{Notation}
\newplaintheorem{cexample}{Counter-Example}

\newcommand{\sref}[1]{\S{}\ref{#1}}

\newcommand{\pair}[2]{\ensuremath{\langle{#1},{#2}\rangle}\xspace}

\newcommand{\tuple}[1]{\ensuremath{\langle{#1}\rangle}\xspace}
\newcommand{\set}[1]{\ensuremath{\{{#1}\}}\xspace}

\newcommand{\defas}{\ensuremath{\stackrel{\text{\tiny def}}{=}}\xspace}
\newcommand{\thus}{\ensuremath{\Longrightarrow}\xspace}

\newcommand\cala{\ensuremath{\mathcal{A}}\xspace}

\newcommand\calm{\ensuremath{\mathcal{M}}\xspace}

\newcommand{\false}{\bot}

\newcommand{\omt}{\ensuremath{\text{OMT}}\xspace}

\newcommand{\lb}{\lbgen{}{}}
\newcommand{\ub}{\ubgen{}{}}

\newcommand{\pivot}{\ensuremath{\mathsf{pivot}}\xspace}

\newcommand{\cost}{\ensuremath{{cost}}\xspace}

\newcommand{\mincost}{\ensuremath{{mincost}}\xspace}

\newcommand\mysout{\bgroup \markoverwith{{-}}\ULon}
\newcommand\nosout{\bgroup \markoverwith{{ }}\ULon}
\definecolor{mygray}{rgb}{0.90,0.90,0.90}
\definecolor{mywhite}{rgb}{1.00,1.00,1.00}

\renewcommand{\mincost}{\ensuremath{\mathsf{min}_\cost}\xspace}

\newcommand{\optimathsat}{\textsc{OptiMathSAT}\xspace}

\newcommand{\satres}{\textsc{sat}\xspace}
\newcommand{\unsatres}{\textsc{unsat}\xspace}

\newcommand{\vi}{\ensuremath{\varphi}\xspace}

\newcommand{\T}{\ensuremath{\mathcal{T}}\xspace}

\newcommand{\smt}{SMT\xspace}

\newcommand{\bv}{\ensuremath{\mathcal{BV}}\xspace}

\newcommand{\smtbv}{\smttt{\bv}}

\newcommand{\mathsat}{\textsc{MathSAT}\xspace}

\newcommand{\zthree}{\textsc{Z3}\xspace}

\newcommand{\mathsatfive}{\textsc{MathSAT5}\xspace}

\ifthenelse{\equal{\longversion}{true}}
{%
  \renewcommand{\ignoreinshort}[1]{{\textcolor{black}{#1}}}
  \renewcommand{\ignoreinlong}[1]{}
  \specialcomment{IGNOREINSHORT}{\begingroup\color{black}}{\endgroup}
  \excludecomment{IGNOREINLONG}
}%
{%
 \renewcommand{\ignoreinshort}[1]{}
 \renewcommand{\ignoreinlong}[1]{#1}
 \excludecomment{IGNOREINSHORT}
  \specialcomment{IGNOREINLONG}{\begingroup\color{black}}{\endgroup}
}
 \excludecomment{IGNORE}

\newcommand{\smtlib}{\textsc{SMT-LIBv2}}

\renewcommand{\cost}{\ensuremath{{obj}}\xspace}

\newcommand{\obvbs}{\textsc{obv-bs}\xspace}

\newcommand{\ofpbs}{\textsc{ofp-bs}\xspace}

\newcommand{\fp}{\ensuremath{FP}\xspace}

\newcommand{\cdcl}{CDCL\xspace}
\newcommand{\sat}{SAT\xspace}

\newcommand{\lra} {\ensuremath{\mathcal{LRA}}\xspace}

\renewcommand{\bv}  {\ensuremath{\mathcal{BV}}\xspace}
\renewcommand{\fp}  {\ensuremath{\mathcal{FP}}\xspace}

\newcommand{\lracupt} {\ensuremath{\lra\,\cup\,\T}\xspace}

\newcommand{\fpcupt}  {\ensuremath{\fp\,\cup\,\T}\xspace}

\renewcommand{\smtbv}{\ensuremath{\text{SMT}(\bv)}\xspace}
\newcommand{\smtfp}  {\ensuremath{\text{SMT}(\fp)}\xspace}

\newcommand{\omtbv}  {\ensuremath{\text{OMT}(\bv)}\xspace}
\newcommand{\omtfp}  {\ensuremath{\text{OMT}(\fp)}\xspace}

\newcommand{\omtlracupt} {\ensuremath{\text{OMT}(\lracupt)}\xspace}

\newcommand{\omtfpcupt}  {\ensuremath{\text{OMT}(\fpcupt)}\xspace}

\newcommand{\restr}[2]{\ensuremath{[\![#1]\!]_{#2}}\xspace}

\newcommand{\costbit}[1]{\ensuremath{\cost[#1]}\xspace}
\newcommand{\costbits}{\ensuremath{[\costbit{0},\,...,\costbit{n-1}]}\xspace}

\newcommand{\attr}{\ensuremath{attr}\xspace}
\newcommand{\attrbit}[1]{\ensuremath{\attr[#1]}\xspace}
\newcommand{\attrbits}{\ensuremath{[\attrbit{0},\,...,\attrbit{n-1}]}\xspace}

\newcommand{\dattr}[1]{\ensuremath{\attr_{#1}}\xspace}
\newcommand{\dattrbit}[2]{\ensuremath{\dattr{#1}[#2]}\xspace}

\newcommand{\attreq}[1]{\ensuremath{\ifthenelse{\equal{#1}{}}{A}{A[#1]}}\xspace}
\newcommand{\dattreq}[2]{\ensuremath{\ifthenelse{\equal{#2}{}}{A_{#1}}{A_{#1}[#2]}}\xspace}

\newcommand{\dcala}[1]{\ensuremath{\ifthenelse{\equal{#1}{}}{\cala_{\varphi}}{\cala_{\varphi}[#1]}}\xspace}

\newcommand{\nan}{\textsc{NaN}\xspace}

\newcommand{\minidx}{\ensuremath{m}\xspace}

\newcommand{\mincostphi}{\ensuremath{\mincost(\varphi)}\xspace}

\newcommand{\minusinf}{\ensuremath{\mathtt{-\infty}}\xspace}
\newcommand{\plusinf}{\ensuremath{\mathtt{+\infty}}\xspace}

\renewcommand{\ub}{{\sf ub}\xspace}
\renewcommand{\lb}{{\sf lb}\xspace}
\renewcommand{\cost}{{\sf obj}\xspace}

\newenvironment{ptchange}{\color{darkviolet}}{\normalcolor}

\lstdefinestyle{box}{
    backgroundcolor=\transparent{0.25}\color{lightgray},
    framexleftmargin=0pt,
    framexrightmargin=0pt,
    framextopmargin=0pt,
    framexbottommargin=0pt,
    frame=tblr,
    framerule=0.25pt,
    basicstyle=\small\ttfamily
}

\lstdefinestyle{bigbox}{
    backgroundcolor=\transparent{0.25}\color{lightgray},
    framexleftmargin=0pt,
    framexrightmargin=0pt,
    framextopmargin=0pt,
    framexbottommargin=0pt,
    frame=tblr,
    framerule=0.25pt,
    basicstyle=\footnotesize\ttfamily,
    numbers=left,
    stepnumber=1,
    numberstyle=\footnotesize\ttfamily,
    numbersep=10pt,
    showstringspaces=false,
    tabsize=1,
    breaklines=true,
    breakatwhitespace=false,
    xleftmargin=3em,
    framexleftmargin=2.5em
}

\lstdefinestyle{numbers}{
    numbers=left,
    stepnumber=1,
    numberstyle=\footnotesize\ttfamily,
    numbersep=10pt,
    showstringspaces=false,
    tabsize=1,
    breaklines=true,
    breakatwhitespace=false,
    xleftmargin=3em,
    framexleftmargin=2.5em
}

\lstdefinestyle{escape}{
    escapeinside={(*}{*)}
}

\lstnewenvironment{pseudocode}{\lstset{language=C++,style=box,style=numbers,style=escape}} {}

\crefformat{footnote}{#2\footnotemark[#1]#3}

\renewcommand\algorithmiccomment[1]{%
  \hfill\#\ \eqparbox{COMMENT}{#1}%
}

\makeatletter
\patchcmd{\algorithmic}{\addtolength{\ALC@tlm}{\leftmargin} }{\addtolength{\ALC@tlm}{\leftmargin}}{}{}
\makeatother

\SetCommentSty{mycommfont}

\ifthenelse{\equal{\withproofs}{true}}
{%
\specialcomment{PROOF}{\begingroup\color{black}}{\endgroup}
\excludecomment{APPENDIX}
}%
{%
\specialcomment{APPENDIX}{\begingroup\color{black}}{\endgroup}
\excludecomment{PROOF}
}

\newcommand{\un}[1]{\underline{#1}}
\renewcommand{\thus}{\Rightarrow}
\newcommand{\isnan}[1]{\mathsf{IsNaN({\ensuremath #1})}\xspace}
\newcommand{\vinonan}{{\ensuremath{\varphi}_{\mathsf{noNaN}}\xspace}}

\renewcommand{\marg}[1]{}

\begin{document}

\pagestyle{plain}
\pagenumbering{arabic}

\begin{IGNOREINLONG}
\title{Optimization Modulo the Theory of Floating-Point Numbers
\thanks{
We would like to thank the anonymous reviewers
for their insightful comments and suggestions, and 
we thank Alberto Griggio for support with \mathsatfive{} code.
}}
\end{IGNOREINLONG}
\begin{IGNOREINSHORT}
\title{Optimization Modulo the Theories of Signed Bit-Vectors and Floating-Point Numbers
\thanks{
We would like to thank the anonymous reviewers
for their insightful comments and suggestions, and 
we thank Alberto Griggio for support with \mathsat{} code.
}}
\end{IGNOREINSHORT}

\author{Patrick Trentin \and Roberto Sebastiani }
\institute{DISI, University of Trento, Italy}

\maketitle


\begin{abstract}
Optimization Modulo Theories (OMT) is an important extension of SMT which
allows for finding models that optimize given objective functions, 
typically consisting in linear-arithmetic or pseudo-Boolean terms.
%
However, many SMT and OMT applications, in particular from SW
and HW verification, 
require handling {\em bit-precise} representations
of numbers, which in SMT are handled by means of the theory of Bit-Vectors
(\bv) for the integers and that of Floating-Point Numbers (\fp)
for the reals respectively.
Whereas an approach for OMT with (unsigned) \bv{} has been proposed
by Nadel \& Ryvchin, unfortunately we are not aware of any existing
approach for OMT with \fp.

In this paper we fill this gap. We present a novel OMT approach, based
on the novel concept of {\em attractor} and {\em dynamic attractor},
which extends the work of Nadel \& Ryvchin to signed \bv and, most
importantly, to \fp. We have implemented some  \omtbv{} and \omtfp{}
procedures on top of \optimathsat{} and tested the latter ones on
modified problems from the SMT-LIB repository.
The empirical results support the validity and feasibility of the novel
approach.

\end{abstract}

\renewcommand{\abstractname}{\ackname}


\section{Introduction}
\label{sec:intro}
Optimization Modulo Theories (\omt)
\ignoreinshort{\cite{nieuwenhuis_sat06,cimattifgss10,oven11,%
st-ijcar12,dilligdma12,cgss_sat13_maxsmt,manoliosp13,li_popl14,bjorner_scss14,%
LarrazORR14,st_tocl14,st_cav15,st_tacas15,araujo16,NadelR16,st_tacas17,albuquer17,%
st_jar18,fazekasbb18,puli18,araujo18}}
\ignoreinlong{\cite{nieuwenhuis_sat06,cimattifgss10,%
li_popl14,bjorner_scss14,%
LarrazORR14,st_tocl14,st_tacas15,NadelR16,%
st_jar18,fazekasbb18}}
is an important extension to Satisfiability Modulo
Theories which allows for finding models that optimize one or more
objectives,  which typically consist in some linear-arithmetic or Pseudo-Boolean
function application.

However, many SMT and OMT applications, in particular from SW
and HW verification, 
require handling {\em bit-precise} representations
of numbers, which in SMT are handled by means of the theory of Bit-Vectors
(\bv) for the integers and that of Floating-Point Numbers (\fp)
for the reals respectively.
(For instance, during the verification process of a piece of software,
one may look for the minimum/maximum value of some {\tt int} \resp{{\tt
  float}} parameter causing an \smtbv \resp{\smtfp} call to return
\satres ---which typically corresponds to the presence of some bug--- so
that to guarantee a safe range for such parameter. ) 

OMT for the theory of (unsigned) bit-vectors (\omtbv{}) was proposed by 
Nadel and Ryvchin \cite{NadelR16}, although a reduction to the problem
to MaxSAT was already implemented in the SMT/OMT solver \zthree{} \cite{bjorner_tacas15}.
The work in \cite{NadelR16} was based on the observation that
\omt{} on unsigned \bv can be seen as lexicographic optimization
over the bits in the bitwise representation of the objective, ordered
from the most-significant bit (MSB) to the least-significant 
bit (LSB).

\smallskip
In this paper we address ---for the first time to the best of our
knowledge--- \omt{} for the theory of signed Bit-Vectors and, most
importantly, for the theory of Floating-Point Arithmetic (\omtfp), by
exploiting some properties of the two's complement encoding for signed
\bv and of the IEEE 754-2008 encoding for \fp respectively. 

We start from introducing the notion of {\em attractor}, which
represent (the bitwise encoding of) the target value for
the objective which the optimization process aims at. This
allows us for easily leverage the procedure of \cite{NadelR16} to work
with both {\em signed} and {\em unsigned} Bit-Vectors, by minimizing
lexicographically the bitwise distance between the objective and the
attractor, that is, by minimizing lexicographically the bitwise-xor between
the objective and the attractor.

Unfortunately there is no such notion of (fixed) attractor for \fp numbers,
because the target value moves as long as the bits of the objective
are updated from the MSB to the LSB, and the optimization process may
have to change dynamically its aim, even at the opposite direction.
(For instance, as soon as the minimization
process realizes there is no solution with a negative value for the
objective and thus sets its MSB to 0, the target value is switched from
$-\infty$ to $0+$, and the search switches direction, from the maximization
of the exponent and the significand to their minimization.)

To cope with this fact, we introduce the notions of {\em dynamic
  attractor} and {\em attractor trajectory}, representing the dynamics
of the moving target value, which are progressively updated as soon as
the bits of the objective are updated from the MSB to the LSB. Based
on these ideas, we present novel \omtfp procedures, which require at
most $n+2$, incremental calls to an \smtfp solver, $n$ being the
number of bits in the representation of the objective.
Notice that  these procedures do not depend on the underlying \smtfp
procedure used, provided the latter allows for accessing and setting the
single bits of the objective. 

We have implemented these \omtbv and \omtfp procedures on top of the
\optimathsat \omt solver \cite{st_jar18}.
We have run an experimental evaluation of the \omtfp  procedures on modified
\smtfp problems from the SMT-LIB library.
The empirical results support the validity and feasibility of the novel
approach.

The rest of the paper is organized as follows.
In \sref{sec:background} we provide the necessary background on \bv
and \fp theories and reasoning. 
In \sref{sec:tf} we provide the novel theoretical definitions and
results.
In \sref{sec:procedures} we describe our novel \omtfp procedures.
In \sref{sec:expeval} we present the empirical evaluation.
In \sref{sec:concl} we conclude, hinting some future directions.
\begin{IGNOREINLONG}
The proofs of the theoretical results from \sref{sec:tf} are made
in the extended version of this paper \cite{st_cade19_extended}.
\end{IGNOREINLONG}

\section{Background}
\label{sec:background}
We assume some basic knowledge  on \sat and \smt and briefly introduce
the reader to the Bit-Vector and Floating-Point theories.


\paragraph{Bit-Vectors.}

A {\em bit} is a Boolean variable that can be interpreted as $0$ or $1$.
A Bit-Vector (\bv) variable $\mathbf{v}^{[n]}$ is a vector of $n$ bits,
where $v[0]$ is the Most Significant Bit (MSB)
and $v[n-1]$ is the Least Significant Bit (LSB).%
\footnotemark{}%
\footnotetext{\label{footnote:msbtolsb}
Although most often in the literature the indexes $i\in[0,...,n-1]$ use
to grow from the LSB to the MSB, in this paper we use the opposite
notation because we always reason from the MSB down to the LSB, so
that to much simplify the explanation.
}
A \bv constant of width $n$ is an interpreted vector of $n$ values in
$\set{0, 1}$. We $\overline{overline}$ a bit value or a \bv value to denote its
  complement (e.g., $\overline{[11010010]}$ is $[00101101]$).
A \bv variable/constant of width $n$ can be \emph{unsigned}, in which case
its domain is $[0, 2^n - 1]$, or \emph{signed}, which we assume to
comply with the \emph{Two's complement} representation, so that  its domain is
$[-2^{(n - 1)},2^{(n - 1)}-1]$.
Therefore, the vector $[11111111]$ can be interpreted either as the unsigned \bv
constant $\mathbf{255}^{[8]}$ or as the signed \bv constant $\mathbf{-1}^{[8]}$.
Following the \smtlib{} standard \cite{smtlib_two_url}, we
may also represent a \bv constant in \emph{binary} (e.g. $\mathbf{28}^{[8]}$ is written
$\#b00011100$) or in \emph{hexadecimal} (e.g. $\mathbf{28}^{[8]}$ is written $\#x1C$) form.
A \bv term is built from \bv constants, variables and interpreted \bv functions
which represent standard RTL operators: word concatenation (e.g.
$\mathbf{3}^{[8]} \circ \mathbf{x}^{[8]}$), sub-word selection (e.g.
$(\mathbf{3}^{[8]}[6:3])^{[4]}$), modulo-n sum and multiplication (e.g.
$\mathbf{x}^{[8]} +_{8} \mathbf{y}^{[8]}$ and 
$\mathbf{x}^{[8]} \cdot_{8} \mathbf{y}^{[8]}$), bit-wise operators
(like, e.g., $\textbf{and}_n$, $\textbf{or}_n$, $\textbf{xor}_n$, $\textbf{nxor}_n$,
$\textbf{not}_n$), left and right shift ${<<}_n$, ${>>}_n$.
A \bv atom can be built by combining \bv terms with interpreted predicates like
$\geq_n$, $<_n$ (e.g. $\mathbf{0}^{[8]} \geq_8 \mathbf{x}^{[8]}$) and equality.
We refer the reader to
\ignoreinshort{\cite{smtlib_two_url,Hadarean15}}
\ignoreinlong{\cite{smtlib_two_url}}
for further details on
the syntax and semantics of Bit-Vector theory.

There are two main techniques for \bv satisfiability, the ``{\em eager}''
and the ``{\em lazy}'' approach, which are substantially complementary
to one another \cite{HadareanBJBT14}.
In the {\em eager} approach, \bv terms and constraints are encoded into \sat
via bit-blasting
\ignoreinshort{\cite{dblp:conf/cav/ganeshd07,bb09,brummayer-phd-09,%
Hadarean15,npbf15,niemetz-phd-17}.}
\ignoreinlong{\cite{dblp:conf/cav/ganeshd07,bb09,%
npbf15}.}
In the {\em lazy} approach, \bv terms are not immediately expanded
--so to avoid any scalability issue-- and the \bv solver is comprised by
a layered set of techniques, each of which deals with a sub-portion of the
\bv theory
\ignoreinshort{\cite{brinkmanndrechsler:aspdac2002,encoding_pdpar05,bruttomessocfghnps07,Hadarean15}.}
\ignoreinlong{\cite{brinkmanndrechsler:aspdac2002,encoding_pdpar05,bruttomessocfghnps07}.}


\paragraph{Floating-Point.}

The theory of \emph{Floating-Point Numbers} (\fp), \cite{smtlib_two_url,%
pt_smt10,BrainTRW15}, is based on the IEEE standard 754-2008 \cite{ieee754}
for floating-point arithmetic, restricted to the binary case.
A \fp sort is an indexed nullary sort identifier of the form
\texttt{(\_ FP <$ebits$> <$sbits$>)} s.t. both $ebits$ and $sbits$
are positive integers greater than one, $ebits$ defines the number
of bits in the exponent and $sbits$ defines the number of bits in
the significand, including the hidden bit.
A \fp variable $\mathbf{v}^{[n]}$ with sort {\tt (\_ FP <$ebits$> <$sbits$>)}
can be indifferently viewed as a vector of $n \defas ebits + sbits$ bits,
where $v[0]$ is the Most Significant Bit (MSB)
and $v[n-1]$ is the Least Significant Bit (LSB), or
as a triplet of Bit-Vectors
$\langle \mathbf{sign}, \mathbf{exp}, \mathbf{sig} \rangle$
s.t. $\mathbf{sign}$ is a \bv of size $1$, $\mathbf{exp}$ is a
\bv of size $ebits$ and $\mathbf{sig}$ is a \bv of size $sbits - 1$.
A \fp constant is a triplet of \bv{} constants.
Given a fixed floating-point sort, i.e. a pair $\pair{ebits}{sbits}$, the following
\fp constants are implicitly defined:\\

\begin{centering}
\begin{tabular}{lll}
{\bf value}           & {\bf Symbol}                            & {\bf \bv Repr.} \\
\emph{plus infinity}  & \texttt{(\_ +oo <$ebits$> <$sbits$>)}   & \texttt{(fp \#b0 \#b1...1 \#b0...0)} \\
\emph{minus infinity} & \texttt{(\_ -oo <$ebits$> <$sbits$>)}   & \texttt{(fp \#b1 \#b1...1 \#b0...0)} \\
\emph{plus zero}      & \texttt{(\_ +zero <$ebits$> <$sbits$>)} & \texttt{(fp \#b0 \#b0...0 \#b0...0)} \\
\emph{minus zero}     & \texttt{(\_ -zero <$ebits$> <$sbits$>)} & \texttt{(fp \#b1 \#b0...0 \#b0...0)} \\
\emph{not-a-number}   & \texttt{(\_ NaN <$ebits$> <$sbits$>)}   & \texttt{(fp t \#b1...1 s)} \\
\end{tabular}
\end{centering}
\\

\noindent
where $\texttt{t}$ is either $0$ or $1$ and $\texttt{s}$ is a \bv which contains at least a $1$.

Setting aside special \fp constants, the remaining \fp values can be classified
to be either normal or subnormal (a.k.a. denormal) \cite{ieee754}. A \fp number
is said to be \emph{subnormal} when every bit in its exponent is equal to zero,
and \emph{normal} otherwise.
The significand of a normal \fp number is always interpreted as if the leading
binary digit is equal $1$, while for denormalized \fp values the leading binary
digit is always $0$. This allows for the representation of numbers that are closer
to zero, although with reduced precision.

\begin{example}
Let $x$ be the normal \fp constant \texttt{(\_ FP \#b0 \#b1100 \#b0101000)}, and
$y$ be the subnormal \fp constant \texttt{(\_ FP \#b0 \#b0000 \#b0101000)}, so that
their corresponding sort is \texttt{(\_ FP <4> <8>)}. Then, according to the semantics
defined in the IEEE standard 754-2008 \cite{ieee754}, the floating-point value of $x$ and $y$
in decimal notation is given by:
\begin{align*}
x = &  \:\: (-1)^{0} \cdot 2^{(12 - 7)} \cdot \bigg( 1 + \sum_{i=1}^{7} \Big(x[4+i] \cdot 2^{-i}\Big) \bigg)
  = 1 \cdot 2^5 \cdot \bigg( 1 + \frac{1}{2^2} + \frac{1}{2^4} \bigg)
  = 42 \\
y = &  \:\: (-1)^{0} \cdot 2^{(0 - 7 \blue{+ 1})} \cdot \bigg( \blue{0} + \sum_{i=1}^{7} \Big(y[4+i] \cdot 2^{-i}\Big) \bigg)
  = 1 \cdot 2^{-6} \cdot \bigg( \frac{1}{2^2} + \frac{1}{2^4} \bigg)
      = \frac{5}{2^{10}}.
\hfill \diamond      
\end{align*}%
\end{example}


The theory of \fp provides a variety of built-in floating-point operations as
defined in the IEEE standard 754-2008.
This includes binary arithmetic operations (e.g. $+, -, \star, \div$), basic unary
operations (e.g. $abs, -$), binary comparison operations (e.g. $\leq, <, \neq, =, >, \geq$),
the remainder operation, the square root operation and more.
Importantly, arithmetic operations are performed \emph{as if with infinite
precision}, but the result is then \emph{rounded} to the ``nearest''
representable \fp number according to the specified \emph{rounding mode}.
Five \emph{rounding modes} are made available, as in \cite{ieee754}.


The most common approach for \fp-satisfiability is to encode \fp expressions into
\bv formulas based on the circuits used to implement floating-point operations,
using appropriate under- and over-approximation schemes --or a mixture of both--
to improve performance
\ignoreinshort{\cite{wahl09,rummer14,rummer17,rummer18}.}
\ignoreinlong{\cite{wahl09,rummer14,rummer18}.}
Then, the \bv{}-Solver is used to deal with the \fp formula, using either the
\emph{eager} or the \emph{lazy} \bv{} approach.
An alternative approach, based on \emph{abstract interpretation}, is presented in
\ignoreinshort{\cite{braindghk13,braindghk14,fmcad12-float}.}
\ignoreinlong{\cite{braindghk13,braindghk14}.}
With this technique, called \emph{Abstract \cdcl} (ACDCL), the set of
feasible solutions is over-approximated with floating-point intervals, so that
intervals-based conflict analysis is performed to decide \fp-satisfiability.


\section{Theoretical Framework}
\label{sec:tf}

We present our generalization of \cite{NadelR16} to the case of
signed/unsigned Bit-Vector Optimization, and then move on to deal with
Floating-Point Optimization.


    \subsection{Bit-Vector Optimization}
    \label{sec:tf-bv}
    Without any loss of generality, we assume that every objective function
$f(...)$ is replaced by a variable $\cost$ of the same type by conjoining
``$\cost = f(...)$'' to the input formula.
We use the symbol $n$ to denote the bit-width of $\cost$, and $\costbit{i}$
to denote the $i$-th bit of $\cost$, where $\costbit{0}$ and $\costbit{n-1}$
are the Most Significant Bit (MSB) and the Least Significant Bit (LSB) of
$\cost$ respectively.\footref{footnote:msbtolsb}

\ignoreinshort{We define the {\it Bit-Vector Optimization problem}
as follows.}

\begin{definition}{(\omtbv).}
Let $\varphi$ be a 
\smtbv{} formula and \cost be a
--signed or unsigned-- \bv variable occurring in $\varphi$.
We call an {\bf Optimization Modulo \bv problem, \omtbv}, the problem
of finding a model $\calm$ for $\varphi$ (if any) whose value of \cost,
denoted with \mincostphi, is minimum wrt. the total order relation $\leq_n$
for signed \bv{}s if \cost is signed, and the one for unsigned \bv{}s otherwise. 
(The dual definition where we look for the {\it maximum} follows straightforwardly)
\end{definition}

Hereafter, we generalize the unsigned \bv maximization procedures described
in \cite{NadelR16} to the case of signed and unsigned \bv optimization.
To this extent, we introduce the novel notion of \bv \emph{attractor}.


\begin{definition}{(Attractor, attractor equalities).}
\label{def:attractor}
When minimizing \resp{maximizing}, we call
{\bf attractor}  for \cost 
the smallest \resp{greatest} \bv-value \attr of the sort of \cost.
We call  {\bf vector of attractor
    equalities}  the vector $\attreq{}$ s.t. 
$\attreq{k} \defas (\costbit{k} = \attrbit{k})$, 
$k\in [0..n-1]$.
\end{definition}


\begin{example}
If $\cost^{[8]}$ is an {\em unsigned} \bv objective of width $8$,
then its corresponding attractor \attr is $\mathbf{0}^{[8]}$, i.e. 
$[00000000]$, when $\cost^{[8]}$ is minimized and it is 
$\mathbf{255}^{[8]}$, i.e. $[11111111]$, when $\cost^{[8]}$ is
maximized.
When $\cost^{[8]}$ is instead a {\em signed} \bv objective, following
the {two's complement} encoding, the corresponding \attr is
$\mathbf{-128}^{[8]}$, i.e. $[10000000]$, for minimization and
$\mathbf{127}^{[8]}$, i.e. $[01111111]$, for maximization.
\hfill $\diamond$
\end{example}

In essence, the \emph{attractor} can be seen as the target value of the
optimization search and therefore it can be used to determine the
desired improvement direction and to guide the decisions taken by the
optimization search.
By construction, if a model \calm satisfies all equalities $\attreq{i}$,
then $\calm(\cost) = \attr$.


\begin{IGNOREINSHORT}

More in general, if $\calm$ is a model of $\varphi$, then the value of $\cost$ in
$\calm$, denoted with $\calm(\cost)$, is given by

\begin{equation}
\label{eq:bv-unsigned}
\tau(\cost) = \:\: \sum_{i=0}^{i=n-1} (2^{n-1-i} \cdot \textsc{ite}(\calm(\attreq{i}), \attrbit{i}, \overline{\attrbit{i}}))
\end{equation}

\noindent
when $\cost$ is an {\em unsigned} \bv objective, and by

\begin{equation}
\label{eq:bv-signed}
\begin{aligned}
\tau(\cost) = & \:\: \sum_{i=1}^{i=n-1} (2^{n-1-i} \cdot \textsc{ite}(\calm(\attreq{i}), \attrbit{i}, \overline{\attrbit{i}})) \\
& \:\: - (2^{n-1}) \cdot \textsc{ite}(\calm(\attreq{0}), \attrbit{0}, \overline{\attrbit{0}})
\end{aligned}
\end{equation}

\noindent
when $\cost$ is a {\em signed} \bv objective, using the \emph{two's complement} representation.
The function $\textsc{ite}$, appearing in both previous equations, returns $\attrbit{i}$
if the \emph{attractor equality} $\attreq{i}$ is true in $\calm$ and $\overline{\attrbit{i}}$
otherwise.

\end{IGNOREINSHORT}

We use the symbol $\mu_k$ to denote a generic (possibly partial)  assignment
which assigns at least the $k$ most-significant bits of $\cost$.
We use the symbol $\tau_k$ to denote an assignment to all and only
the $k$ most-significant bits of $\cost$.
Given $i<k$, we denote by $\mu_k[i]$ \resp{$\tau_k[i]$} the value in
\set{0,1} assigned to 
$\cost[i]$ by $\mu_k$ \resp{$\tau_k$}.
Moreover, we use the expression $\restr{\mu_k}{i}$ where $i \leq k$ to denote the restriction
of $\mu_k$ to all and only the $i$ most-significant bits of $\cost$, $\costbit{0},...,\costbit{i-1}$.
Given a model $\calm$ of $\varphi$ and a variable $v$, we denote by
$\calm(v)$ the evaluation of $v$ in $\calm$. 
With a little abuse of notation, and when this does not
  cause ambiguities, we sometimes use an
  attractor equality $\attreq{i} \defas (\costbit{i} = \attrbit{i})$
to denote the single-bit assignment $\costbit{i} := \attrbit{i}$ and its
negation $\neg\attreq{i}$ to denote the assignment to the complement value 
$\costbit{i} := \overline{\attrbit{i}}$.


\begin{definition}(lexicographic maximization)
\label{def:lexoptimizes}
Consider an \omt instance $\pair{\varphi}{\cost}$ and 
the vector of attractor equalities $\attreq{}$.
We say that an assignment $\tau_n$ 
to \cost{} {\bf lexicographically maximizes $\attreq{}$
 wrt. $\varphi$} iff, for every $k\in[0..{n-1}]$, 
\begin{itemize}
\item $\tau_n[k] = \overline{\attr{}[k]}$\ \ if $\varphi \wedge
  \restr{\tau_n}{k}\wedge \attreq{k}$ is unsatisfiable,
\item $\tau_n[k] = \attr{}[k]$\ \ otherwise.
\end{itemize}
where $\attreq{k}$ is the attractor equality $(\costbit{k}= \attr{}[k])$.
\ignoreinshort{ (The dual definition of ``{\em lexicographically
  minimizes}'' comes by switching
$\attr{}[k]$ with $\overline{\attr{}[k]}$.)}
Given a model \calm for \vi, we say that \calm{} {
  lexicographically maximizes
$\attreq{}$ wrt. \vi} iff its restriction to \cost lexicographically
maximizes $\attreq{}$ wrt. $\vi$.
  \end{definition}

Starting from the MSB to the LSB, 
$\tau_n$ \resp{$\calm$} in Definition~\ref{def:lexoptimizes} assigns
to each $\cost[k]$ the value $\attrbit{k}$ 
unless it is inconsistent wrt. $\vi$ and the assignments to the
previous  $\cost[i]$s, $i \in [0..k-1]$. 
Notice that  this corresponds to minimize \resp{maximize} the value
$\sum_{k=0}^{n-1}2^{n-1-k}\cdot(\costbit{k}\:\mathbf{xor}_1\:\attrbit{k})$
\resp{$\sum_{k=0}^{n-1}2^{n-1-k}\cdot{(\costbit{k}\:\mathbf{nxor}_1\:\attrbit{k})}$},
 ---where
$\mathbf{xor}_n$ is the bitwise-xor operator and $\mathbf{nxor}_n$ is its complement--- because 
$2^{n-1-i}>\sum_{k=i+1}^{n-1}2^{n-1-k}$.


The following fact derives from the
above definitions and the properties of two's complement
representation adopted by the \smtlib{} standard%
\ignoreinshort{\footnote{If the standard adopted were the sign-and-magnitude binary encoding,
then Theorem~\ref{thm:bvoptnasc} would not hold. Nevertheless, in
such a case we could adopt a simplified version of the technique for \fp 
optimization described in \sref{sec:tf-fp}.}
}
for signed \bv.~

\begin{theorem}{}\label{thm:bvoptnasc}
An optimal solution of an \omtbv problem $\pair{\varphi}{\cost}$ is any model
$\calm$ of $\varphi$ which {\em lexicographically maximizes} the vector of
{\it attractor equalities} $\attreq{}$.
\end{theorem}

\begin{PROOF}
\begin{proof}
(We investigate the  minimization case, since the  maximization case 
is dual.)

In the case of minimization 
with 
{\em unsigned \bv},  \attr is $[00...00]$%
, 
so that 
the lexicographic optimization corresponds to minimize
$\sum_{k=0}^{n-1}2^{n-1-k}\cdot\costbit{k}$
which is
the standard minimization for unsigned \bv. 

In the case of minimization 
with 
{\em signed \bv},  \attr is $[10...00]$,
so that 
the lexicographic optimization corresponds to minimize 
$2^{n-1}\cdot\overline{\costbit{0}}+\sum_{k=1}^{n-1}2^{n-1-k}\cdot\costbit{k}$
which ---by means of subtracting the constant value $2^{n-1}$--- 
is equivalent to minimize 
$-2^{n-1}\cdot\costbit{0}+\sum_{k=1}^{n-1}2^{n-1-k}\cdot\costbit{k}$,
which is the standard minimization for two's complement \bv.
%
%
\hfill $\Box$
\end{proof}  

\end{PROOF}

Definitions~\ref{def:attractor} and \ref{def:lexoptimizes} with
Theorem~\ref{thm:bvoptnasc} suggest thus a direct extension to
the minimization/maximization of 
{\em signed} \bv of the
algorithm for unsigned \bv in \cite{NadelR16}: {\em apply the
unsigned-\bv maximization \resp{minimization} algorithm of
\cite{NadelR16} to the objective 
$\cost' \defas {(\cost\:\mathbf{nxor}_n\:\attr)}$ 
\resp{$\cost' \defas (\cost\:\mathbf{xor}_n\:\attr)$} instead than simply to
\cost \resp{$\overline{\cost}$}.}


\begin{example}
Let $\cost^{[3]}$ be a signed \bv goal of $3$ bits to be minimized and
$\attr \defas [100]$ be its attractor, so that the corresponding
vector of attractor equalities $\attreq{}$ is equal to
$[\cost[0] = 1, \cost[1] = 0, \cost[2] = 0]$.

An assignment $\tau_3 \defas \set{\attreq{0}, \neg \attreq{1}, \neg \attreq{2}}$
(for which $\cost^{[3]} = \mathbf{-1}^{[3]}$) is lexicographically
better than $\tau_3' \defas \set{\neg \attreq{0}, \attreq{1}, \attreq{2}}$
(for which $\cost^{[3]} = \mathbf{0}^{[3]}$), because the former
satisfies the \emph{attractor equality} corresponding to the MSB
while the latter does not.
Moreover, the assignment $\tau_3$ is lexicographically worse
than the assignment $\tau_3'' \defas \set{\attreq{0}, \neg \attreq{1}, \attreq{2}}$
(for which $\cost^{[3]} = \mathbf{-2}^{[3]}$), because --all the rest
being equal-- the latter assignment makes the \emph{attractor equality}
$(\cost[2] = 0)$ true.
\hfill $\diamond$
\end{example}


    \subsection{Floating-Point Optimization}
    \label{sec:tf-fp}

We define the {\it Floating-Point Optimization problem} as follows.

\begin{definition}{(\omtfp).}
\label{def:omtfp}
Let $\varphi$ be a 
\smtfp{} formula and \cost be a
\fp variable occurring in $\varphi$.
We call an {\bf Optimization Modulo \fp problem}, the problem
of finding a model $\calm$ for $\varphi$ (if any) whose value of
\cost, denoted with \mincostphi, is either
\begin{itemize}
\item minimum wrt. the usual total order relation $\leq$ for \fp numbers,
      if $\varphi$ is satisfied by at least one model $\calm'$
      s.t. $\calm'(\cost)$ is not $\nan$,
\item some binary representation of $\nan$, otherwise.
\end{itemize}
(The dual definition where we look for the {\it maximum} follows straightforwardly.)
\end{definition}
 

Definition~\ref{def:omtfp} is made necessarily convoluted
by the fact that \cost{} can be $\nan$.
In fact, in the \smtlib{} standard the comparisons \set{\le,<,\ge,>}
between $\nan$ and any other \fp value are always evaluated false
because $\nan$ has multiple representations at the binary
level\ignoreinshort{~(see Table~\ref{tab:fp-values})}.
Also, requiring the optimal solution to be always different from $\nan$
makes the resulting \omtfp problem $\pair{\vi\wedge\neg\isnan{\cost}}{\cost}$
unsatisfiable when $\varphi$ is satisfied only by models \calm s.t.
$\calm(\cost)$ is \nan. 
For these reasons, we admit $\nan$ as the optimal solution value for
\cost  if and only if $\varphi$ is satisfied only by models \calm
s.t. $\calm(\cost)$ is \nan. 


In the rest of this section we assume that we have already checked,
in sequence, that 
\begin{itemize}
  \item[$i)$] the input formula \vi is satisfiable ---by invoking an
              \smtfp solver on \vi. If the solver returns \unsatres,
              then there is no need to proceed;
 \item[$ii)$] $\varphi$ is satisfied by at least one model $\calm'$
              s.t. $\calm'(\cost)$ is not $\nan$ ---by invoking an \smtfp
              solver on $\vi\wedge\neg\isnan{\cost}$ if the model \calm
             returned by the previous \smt call is s.t. $\calm(\cost)$ is
             $\nan$. If the solver returns \unsatres, then we conclude that the
             minimum is \nan.
\end{itemize}
After that, we can safely focus our investigation on the restricted \omtfp
problem \pair{\vinonan}{\cost}, where
$\vinonan\defas\vi\wedge\neg\isnan{\cost}$, knowing it is satisfiable.


\begin{IGNOREINSHORT}
In Section~\sref{sec:tf-bv}, we have introduced the concept of a \bv objective
attractor, and we have shown how this value can be used to drive the optimization
search towards the optimum value, when minimizing or maximizing a \emph{signed} or
\emph{unsigned} \bv goal.
However, in the case of floating-point optimization, it is not possible to
statically determine the attractor value in advance, before the search is
even started.
This is due to the more complex representation of \fp variables, which
uses three separate Bit-Vectors (i.e. sign, exponent and significand), and
the presence of various classes of special values (i.e. zeros, infinity, NaN),
which make definition~\ref{def:attractor} ambiguous for \fp optimization.
We illustrate this problem with the following example.


\begin{example}
\label{ex:fp-sa-bad}
Let $\pair{\vinonan}{\cost}$ be an \omtfp{} problem where \cost is a \fp objective,
of sort {\tt (\_ FP 3 5)}, to be minimized.
To make our explanation easier to follow, we show in Table~\ref{tab:fp-values} a short
list of sample values for an \fp variable of the same sort as \cost. Each \fp value is
represented as a triplet of Bit-Vectors $\langle \mathbf{sign}, \mathbf{exp},
\mathbf{sig} \rangle$ --following the \smtlib{} conventions described in
Section~\sref{sec:background}-- and also in decimal notation.

\begin{table}[h]
\centering
\begin{tabular}{l|lll|r}
\textbf{}   & \textbf{sign} & \textbf{exp} & \textbf{sig} & \textbf{value}     \\
\hline
\textbf{1}  & \#b0          & \#b111       & \#b1111      & \nan               \\
\textbf{}   & ...           & ...          & ...          & \nan               \\
\textbf{2}  & \#b0          & \#b111       & \#b0000      & \plusinf           \\
\textbf{3}  & \#b0          & \#b110       & \#b1111      & $\frac{31}{2}$     \\
\textbf{}   & ...           & ...          & ...          & ...                \\
\textbf{4}  & \#b0          & \#b000       & \#b0001      & $\frac{1}{64}$    \\
\textbf{5}  & \#b0          & \#b000       & \#b0000      & $+0$               \\
\hline
\textbf{6}  & \#b1          & \#b000       & \#b0000      & $-0$               \\
\textbf{7}  & \#b1          & \#b000       & \#b0001      & $-\frac{1}{64}$   \\
\textbf{}   & ...           & ...          & ...          & ...                \\
\textbf{8}  & \#b1          & \#b110       & \#b1111      & $-\frac{31}{2}$    \\
\textbf{9}  & \#b1          & \#b111       & \#b0000      & \minusinf          \\
\textbf{}   & ...           & ...          & ...          & \nan               \\
\textbf{10} & \#b1          & \#b111       & \#b1111      & \nan              
\end{tabular}
\caption[Sample \fp variable values]{
\label{tab:fp-values}
Sample values for a \fp variable with sort {\tt (\_ FP 3 5)}.
}
\end{table}

From Table~\ref{tab:fp-values}, we immediately notice that the binary representation
of both the exponent and the significant of a Floating-Point number grows in opposite
directions in the positive and in the negative domains.
In addition, by sorting the values according to their binary representation, we
observe that \minusinf [resp. \plusinf] is not the smallest [resp. greatest]
representable \fp value in the negative [resp. positive] domain.
In fact, both extreme ends of the table are occupied by \nan, which has multiple
binary representations.

In what follows, we temporarily disregard the effects of unit-propagation, which
might assign some (or all) bits of $\cost$ as a result of some constraints in
$\vinonan$, and pick some values as candidate attractors for an \fp goal
to be minimized.

Suppose that the attractor is chosen to be equal to the value \minusinf listed
at row $9$ in Table~\ref{tab:fp-values}, which is the smallest \fp value wrt.
total order relation $\leq$ for \fp numbers.
Assume that the optimal value of the \fp goal is the sub-normal \fp value
\texttt{(fp \#b1 \#b000 \#b1111)} (i.e. $\frac{-15}{64}$). Then, it can be
seen that after both the sign and the exponent bits have been decided to
be equal \texttt{\#b1} and \texttt{\#b000} respectively, the remaining bits
of the attractor pull the search in the wrong direction, that is, towards
$-0$.

Selecting a different \fp value as candidate attractor does not really
solve the problem, or rather, it results in a different set of issues.

For instance, an attractor equal to the \nan value listed at row $10$ in
Table~\ref{tab:fp-values}, which is the smallest representable \fp value
according to the binary ordering, would solve the problem for the previous
case in which the optimum \fp value is \texttt{(fp \#b1 \#b000 \#b1111)}.
However, this attractor would remain an unsuitable choice for an \omtfp
instance where the \fp goal is forced to be positive, because after the sign
bit of the objective function has been decided to be equal {\tt\#b0} the
remaining bits of the attractor drive the search in the wrong direction,
that is, towards \plusinf.
\hfill $\diamond$
\end{example}


Since there is no statically-determined \fp value that can be used as an attractor
when dealing with floating-point optimization, we introduce the new concept of
{\em dynamic attractor}.
\end{IGNOREINSHORT}


\begin{definition}{(Dynamic Attractor.)}
\label{def:dattr}
Let $\pair{\vinonan}{\cost}$ be a restricted \omtfp problem, where
$\vinonan\defas\vi\wedge\neg\isnan{\cost}$ is a satisfiable
\smtfp{} formula and \cost is a \fp objective to be minimized
[resp. maximized].
Let $k\in[0..n]$ and $\tau_k$ be an assignment to the $k$
most-significant bits of \cost.

Then, we say that an \fp-value \dattr{\tau_k} for \cost is a
{\bf dynamic attractor for \cost wrt. $\tau_k$} iff it is the smallest
[resp. largest] \fp value different from \nan s.t.
      the $k$ most-significant bits of \dattr{\tau_k} have the same value of the
      $k$ most-significant bits of \cost in $\tau_k$.
We call  {\bf vector of attractor equalities} the vector $\dattreq{\tau_k}{}$ s.t. 
$\dattreq{\tau_k}{i} \defas (\costbit{i} = \dattrbit{\tau_k}{i})$,
$i\in [0..n-1]$.
\end{definition}  


The following fact derives from the above definitions and the properties
of IEEE 754-2008 standard representation adopted by \smtlib{} standard
for  \fp.

\begin{lemma}\label{prop:codifica_fp}
Let $\pair{\vinonan}{\cost}$ be a restricted  minimization \resp{maximization}
\omtfp problem,
let $\tau_k$ be an assignment to
$\costbit{0}...\costbit{k-1}$ and
$\dattr{\tau_k}$ be its corresponding dynamic attractor, for some $k\in[0..n-1]$.
Let $\tau_{k+1}\defas\tau_k\cup\set{\costbit{k}:=\dattrbit{\tau_k}{k}}$
and
$\tau'_{k+1}\defas\tau_k\cup\set{\costbit{k}:=\overline{\dattrbit{\tau_k}{k}}}$,
and let $\calm$, $\calm'$ two models for $\vinonan$ which extend
$\tau_{k+1}$ and $\tau'_{k+1}$ respectively. 

Then $\calm(\cost)\le\calm'(\cost)$ 
\resp{$\calm(\cost)\ge\calm'(\cost)$}. 
\end{lemma}


\begin{PROOF}

\begin{proof}
(We prove the case of minimization, since that of maximization is dual wrt.
the value of the sign bit.)
We distinguish three cases based on the value of $k$.


{\bf Case $k = 0$ (sign bit)}.
Then $\dattrbit{\tau_0}{0} = 1$, $\tau_{1} = \set{\costbit{0}=1}$
and $\tau'_{1} = \set{\costbit{0}=0}$, where $\costbit{0}$ is the MSB of $\cost$ and
represents the sign of the floating-point value. Then $\cost$ is smaller or equal
zero in every model $\calm$ and larger or equal zero in every model $\calm'$ of
$\vinonan$, so that $\calm(\cost)\le\calm'(\cost)$ is verified.


{\bf Case $k \in [1..ebits]$ (exponent bits)},
where $ebits$ is the number of bits in the exponent
of $\cost$. Then, $\dattrbit{\tau_k}{k}$ is  $1$ if $\tau_k[0]=1$
and  $0$ otherwise.

In the first case, $\cost$ can only be negative-valued in both $\calm$
and $\calm'$.
More precisely, $\calm(\cost)$ can be either $\minusinf$ or a normal
negative value, whereas $\calm'(\cost)$ can be either a normal or a sub-normal
negative value.
Hereafter, we consider only the case in which both have a normal negative value,
because the case in which $\calm(\cost) = \minusinf$ or $\calm'(\cost)$ is
sub-normal are both trivial, given that the absolute value of any sub-normal
\fp number is smaller than the absolute value of any normal \fp number.
Furthermore, we disregard the significand bits in $\calm$ and $\calm'$ because
their contribution to the value of $\cost$ is always less significant than that
of the bits in the exponent.
Given these premises, the exponent value of $\cost$ in every possible
$\calm$ is larger than the exponent of $\cost$ in every possible $\calm'$
by a value equal to $2^{ebits-k}$ and therefore, given that both $\calm(\cost)$
and $\calm'(\cost)$ are negative-valued, $\calm(\cost)\le\calm'(\cost)$.

The case in which $\tau_k[0]=0$, that is when $\cost$ can only be positive-valued
in both $\calm$ and $\calm'$, is dual.


{\bf Case $k > ebits$  (significand bits)}. 
Then there are three sub-cases.

If for every $i \in [1..ebits]$ the value of $\tau_k[i]$ is equal $1$,
then the only possible value of $\calm(\cost)$ for every possible
$\calm$ is $\plusinf$, and therefore $\dattrbit{\tau_k}{k} = 0$.
On the other hand, there exists no possible model $\calm'$ of $\vinonan$,
because the assignment $\costbit{k} = 1$ would imply $\cost$ being equal
to $\nan$, so the statement $\calm(\cost)\le\calm'(\cost)$ is vacuously
true.

If instead there is some $i \in [1..ebits]$ s.t. $\tau_k[i] = 0$, then
$\dattrbit{\tau_k}{k}$ is  $1$ if $\tau_k[0] = 1$ (i.e. 
\cost is negative-valued) and $0$ otherwise (i.e. \cost
is positive-valued).
In both cases, we can disregard the exponent bits in $\calm$ and $\calm'$
because their contribution to the value of $\cost$ is the same in either
model.
For the same reasons, since $\calm(\cost)$ and $\calm'(\cost)$ can only
be either both normal or both sub-normal, we can ignore the contribution
of the leading hidden bit and focus on the bits of the significand.

When $\tau_k[0] = 1$ and \cost must be negative-valued, the decimal value
of the significand in $\calm$ is larger than the decimal value of every
possible significand in $\calm'$ by exactly $2^{{}-(k-ebits)}$.
Given that both $\calm(\cost)$ and $\calm'(\cost)$ are negative-valued,
we have that $\calm(\cost)\le\calm'(\cost)$.

The case in which $\tau_k[0]=0$, that is when \cost can only be
positive-valued in both $\calm$ and $\calm'$, is dual.
\hfill $\Box$
\end{proof}


\end{PROOF}


%
Lemma~\ref{prop:codifica_fp} states that, given the current
assignment $\tau_k$ to the $k$ most-significant-bits of \cost,
$\costbit{k}=\dattrbit{\tau_k}{k}$ is always the best extension of
$\tau_k$ to the next bit (when consistent).
A dynamic attractor \dattr{\tau_k} can thus be used by the
optimization search to guide the assignment of the $k+1$-th bit of $\cost$
towards the direction of maximum gain which is allowed by $\tau_k$, so that
to obtain the ``best'' extension $\tau_{k+1}$ of $\tau_k$.
Once the (new) assignment $\tau_{k+1}$ is found, the \omt solver can compute
the dynamic attractor $\dattr{\tau_{k+1}}$ for $\cost$ wrt. $\tau_{k+1}$ 
and then use it to 
assign the $k+2$-th bit of $\cost$, and so on.


Let $\pair{\vinonan}{\cost}$ be an \omtfp instance, s.t. $\cost$ is a
\fp variable of $n$ bits, and $\tau_0$ be an initially empty
assignment. If at each step of the optimization search the
assignment of the $k$-th bit of $\cost$ is guided by the dynamic
attractor for $\cost$ wrt. $\tau_k$, then the corresponding sequence
of $n$ dynamic attractors (of increasing order $k$) is unique and
depends exclusively on $\vinonan$.
Intuitively, this is the case because the (current) dynamic attractor
always points in the direction of maximum gain.
We illustrate this in the following example.

\begin{example}
\label{ex:fp-da-good}
Let $\pair{\vinonan}{\cost}$ be an \omtfp{} problem where \cost is a \fp objective,
of sort {\tt (\_ FP 3 5)}, to be minimized\ignoreinshort{, as in Example~\ref{ex:fp-sa-bad}}.
At the beginning of the search, nothing is known about the structure of the solution.
Therefore, $\tau_0 = \emptyset$ and, since \cost is being minimized, the \emph{dynamic
attractor} for $\cost$ wrt. $\tau_0$ (i.e. $\dattr{\tau_0}$) is equal to
\texttt{(fp \#b1 \#b111 \#b0000)} (i.e. \minusinf), which gives a preference
to any feasible value of \cost in the negative domain.

If at some point of the optimization search we discover that the domain of the
objective function can only be positive, so that the first bit of $\cost$ is
permanently set to $0$ in $\tau_1$, then the new dynamic attractor for \cost
wrt. $\tau_1$ (i.e. $\dattr{\tau_1}$) is equal to
\texttt{(fp \#b0 \#b000 \#b0000)} (i.e. $+0$).

Furthermore, if later on we also find out that at least one bit in the exponent
of \cost can be assigned to $0$ in a feasible solution of the problem that extends
$\tau_i$, for some $i$, then we can remove \plusinf from the optimization search
interval.
\hfill $\diamond$
\end{example}


\begin{definition}{(Attractor Trajectory \dcala{}).}
\label{def:cala}
Consider the restricted \omtfp problem $\pair{\vinonan}{\cost}$ s.t. 
$\vinonan\defas\vi\wedge\neg\isnan{\cost}$ as in Definition~\ref{def:dattr},
a triplet of inductively-defined sequences 
\tuple{
\set{\tau_{0}, \tau_{1}, ..., \tau_{n}},
\set{\dattr{\tau_0}, \dattr{\tau_1}, ...., \dattr{\tau_n}},
\set{\dattreq{\tau_0}{}, \dattreq{\tau_1}{}, ..., \dattreq{\tau_n}{}}
}
---where 
each $\tau_k$ is an assignment to the
first $k$ most-significant bits of $\cost$ s.t. $\tau_k \subset \tau_{k+1}$,
\dattr{\tau_k} is its corresponding dynamic attractor and \dattreq{\tau_k}{}
is its corresponding \emph{vector of attractor equalities}--- so that, for every
$k\in[0..n-1]$:
\begin{itemize} 
\item [(i)] $\tau_{k+1}[k]= \overline{\dattrbit{\tau_k}{k}} $
\ \ if
    $\vinonan \wedge \tau_k \wedge  \dattreq{\tau_k}{k}$ is unsatisfiable,
\item[(ii)] $\tau_{k+1}[k] = \dattrbit{\tau_k}{k}$  otherwise.
\end{itemize}
\noindent
Then we define the {\bf attractor trajectory \dcala{}} as the vector
$[\dattreq{\tau_0}{0}, ..., \dattreq{\tau_{n-1}}{n-1}]$.
\end{definition}


\noindent
The attractor trajectory \dcala{} contains those attractor equalities
$(\costbit{k} = \dattrbit{\tau_k}{k})$ which are of critical importance
for the decisions taken by the optimization search.
Intuitively, this is the case because the value of the $k$-th
bit of \cost (i.e. \costbit{k}) is still undecided in $\tau_k$.


\begin{example}
\label{ex:fp-da-good-ext1}

\begin{figure}[t]
\vspace{-20pt}%
{\scriptsize 
\begin{align*}
\tau_0 &= \emptyset                         & \dattr{\tau_0} &= \texttt{(fp \#b1 \#b111 \#b0000)} = [\un{1}.111.1111] & [\text{i.e.\:} \minusinf]\ &\thus\unsatres \\\
\tau_1 &= \tau_0 \cup \{ \costbit{0} = 0 \} & \dattr{\tau_1} &= \texttt{(fp \#b0 \#b000 \#b0000)} = [0.\un{0}00.0000] & [\text{i.e.\:} \mbox{+0}]\ &\thus\unsatres \\
\tau_2 &= \tau_1 \cup \{ \costbit{1} = 1 \} & \dattr{\tau_2} &= \texttt{(fp \#b0 \#b100 \#b0000)} = [0.1\un{0}0.0000] & [\text{i.e.\:} \mbox{+2}]\ &\thus\unsatres \\
\tau_3 &= \tau_2 \cup \{ \costbit{2} = 1 \} & \dattr{\tau_3} &= \texttt{(fp \#b0 \#b110 \#b0000)} = [0.11\un{0}.0000] & [\text{i.e.\:} \mbox{+8}]\ &\thus\satres\\ 
\tau_4 &= \tau_3 \cup \{ \costbit{3} = 0 \} & \dattr{\tau_4} &= \texttt{(fp \#b0 \#b110 \#b0000)} = [0.110.\un{0}000] & [''\ \  \ ''\   ]\ &\thus\unsatres \\
\tau_5 &= \tau_4 \cup \{ \costbit{4} = 1 \} & \dattr{\tau_5} &= \texttt{(fp \#b0 \#b110 \#b1000)} = [0.110.1\un{0}00] & [\text{i.e.\:} \mbox{+12}]\ &\thus\unsatres \\
\tau_6 &= \tau_5 \cup \{ \costbit{5} = 1 \} & \dattr{\tau_6} &= \texttt{(fp \#b0 \#b110 \#b1100)} = [0.110.11\un{0}0] & [\text{i.e.\:} \mbox{+14}]\ &\thus\satres \\
\tau_7 &= \tau_6 \cup \{ \costbit{6} = 0 \} & \dattr{\tau_7} &= \texttt{(fp \#b0 \#b110 \#b1100)} = [0.110.110\un{0}] & [ ''\ \  \ ''\   ]\ &\thus\unsatres \\
\tau_8 &= \tau_7 \cup \{ \costbit{7} = 1 \} & \dattr{\tau_8} &= \texttt{(fp \#b0 \#b110 \#b1101)} = [0.110.1101] & [\text{i.e.\:} \nicefrac{29}{2}]\ &
\end{align*}
\vspace{-10pt}
\begin{align*}
\dattreq{\tau_0}{} &= [ \un{\costbit{0} = 1}, \costbit{1} = 1, \costbit{2} = 1, \costbit{3} = 1, \costbit{4} = 0, \costbit{5} = 0, \costbit{6} = 0, \costbit{7} = 0 ] \\
\dattreq{\tau_1}{} &= [ \costbit{0} = 0, \un{\costbit{1} = 0}, \costbit{2} = 0, \costbit{3} = 0, \costbit{4} = 0, \costbit{5} = 0, \costbit{6} = 0, \costbit{7} = 0 ] \\
\dattreq{\tau_2}{} &= [ \costbit{0} = 0, \costbit{1} = 1, \un{\costbit{2} = 0}, \costbit{3} = 0, \costbit{4} = 0, \costbit{5} = 0, \costbit{6} = 0, \costbit{7} = 0 ] \\
\dattreq{\tau_3}{} &= [ \costbit{0} = 0, \costbit{1} = 1, \costbit{2} = 1, \un{\costbit{3} = 0}, \costbit{4} = 0, \costbit{5} = 0, \costbit{6} = 0, \costbit{7} = 0 ] \\
\dattreq{\tau_4}{} &= [ \costbit{0} = 0, \costbit{1} = 1, \costbit{2} = 1, \costbit{3} = 0, \un{\costbit{4} = 0}, \costbit{5} = 0, \costbit{6} = 0, \costbit{7} = 0 ] \\
\dattreq{\tau_5}{} &= [ \costbit{0} = 0, \costbit{1} = 1, \costbit{2} = 1, \costbit{3} = 0, \costbit{4} = 1, \un{\costbit{5} = 0}, \costbit{6} = 0, \costbit{7} = 0 ] \\
\dattreq{\tau_6}{} &= [ \costbit{0} = 0, \costbit{1} = 1, \costbit{2} = 1, \costbit{3} = 0, \costbit{4} = 1, \costbit{5} = 1, \un{\costbit{6} = 0}, \costbit{7} = 0 ] \\
\dattreq{\tau_7}{} &= [ \costbit{0} = 0, \costbit{1} = 1, \costbit{2} = 1, \costbit{3} = 0, \costbit{4} = 1, \costbit{5} = 1, \costbit{6} = 0, \un{\costbit{7} = 0} ] \\
\dattreq{\tau_8}{} &= [ \costbit{0} = 0, \costbit{1} = 1, \costbit{2} = 1, \costbit{3} = 0, \costbit{4} = 1, \costbit{5} = 1, \costbit{6} = 0, \costbit{7} = 1 ] \\
\end{align*}
}
\vspace{-30pt}
\caption[\fp optimization with Dynamic Attractor]{
\label{fig:fp-good}
An example of \fp optimization using the dynamic attractor.
(``$\thus \satres/\unsatres$'' denotes the satisfiability of
$\vinonan\wedge\tau_k\wedge\dattreq{\tau_k}{k}$,
the symbols ``$''\ \ ''$'' stand for ``the same as above''.
For ease of illustration, we have \un{underlined} the critical bit $\dattr{\tau_k}[k]$
in the attractors and each attractor equality of the attractor trajectory \dcala{}
inside the vectors of attractor equalities.)
}
\end{figure}


Let $\pair{\vinonan}{\cost}$ be a restricted \omtfp{} problem where \cost is a \fp objective,
of sort {\tt (\_ FP 3 5)}, to be minimized\ignoreinshort{, as in Example~\ref{ex:fp-sa-bad}}.
We consider the case in which the input formula $\vinonan$ requires $\cost$ to be
larger or equal $\nicefrac{29}{2}$ and it does not impose any other constraint on
the value of $\cost$.
Given the sequence of (partial) assignments $\tau_{0}, ..., \tau_{8}$ in
Figure~\ref{fig:fp-good}, the corresponding list of dynamic attractors and the
corresponding vectors of attractor equalities, then the attractor trajectory
$\dcala{}$ is equal to the vector $[\costbit{0} = 1, \costbit{1} = 0,
\costbit{2} = 0, \costbit{3} = 0, \costbit{4} = 0, \costbit{5} = 0,
\costbit{6} = 0, \costbit{7} = 0]$.
\hfill $\diamond$
\end{example}


\begin{lemma}{}
\label{thm:tau}
Consider 
$\pair{\vinonan}{\cost}$,
$\tau_0,  ..., \tau_n$, 
$\dattr{\tau_0},  ...., \dattr{\tau_n}$,
$\dattreq{\tau_0}{},  ..., \dattreq{\tau_n}{}$,
and $\dcala{}$  as in
definition~\ref{def:cala}.
Then $\tau_n$ lexicographically maximizes $\dcala{}$ wrt. $\vinonan$.
\end{lemma}


\begin{PROOF}
\begin{proof}
By Definition~\ref{def:cala}, we have that, for each $k\in[0..n-1]$, 
\begin{itemize}
\item[$(i)$] $\tau_{k+1}[k]= \overline{\dattrbit{\tau_k}{k}}$\ if
    $\vinonan \wedge \tau_k \wedge \dattreq{\tau_k}{k}$\ is
    unsatisfiable,  
\item[$(ii)$] $\tau_{k+1}[k] = \dattrbit{\tau_k}{k}$ otherwise.
\end{itemize}
\noindent
By construction, $\tau_k = \restr{\tau_n}{k}$. Therefore, we can replace
$\tau_k$ with $\restr{\tau_n}{k}$ so that 
\begin{itemize}
\item[$(i)$] $\restr{\tau_n}{k+1}[k]= \overline{\dattrbit{\restr{\tau_n}{k}}{k}}$ if
    $\vinonan \wedge \restr{\tau_n}{k} \wedge \dattreq{\restr{\tau_n}{k}}{k}$ is unsatisfiable,

\item[$(ii)$] $\restr{\tau_n}{k+1}[k] = \dattrbit{\restr{\tau_n}{k}}{k}$ otherwise.
\end{itemize}
\noindent
We notice the following facts.
For each $k\in[0..n-1]$, $\restr{\tau_n}{k} \subset \tau_n$.
Furthermore, for each $k\in[0..n-1]$, $\dcala{k} = \dattreq{\restr{\tau_n}{k}}{k}$ because
$\dcala{k} = \dattreq{\tau_k}{k}$ by the definition of attractor trajectory, and
$\dattreq{\tau_k}{k} = \dattreq{\restr{\tau_n}{k}}{k}$ by the equality
$\tau_k = \restr{\tau_n}{k}$.
Thus, we can replace $\restr{\tau_n}{k+1}$ with $\tau_n$ and 
$\dattreq{\restr{\tau_n}{k}}{k}$ with $\dcala{k}$, as follows. For each $k\in[0..n-1]$, 
\begin{itemize}
\item[$(i)$]     $\tau_{n}[k]= \overline{\dattrbit{\tau_n}{k}}$ if
    $\vinonan \wedge \restr{\tau_n}{k} \wedge \dcala{k}$ is unsatisfiable,
\item[$(ii)$] $\tau_{n}[k] = \dattrbit{\tau_n}{k}$ otherwise.
\end{itemize}

\noindent
Hence, $\tau_n$ lexicographically maximizes $\dcala{}$ wrt. $\vinonan$.
\hfill $\Box$
\end{proof} 

\end{PROOF}


\begin{IGNOREINSHORT}
Finally, we make the following two observations.
The first is that the sequence $\tau_0, \tau_1, ..., \tau_{n}$ in
definition~\ref{def:cala} can be iteratively constructed using its
list of requirements, for instance, by means of a sequence of
incremental calls to an \smt solver.
The second, more important, observation is that $\tau_n$ corresponds
to the assignment of values which makes $\cost$ optimal in $\vinonan$.
\end{IGNOREINSHORT}


\begin{IGNOREINSHORT}
Using the above definitions, we show that the following fact holds.
\end{IGNOREINSHORT}


\begin{theorem}{}\label{thm:fpopt1}
Let
$\pair{\vinonan}{\cost}$,
$\tau_0,  ..., \tau_n$, 
$\dattr{\tau_0},  ...., \dattr{\tau_n}$,
$\dattreq{\tau_0}{},  ..., \dattreq{\tau_n}{}$,
and $\dcala{}$ be as in
definition~\ref{def:cala}.
Then, any model $\calm$ of $\vinonan$ which lexicographically maximizes
the attractor trajectory \dcala{}  is an optimal solution for the \omtfp
problem $\pair{\vinonan}{\cost}$.
\end{theorem}


\begin{PROOF}
\begin{proof}
(We prove the case of minimization, since that of maximizations is dual.)\\
By Lemma~\ref{thm:tau} we have that $\tau_n$  lexicographically
maximize $\dcala{}$. 
Let \calm be a model of $\vinonan$ which
lexicographically maximizes  \dcala{}, and let $\mu$ be its
restriction to \cost.  
Since both $\tau_n$ and $\calm$ lexicographically maximize $\dcala{}$, 
for the uniqueness of $\tau_n$, we
immediately notice that $\mu = \tau_n$, so that  $\tau_k =
\restr{\mu}{k}$ for each $k\in[0..n]$ and $\mu$ lexicographically maximize $\dcala{}$. 

By definition, $\calm$ is an optimal solution for
$\pair{\vinonan}{\cost}$ iff 
there exists no other model $\calm'$ for it s.t. 
$\calm'(\cost) < \calm(\cost)$. Hence, we 
show by contradiction that no such $\calm'$ can exist.

Assume (for the sake of contradiction), that there exists a model
$\calm'$ for $\vinonan$,
s.t. $\calm'(\cost) < \calm(\cost)$, and 
let $\mu'$ be the restriction of $\calm'$ to \cost.
Then there must be at least one index $i$ for
which $\mu[i] \neq \mu'[i]$. 
Let \minidx be the smallest such index. 
Recalling that $\tau_{\minidx} = \restr{\mu}{\minidx}$ and
$\tau_{\minidx+1} = \restr{\mu}{\minidx+1}$,
we set 
$\tau_{\minidx+1}' \defas \restr{\mu'}{\minidx+1}$.
Then, $\tau_{\minidx} \subset \tau_{\minidx+1}$, $\tau_{\minidx} \subset \tau_{\minidx+1}'$,
$\tau_{\minidx+1} \neq \tau_{\minidx+1}'$. 
In particular,
$\tau_{\minidx+1}[\minidx] = \overline{\tau_{\minidx+1}'[\minidx]}$
and therefore $\tau_{\minidx+1}[\minidx] = \dattrbit{\tau_{\minidx}}{\minidx}$ if
$\tau_{\minidx+1}'[\minidx] = \overline{\dattrbit{\tau_{\minidx}}{\minidx}}$,
and vice versa.

Then, we distinguish two cases.


In the first case,
$\tau_{\minidx+1}[\minidx] = \overline{\dattrbit{\tau_{\minidx}}{\minidx}}$
and
$\tau_{\minidx+1}'[\minidx] = \dattrbit{\tau_{\minidx}}{\minidx}$
.
From
$\tau_{\minidx+1}[\minidx] = \overline{\dattrbit{\tau_{\minidx}}{\minidx}}$
and the fact that $\mu$ lexicographically maximizes $\dcala{}$, we derive that
$\vinonan \wedge \tau_{\minidx} \wedge \dcala{\minidx}$ is
unsatisfiable, where
$\dcala{\minidx} \defas (\costbit{\minidx} = \dattrbit{\tau_{\minidx}}{\minidx})$.
Since $\tau_{\minidx} \subset \tau_{\minidx+1}' \subseteq \mu'$ and
$\tau_{\minidx+1}'[\minidx] = \dattrbit{\tau_{\minidx}}{\minidx}$,
we conclude that $\vinonan \wedge \mu' \models \false$, so that
$\calm'$ cannot be a model of $\vinonan$, contradicting the
initial assumption.


In the second case,
$\tau_{\minidx+1}[\minidx] = \dattrbit{\tau_{\minidx}}{\minidx}$
and
$\tau_{\minidx+1}[\minidx] = \overline{\dattrbit{\tau_{\minidx}}{\minidx}}$
.
Therefore, by Lemma~\ref{prop:codifica_fp}, for every pair of
models $\calm_1$, $\calm_2$ for $\vinonan$ which extend respectively 
$\tau_{\minidx+1}$ and $\tau_{\minidx+1}'$ we have that 
$\calm_1(\cost) \leq \calm_2(\cost)$.
Since $\tau_{\minidx+1} = \restr{\mu}{\minidx+1}$ and
$\tau_{\minidx+1}' = \restr{\mu'}{\minidx+1}$, it follows that
$\calm'(\cost) \not< \calm(\cost)$, contradicting the initial
assumption. \hfill $\Box$
\end{proof}

\end{PROOF}


\section{\omtfp{} Procedures}
\label{sec:procedures}
In this paper, we consider two approaches for dealing with \omtfp: a basic
linear/binary search, based on the inline \omt schema for \omtlracupt{} presented
in \cite{st_tocl14}, and {\it Floating-Point Optimization with Binary
Search} (\ofpbs), a brand-new engine inspired by the \obvbs algorithm
for unsigned Bit-Vectors in  \cite{NadelR16} and by
Theorem~\ref{thm:fpopt1}  and relative definitions in \sref{sec:tf-fp}.


    \subsection{\omt-based Approach}
    \label{sec:omtbased-fp}

The \omt-based approach for \omtfp{} adapts the linear- and binary-search
schemata for \omtlracupt{} presented in \cite{st_tocl14}
to deal with \fp objectives.

In the basic linear-search schema, the optimization search is advanced by
means of a sequence of linear cuts, each of which forces the \omt solver
to look for a new model $\calm'$ which improves the value of $\cost$ wrt.
the most recent model $\calm$.
In the binary-search schema, instead, the \omt solver learns an incremental
sequence of cuts which bisect the current domain of the objective function.
\begin{IGNOREINSHORT}
For clarity, we recap here the essential elements of the binary-search schema
presented in \cite{st-ijcar12,st_tocl14}.
At the beginning of the optimization search and following each update of
the lower- ($lb$) and upper- ($ub$) bounds of $\cost$, the \omt solver
computes a pivoting value $\pivot \defas {\tt floor}(\rho \cdot ub + (1 - \rho) \cdot lb)$,
for some value of $\rho$ (e.g. $\frac{1}{2}$). If $\pivot$ lies inside the
range $]lb, ub]$, a cut of the form $(\cost < \pivot)$ is learned. Otherwise,
if --due to rounding side-effects of \fp operations-- $\pivot$ lies outside
the range $]lb, ub]$, a cut of the form $(\cost < \ub)$ is learned instead.
If the cut is satisfiable, the upper-bound of $\cost$ is updated with a new model
value of $\cost$. Otherwise, the lower-bound is made equal to $\pivot$ [resp. $\ub$].
The algorithm terminates when the search interval $[lb, ub[$ becomes empty.
\end{IGNOREINSHORT}
In general, it is reasonable to expect the binary-search schema to converge
towards the optimal solution faster than the linear-search schema, because
the feasible domain of a \fp goal can be comprised by an exponentially
large number of values (wrt. the bit-width of the cost function).

In either schema, whenever the optimization engine encounters for the
first time a solution
s.t. $\cost = \nan$, the \omt solver learns a unit-clause of the form
$\neg(\textsc{isNaN}(\cost))$ so as to look for an optimal solution
different from $\nan$ (if any).

When dealing with \fp objectives, differently from the case of \lra in
\cite{st_tocl14}, it is not necessary to implement a specialized
optimization procedure within the \fp{}-Solver in order to guarantee the
termination of the optimization search.
\begin{IGNOREINSHORT}
Indeed, such procedure is not
available when Floating-Point terms are bit-blasted into Bit-Vectors
{\it eagerly}, or when the {\sc acdcl} \fp-Solver is used, because by
the time the optimization procedure is called the domain interval of any
\fp term contains a singleton value.
Conversely, such a minimization procedure could be envisaged when the
\omt solver uses a {\it lazy} \fp{}-Solver as back-end, so as to speed-up
the convergence towards the optimal solution%
\footnote{
Currently, there is no such specialized optimization procedure embedded
within the {\it lazy} \fp-Solver of \optimathsat, so we won't describe
this approach any further.
}.
\end{IGNOREINSHORT}

    \subsection{Floating-Point Optimization with Binary Search}
    \label{sec:ofpbs}
    The {\it Floating-Point Optimization with Binary Search} algorithm is a
new engine for \omtfp which is inspired by the \obvbs algorithm
for \omtbv \cite{NadelR16} and is a direct implementation of
Definition~\ref{def:cala} and 
Theorem~\ref{thm:fpopt1}.

The optimization search tries to lexicographically maximize an
implicit \emph{attractor trajectory} vector $\dcala{}$, which is
incrementally derived from the current value of the dynamic attractor.
The raw value of the dynamic attractor's bits drive the optimization
search towards the direction of maximum gain at any given point in time,
without disrupting any decision that has been already made.
The dynamic attractor is incrementally updated along the search,
based on the outcome of the previous rounds of the optimization search.
At each round, one bit of the objective function is assigned its
final value. The first round decides the sign, the next batch of rounds decides the exponent and the remaining
rounds decide the fine-grained details of the significand.

\begin{figure}[t]
  \centering
  \begin{minipage}[c]{0.9\linewidth}
  \newcommand{\BREAK}{\STATE \textbf{break}}
  \newcommand{\CONTINUE}{\STATE \textbf{continue}}
  \newcommand{\assign}{\ensuremath{\coloneqq}}
  \algsetup{indent=2em}
  \DontPrintSemicolon
  {\bf function} \ofpbs ($\varphi$, $\cost$)
  \begin{algorithmic}[1]
  \STATE $\pair{res}{\calm} \assign \textsc{\smt.check\_{}under\_{}assumptions}(\varphi, \emptyset)$
  \IF {$(res == \unsatres)$}
    \RETURN $\pair{res}{\emptyset}$ \tcp*{$\varphi$ is unsatisfiable}
  \ENDIF
  \IF {$(\calm(\cost) == \nan)$}
      \STATE $\pair{res}{\calm'} \assign \textsc{\smt.check\_{}under\_{}assumptions}(\vi\wedge\neg\isnan{\cost}, \emptyset)$
      \IF {$(res == \unsatres)$}
        \RETURN $\pair{\satres}{\calm}$ \tcp*{$\cost$ can only be $\nan$}
      \ELSE
        \STATE $\calm \assign \calm'$
        \STATE $\varphi \assign \vi\wedge\neg\isnan{\cost}$ 
      \ENDIF
  \ENDIF
  \STATE $\tau \assign \emptyset$ \tcp*{from now on, \cost cannot be equal \nan}
  \STATE $\dattr{\tau} \assign \textsc{update\_{}dynamic\_{}attractor}(\tau)$
  \STATE $\textsc{\smt.set\_branching\_preference}(\cost)$
  \STATE $\textsc{\smt.update\_{}bits\_{}polarity\_{}to}(\cost, \dattr{\tau})$

  \FOR{$i \assign 0$ \textbf{up to} $n-1$}
    \STATE $eq \assign (\costbit{i} = \dattrbit{\tau}{i})$ \tcp*{attractor equality $\dattreq{\tau}{i}$}
    \IF {$(\calm \models eq)$}
      \STATE $\tau \assign \tau \cup \{ eq \}$
    \ELSE
      \STATE $\textsc{\smt.set\_branching\_preference}(\cost)$
      \STATE $\textsc{\smt.update\_{}bits\_{}polarity\_{}to}(\cost, \dattr{\tau})$
      \STATE $\pair{res}{\calm'} \assign \textsc{\smt.check\_{}under\_{}assumptions}(\varphi, \tau \cup \{ eq \})$
      \IF {$(res == \satres)$}
        \STATE $\tau \assign \tau \cup \{ eq \}$
        \STATE $\calm \assign \calm'$
      \ELSE
        \STATE $\tau \assign \tau \cup \{ \neg eq \}$
        \STATE $\dattr{\tau} \assign \textsc{update\_{}dynamic\_{}attractor}(\tau)$
      \ENDIF
    \ENDIF
  \ENDFOR
  \RETURN $\pair{\satres}{\calm}$
  \end{algorithmic}
  \end{minipage}\\
\caption[The \ofpbs{} algorithm]{
\label{alg:ofpbs}
\ofpbs Algorithm for Floating-Point optimization.
}
\end{figure}

\ignoreinshort{
  \begin{figure}[t]
  \centering
  \begin{minipage}[c]{0.9\linewidth}
  \newcommand{\BREAK}{\STATE \textbf{break}}
  \newcommand{\CONTINUE}{\STATE \textbf{continue}}
  \newcommand{\assign}{\ensuremath{\coloneqq}}
  \algsetup{indent=2em}
  \DontPrintSemicolon
  {\bf function} \textsc{update\_{}dynamic\_{}attractor} ($\tau$)
  \begin{algorithmic}[1]
  \STATE $\mathbf{static} \:\:\dattr{\tau}{} = \minusinf$ \tcp*{track $\minusinf$}
  \IF {$(\tau \neq \emptyset)$}
    \STATE $k \assign \textsc{size}(\tau) - 1$
    \STATE $\dattrbit{\tau}{k} = (1 - \dattrbit{\tau}{k})$ \tcp*{flip current bit}
    \IF {$(\tau[0] == 0)$}
        \FOR{$i \assign k+1$ \textbf{up to} $n-1$}
            \STATE $\dattrbit{\tau}{i} = 0$ \tcp*{track smallest positive value}
        \ENDFOR
    \ELSE
        \IF {$(k \le ebits)$}
            \FOR{$i \assign k+1$ \textbf{up to} $n-1$}
                \STATE $\dattrbit{\tau}{i} = 1$ \tcp*{track largest negative value}
            \ENDFOR
        \ENDIF
    \ENDIF
  \ENDIF
  \RETURN $\dattr{\tau}{}$
  \end{algorithmic}
  \end{minipage}\\
\caption[The function \textsc{update\_{}dynamic\_{}attractor()}]{
\label{alg:gda}
The function \textsc{update\_{}dynamic\_{}attractor()}.
}
\end{figure}

}

The pseudo-code of \ofpbs is shown in Figure \ref{alg:ofpbs}.
The arguments of the algorithm are the input formula $\varphi$ and the
\fp objective $\cost$, where $\cost$ is a \fp variable with $ebits$
bits in the exponent, $sbits - 1$ in the significand and
$n \defas ebits + sbits$ bits overall.

The procedure starts by checking whether the input formula $\varphi$ is
satisfiable and immediately terminates if that is not the case (lines $1$-$3$).
If $\cost = \nan$ in $\calm$ then the procedure checks whether there
exists a model $\calm'$ for $\varphi \wedge \neg\isnan{\cost}$ (lines $4$-$5$).
If this is not the case, the procedure terminates immediately and returns
the pair $\pair{\satres}{\calm}$ (line $7$). Otherwise, the model $\calm$
is updated with the new model $\calm'$, and $\varphi$ is permanently
extended with the constraint $\neg\isnan{\cost}$ (lines $9$-$10$).

At this point, the procedure initializes the value of the dynamic attractor
by invoking an external function {\sc update\_dynamic\_attractor()} with the
empty assignment $\tau$ as parameter, so that the returned value is equal
to \minusinf when minimizing and \plusinf when maximizing (lines $11$-$12$).
Then, the execution moves to the section of code implementing the core part
of the \ofpbs algorithm (lines $15$-$28$), which consists of a loop over the
bits of \cost, starting from the MSB $\costbit{0}$ down to the LSB $\costbit{n-1}$.

Inside this loop, \ofpbs first checks whether the value of $\costbit{i}$ in \calm
matches the $i$-th bit of the (current) dynamic attractor $\dattr{\tau}$.
If this is the case, then the $i$-th bit is already set to its ``best'' value
in $\calm$. Thus, the assignment $\tau$ is extended so as to permanently
set $\costbit{i} = \dattrbit{\tau}{i}$ (line $16$), and the
optimization search moves to the next iteration of the loop.
If instead $\costbit{i} \neq \dattrbit{\tau}{i}$ in $\calm$, we need to verify
whether the value of the objective function in $\calm$ can be improved by forcing
the $i$-th bit of $\cost$ equal to the $i$-th bit of the dynamic attractor.
To do so, we incrementally invoke the underlying \smt solver, this time
checking the satisfiability of $\varphi$ under the list of assumptions
$\tau \cup \{ \costbit{i} = \dattrbit{\tau}{i} \}$ (line $22$).
If the \smt solver returns $\satres$, then the value of the objective
function has been successfully improved. Hence, $\tau$ is extended
with an assignment setting $\costbit{i}$ equal to $\dattrbit{\tau}{i}$,
and $\calm$ is replaced with the new model $\calm'$ (lines $23$-$25$).
Otherwise, it is not possible to improve the objective function by toggling
the value of $\costbit{i}$, and $\tau$ is extended so as to permanently
set $\costbit{i} \neq \dattrbit{\tau}{i}$ (line $27$).
At this point, there is a mismatch between the value of the first $i+1$
bits of \cost in \calm, corresponding to the assignment $\tau$, and
those of the current dynamic attractor. This mismatch is resolved by calling
the function {\sc update\_dynamic\_attractor()} with the updated assignment
$\tau$ as parameter (line $28$).
In either case, the execution moves to the next iteration of loop.

After exactly $n$ iterations of the loop, the optimization search
terminates with the pair $\pair{\satres}{\calm}$, where $\calm$ is
the optimum model of the given \omtfpcupt{} instance.
The \ofpbs algorithm requires at most $n+2$ incremental calls to an
underlying \smtfp solver. The test in rows 17-18 allows for saving lots
of such SMT calls when the current model already assigns 
\costbit{i}  to
its corresponding value in the attractor.


\smallskip

The function {\sc update\_dynamic\_attractor()} takes as input $\tau$, a (partial)
assignment over the $k$ most-significant bits of $\cost$ and, when $\cost$ is
minimized
\footnote{
The implementation of {\sc update\_dynamic\_attractor()} is dual when $\cost$ is maximized.
}, and it essentially works as follows.
If $\tau = \emptyset$, then nothing is known about the solution of the problem,
so \minusinf is returned.
Otherwise, the procedure must compute the smallest \fp
value different from \nan (if any) which extends $\tau$.
Since $\tau \neq \emptyset$ then we know that the sign of the
objective function has been permanently decided in $\tau$.
If $\costbit{0} = 0$ in $\tau$, i.e. $\cost$ must be positive, the procedure
must return the smallest positive \fp value admitted by $\tau$. Hence, we
extend $\tau$ with $\bigcup_{i=|\tau|}^{i=n-1} \costbit{i} = 0$ and return
the corresponding \fp value.
If $\costbit{0} = 1$ in $\tau$, i.e. $\cost$ can be negative values, the
procedure must return the largest negative \fp value admitted by $\tau$. We
first check whether there exists a bit in the exponent of $\cost$ which is
assigned to $0$ in $\tau$. If that is the case, we extend $\tau$ with
$\bigcup_{i=|\tau|}^{i=n-1} \costbit{i} = 1$ and return the corresponding \fp
value. Otherwise, the procedure returns the value \minusinf, which is still
a viable extension of $\tau$.

    \subsection{Search Enhancements}
    \label{sec:enhancements}

Given a \fp value $\attr$ and a \fp goal $\cost$,
(a combination of) the following techniques can be used to adjust the
behavior of the optimization search, similarly what has been proposed
for the case of \omtbv{} by Nadel et al. in \cite{NadelR16}.

\begin{itemize}
\item {\bf branching preference}: the bits of the \fp objective
\cost are marked, inside the \omt solver, as preferred variables for branching
starting from the MSB down to the LSB. This ensures that conflicts involving the
value of the objective function are handled as early as possible, possibly
reducing the amount of work that needs to be redone after each back-jump.
\item {\bf polarity initialization}: the phase-saving value of each $\costbit{i}$
is initialized with the value of $\attrbit{i}$. This encourages the \omt
solver to assign the bits of $\cost$ so as to reassemble the bits of $\attr$,
thus possibly speeding-up the convergence towards the optimal value.
\end{itemize}


In the case of the basic \omt{} schema described in Section~\sref{sec:omtbased-fp},
the effectiveness of either technique depends on the initial choice for $\attr$.
In the lucky case, the value of $\attr$ pulls the optimization search in the
right direction and speeds up the search. In the unlucky case, when $\attr$
pulls in the wrong direction, there is no visible effect or an overall slow
down.
For instance, in the case of the \emph{linear-search} optimization schema,
enabling both options with an unlucky choice of $\attr$ can cause the \omt
solver to start the search from the furthest possible point from the optional
solution, and thus enumerate an exponential number of intermediate solutions.
\begin{IGNOREINSHORT}
Naturally, the \omt-based optimization search algorithm is still guaranteed to
terminate  even in the worst-case scenario, but the unpredictable performance
makes using either technique a generally unsuitable option in practice.
\end{IGNOREINSHORT}


In the case of the \ofpbs{} algorithm described in Section~\sref{sec:ofpbs},  we use
the latest value of the dynamic attractor $\dattr{\tau}$ for both the
{\em branching preference} (lines $11$ and $18$ of Figure~\ref{alg:ofpbs})
and the {\em polarity initialization} (rows $12$ and $19$ of Figure~\ref{alg:ofpbs})
techniques.
We observe that the value of every bit in the dynamic attractor
can change after the sign of the objective function has been decided.
Furthermore, the value of all the significand's bits in the dynamic attractor
can also change during the process of determining the optimal exponent value
of the objective function\ignoreinshort{~(see, e.g., Example~\ref{ex:fp-sa-bad})}.
As a consequence, if the \omt solver applies either enhancement before the correct
improving direction is known, this may cause the underlying \omt engine to advance
the search starting from a sub-optimal set of initial decisions.
Enabling both enhancements at the same time could make things even worse.
In order to mitigate this issue, we have designed a variant of our
optimization-search approach which does not apply either enhancement
on those bits of the objective function for which the best improving
direction is not yet known. We have called this variant {\bf safe bits
restriction}.


\section{Experimental Evaluation}
\label{sec:expeval}
We assess the performance of \optimathsat{} (v. 1.6.2) on a set of \omtfp{}
formulas that have been automatically generated using the \smtfp{}
benchmark-set of \cite{smtlib_two_url}.
The formulas, the results and the scripts necessary to reproduce these
results are made publicly available and can be downloaded from
\cite{fp-test-url}.


\paragraph*{Experiment Setup.}
This experiment has been performed on an {\it i7-6500U 2.50GHz Intel Quad-Core}
machine with $16GB$ of ram and running \emph{Ubuntu Linux} $17.10$. 
For each formula being tested we used a timeout of $600$ seconds.
The \omtfp{} instances used in this experiment have been automatically generated
starting from the satisfiable formulas included in the \smtfp{} benchmark-set
of \cite{smtlib_two_url}. We did not consider any of the unsatisfiable instances that are present in the remote repository.
\begin{IGNOREINLONG}
Since the majority of the original \smtfp{} formulas admits only one
solution, in order to increase the significance of the resulting
\omtfp{} benchmark set, we
relaxed or removed some of the constraints in these formulas so as to broaden the
set of feasible solutions.
\end{IGNOREINLONG}

\begin{IGNOREINSHORT}
For each of the original \smtfp{} formulas we applied the following transformations.
First, we either relaxed or removed some of the constraints in the original problem,
so as to broaden the set of feasible solutions. This step is necessary because the majority
of the original \smtfp{} formulas admits only one solution. However, this is not 
necessarily the ideal situation when comparing different optimization approaches.
Second, for each \fp variable $v$ appearing inside a \smtfp{} problem we generated
a pair of \omtfp{} instances, one for the minimization and another for the maximization
of $v$. At the end of this step, we obtained $39536$ \omtfp{} formulas.
Third, we randomly selected up to $300$ \omtfp{} instances from each of the five
groups of problems in the \omtfp{} benchmark-set. This filtering step yielded a
total of $1120$ \smtlib{} formulas.
\end{IGNOREINSHORT}

We consider two \omt-based baseline configurations,
{\sc \optimathsat{}(omt+lin)} and
{\sc \optimathsat{}(omt+bin)}, that run the linear-
and the binary-search respectively. These configurations
have been tested using both the {\em eager} and the {\em lazy}
\fp approaches.
The third baseline approach, named {\sc \optimathsat{}(eager+\obvbs)},
is based on a reduction of the \omtfp{} problem to \omtbv{} and it
uses \optimathsat{}'s implementation of the \obvbs{} engine%
\footnote{
The binaries of the original \omtbv{} tools presented in \cite{NadelR16} are
not publicly available.
} presented by Nadel et al. in \cite{NadelR16}.
For this test, we have generated an \omtbv{} benchmark-set
using a \bv encoding that mimics the essential aspects of
the \ofpbs{} algorithm described Section~\sref{sec:ofpbs}.

We compared these baseline approaches with a configuration using the
\ofpbs{} algorithm and the {\em eager} \fp approach,
namely {\sc \optimathsat{}(eager+\ofpbs)}.

We have separately tested the effect of enabling the
{\em branching preference} ({\sc bp}), the {\em polarity initialization} ({\sc pi})
and the {\em safe bits restriction} ({\sc so}) enhancements described in
Section~\sref{sec:tf-fp}, whenever these options were supported by the
given configuration.

\begin{IGNOREINSHORT}
Last, in order to assess the significance of the optimization problems
used in this experiment, we have collected the run-time statistics of
\optimathsat{} on the \smt formulas obtained by stripping the objective
function from each \omt instance.
We named this configuration {\sc \optimathsat{}(eager+smt)}.
\end{IGNOREINSHORT}

We have not included other tools in our experiment because we are not aware
of any other \omtfp{} solver.
For all problem instances, we verified the correctness of the optimal solution found
by each configuration with an \smt solver (\mathsatfive). When terminating,
all tools returned the same optimum value. 
\begin{IGNOREINSHORT}
In order to perform this cross-check as efficiently as possible, we enabled
\emph{model generation} on every configuration so that the optimum model
could be extracted and verified.
\end{IGNOREINSHORT}


\paragraph*{Experiment Results.}

\begin{IGNOREINSHORT}
\begin{table}[tb]
\centering
\begin{tabularx}{\textwidth}{|X|rrr|rrr|r|}
\hline
{\bf tool, configuration \& encoding} &
{\bf inst.} &
{\bf term.} &
{\bf t.o.} &
{\bf u} &
{\bf bt} &
{\bf st} &
{\bf time (s.)} \\
\hline
{\sc \optimathsat{}(eager+omt+lin)}            & 1120 & 1003 & 117 & 0 &   5 &  73 & 76375 \\ 
{\sc \optimathsat{}(eager+omt+lin+pi)}         & 1120 & 1003 & 117 & 0 &   5 &  71 & 76785 \\ 
{\sc \optimathsat{}(eager+omt+lin+bp)}         & 1120 &  956 & 164 & 0 &   6 & 105 & 77480 \\ 
{\sc \optimathsat{}(eager+omt+lin+bp+pi)}      & 1120 &  873 & 247 & 0 &  77 & 217 & 54859 \\ 
\hline
\hline
{\sc \optimathsat{}(lazy+omt+lin)}            & 1120 &  868 & 252 & 0 &  93 & 203 & 29832 \\ 
\hline
\hline
{\sc \optimathsat{}(eager+omt+bin)}            & 1120 & 1014 & 106 & 0 &  11 & 281 & 67834 \\ 
{\sc \optimathsat{}(eager+omt+bin+pi)}         & 1120 &  970 & 150 & 0 &   8 & 285 & 69765 \\ 
{\sc \optimathsat{}(eager+omt+bin+bp)}         & 1120 & 1016 & 104 & 0 &  14 & 205 & 68255 \\ 
{\sc \optimathsat{}(eager+omt+bin+bp+pi)}      & 1120 &  991 & 129 & 0 &  65 & {\bf 321} & 56941 \\ 
\hline
\hline
{\sc \optimathsat{}(lazy+omt+bin)}            & 1120 &  900 & 220 & 0 &  90 & 243 & 33260 \\ 
\hline
\hline
{\sc \optimathsat{}(eager+obvbs) [reduction]} & 1120 & 1013 & 107 &  0 &  14 & 141 & 65954 \\ 
\hline
\hline
{\sc \optimathsat{}(eager+ofpbs)}             & 1120 & 1017 & 103 &  0 &   9 & 171 & 70732 \\ 
{\sc \optimathsat{}(eager+ofpbs+pi)}          & 1120 & {\blue {\bf 1019}} & 101 &  0 & 34 & 280 & 64896 \\ 
{\sc \optimathsat{}(eager+ofpbs+pi+so)}       & 1120 & 1018 & 102 &  0 &   7 & 179 & 71430 \\ 
{\sc \optimathsat{}(eager+ofpbs+bp)}          & 1120 &  975 & 145 &  0 &   2 & 145 & 65543 \\ 
{\sc \optimathsat{}(eager+ofpbs+bp+so)}       & 1120 & 1000 & 120 &  0 &   3 & 124 & 68390 \\ 
{\sc \optimathsat{}(eager+ofpbs+bp+pi)}       & 1120 & 1001 & 119 &  0 &  77 & 273 & 60365 \\ 
{\sc \optimathsat{}(eager+ofpbs+bp+pi+so)}    & 1120 & 1006 & 114 & {\bf 19} &  32 &   245 & 59463 \\ 
\hline
\hline
{\sc virtual best}              & 1120 & {\bf 1074} & 46 & - & 559 & 1074 & 27788 \\
\hline
\hline
{\sc \optimathsat{}(eager+smt) [no optimization]} & 1120 & 1048 &  72 & - & - & - &  9259 \\
\hline
\end{tabularx}
\caption[Comparison on \omtfp{} formulas]{
\label{tab:fp}
Comparison among various \optimathsat configurations on the \omtfp{} benchmark-set.
The columns list the
total number of instances (inst.),
the number of instances solved (term.),
the number of timeouts (t.o.),
the number of instances uniquely solved by the given configuration (u),
the number of instances solved faster than any other configuration (bt),
the total number of instances solved in the shortest amount of time (st) and
the total solving time for all solved instances (time).
}
\end{table}

\begin{figure}[tb]
\centering
\scalebox{0.84}{
    \includegraphics{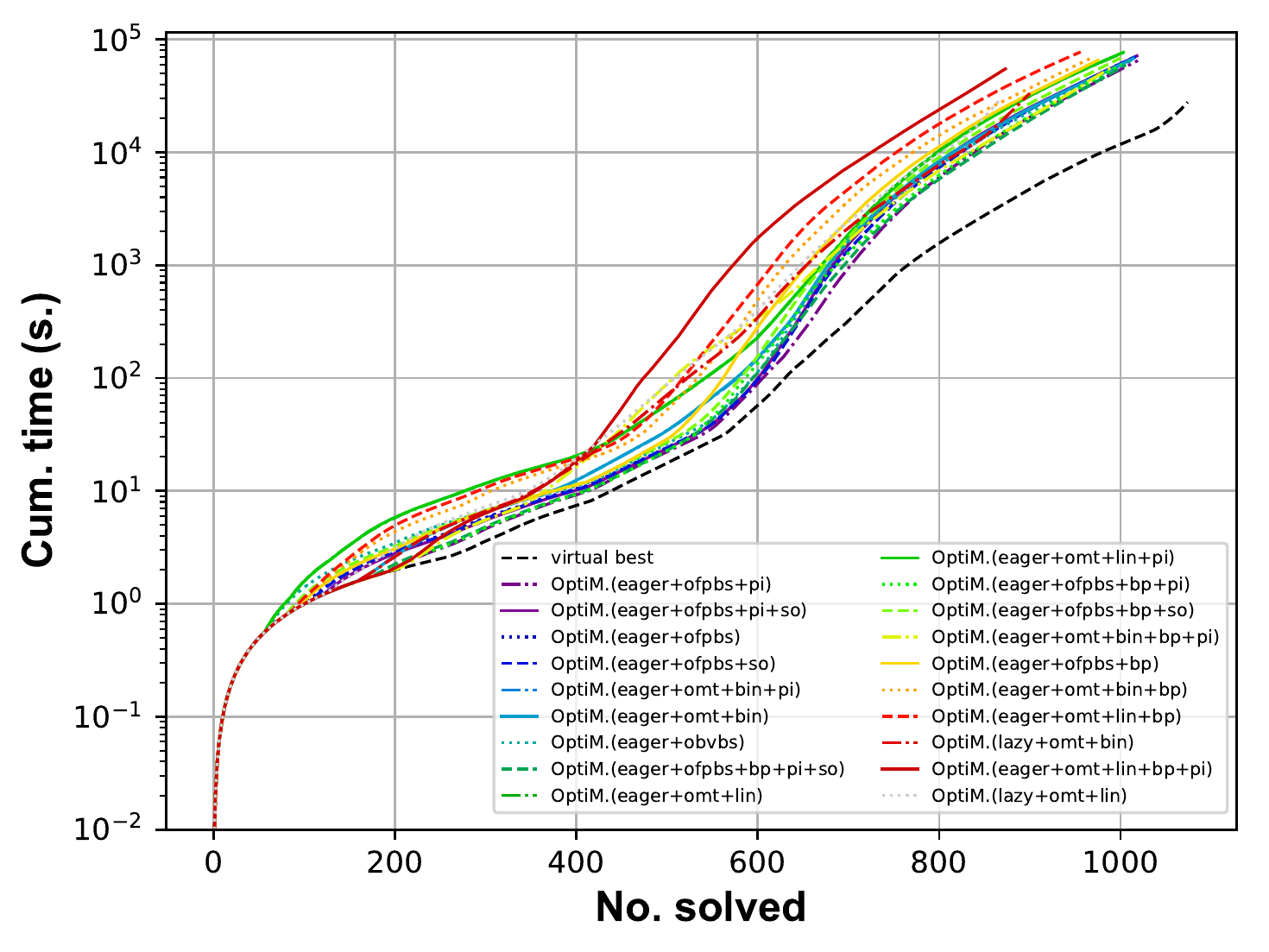}
} 
\caption[Comparison on \omtfp{} formulas (Cactus Plot)]{
\label{fig:fp-cp}
Comparison among various \optimathsat configurations on a subset of $1120$ \omtfp{}
formulas generated from the \smtfp{} formulas of \cite{smtlib_two_url}.
}
\end{figure}
\end{IGNOREINSHORT}

\begin{IGNOREINLONG}
\begin{table}[tb]
\centering
\begin{tabularx}{\textwidth}{|X|rrr|rrr|r|}
\hline
{\bf tool, configuration \& encoding} &
{\bf inst.} &
{\bf term.} &
{\bf t.o.} &
{\bf u} &
{\bf bt} &
{\bf st} &
{\bf time (s.)} \\
\hline
{\sc \optimathsat{}(eager+omt+lin)}            & 1120 & 1003 & 117 & 0 &   5 &  73 & 76375 \\ 
{\sc \optimathsat{}(eager+omt+lin+pi)}         & 1120 & 1003 & 117 & 0 &   5 &  71 & 76785 \\ 
{\sc \optimathsat{}(eager+omt+lin+bp)}         & 1120 &  956 & 164 & 0 &   6 & 105 & 77480 \\ 
{\sc \optimathsat{}(eager+omt+lin+bp+pi)}      & 1120 &  873 & 247 & 0 &  77 & 217 & 54859 \\ 
\hline
\hline
{\sc \optimathsat{}(eager+omt+bin)}            & 1120 & 1014 & 106 & 0 &  11 & 281 & 67834 \\ 
{\sc \optimathsat{}(eager+omt+bin+pi)}         & 1120 &  970 & 150 & 0 &   8 & 285 & 69765 \\ 
{\sc \optimathsat{}(eager+omt+bin+bp)}         & 1120 & 1016 & 104 & 0 &  14 & 205 & 68255 \\ 
{\sc \optimathsat{}(eager+omt+bin+bp+pi)}      & 1120 &  991 & 129 & 0 &  65 & {\bf 321} & 56941 \\ 
\hline
\hline
{\sc \optimathsat{}(lazy+omt+lin)}            & 1120 &  868 & 252 & 0 &  93 & 203 & 29832 \\ 
\hline
\hline
{\sc \optimathsat{}(lazy+omt+bin)}            & 1120 &  900 & 220 & 0 &  90 & 243 & 33260 \\ 
\hline
\hline
{\sc \optimathsat{}(eager+obvbs) [reduction]} & 1120 & 1013 & 107 &  0 &  14 & 141 & 65954 \\ 
\hline
\hline
{\sc \optimathsat{}(eager+ofpbs)}             & 1120 & 1017 & 103 &  0 &   9 & 171 & 70732 \\ 
{\sc \optimathsat{}(eager+ofpbs+pi)}          & 1120 & {\blue {\bf 1019}} & 101 &  0 & 34 & 280 & 64896 \\ 
{\sc \optimathsat{}(eager+ofpbs+pi+so)}       & 1120 & 1018 & 102 &  0 &   7 & 179 & 71430 \\ 
{\sc \optimathsat{}(eager+ofpbs+bp)}          & 1120 &  975 & 145 &  0 &   2 & 145 & 65543 \\ 
{\sc \optimathsat{}(eager+ofpbs+bp+so)}       & 1120 & 1000 & 120 &  0 &   3 & 124 & 68390 \\ 
{\sc \optimathsat{}(eager+ofpbs+bp+pi)}       & 1120 & 1001 & 119 &  0 &  77 & 273 & 60365 \\ 
{\sc \optimathsat{}(eager+ofpbs+bp+pi+so)}    & 1120 & 1006 & 114 & {\bf 19} &  32 &   245 & 59463 \\ 
\hline
\hline
{\sc virtual best}              & 1120 & {\bf 1074} & 46 & - & 559 & 1074 & 27788 \\
\hline
\end{tabularx}
\caption[Comparison on \omtfp{} formulas]{
\label{tab:fp}
Comparison among various \optimathsat configurations on the \omtfp{} benchmark-set.
The columns list the
total number of instances (inst.),
the number of instances solved (term.),
the number of timeouts (t.o.),
the number of instances uniquely solved by the given configuration (u),
the number of instances solved faster than any other configuration (bt),
the total number of instances solved in the shortest amount of time (st) and
the total solving time for all solved instances (time).
}%
\end{table}%
\end{IGNOREINLONG}


\begin{IGNOREINSHORT}
\begin{figure}[tbp]
    \centering
    \footnotesize
    \begin{tabular}{ccc}%
&\includegraphics[scale=0.47]{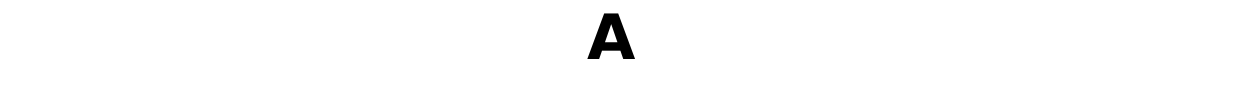}&\includegraphics[scale=0.47]{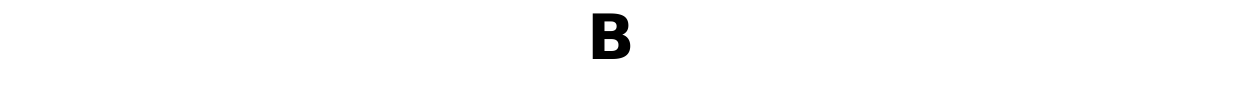}
\\%
        \includegraphics[scale=0.47]{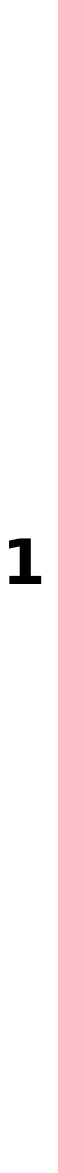}
&
        \includegraphics[scale=0.47]{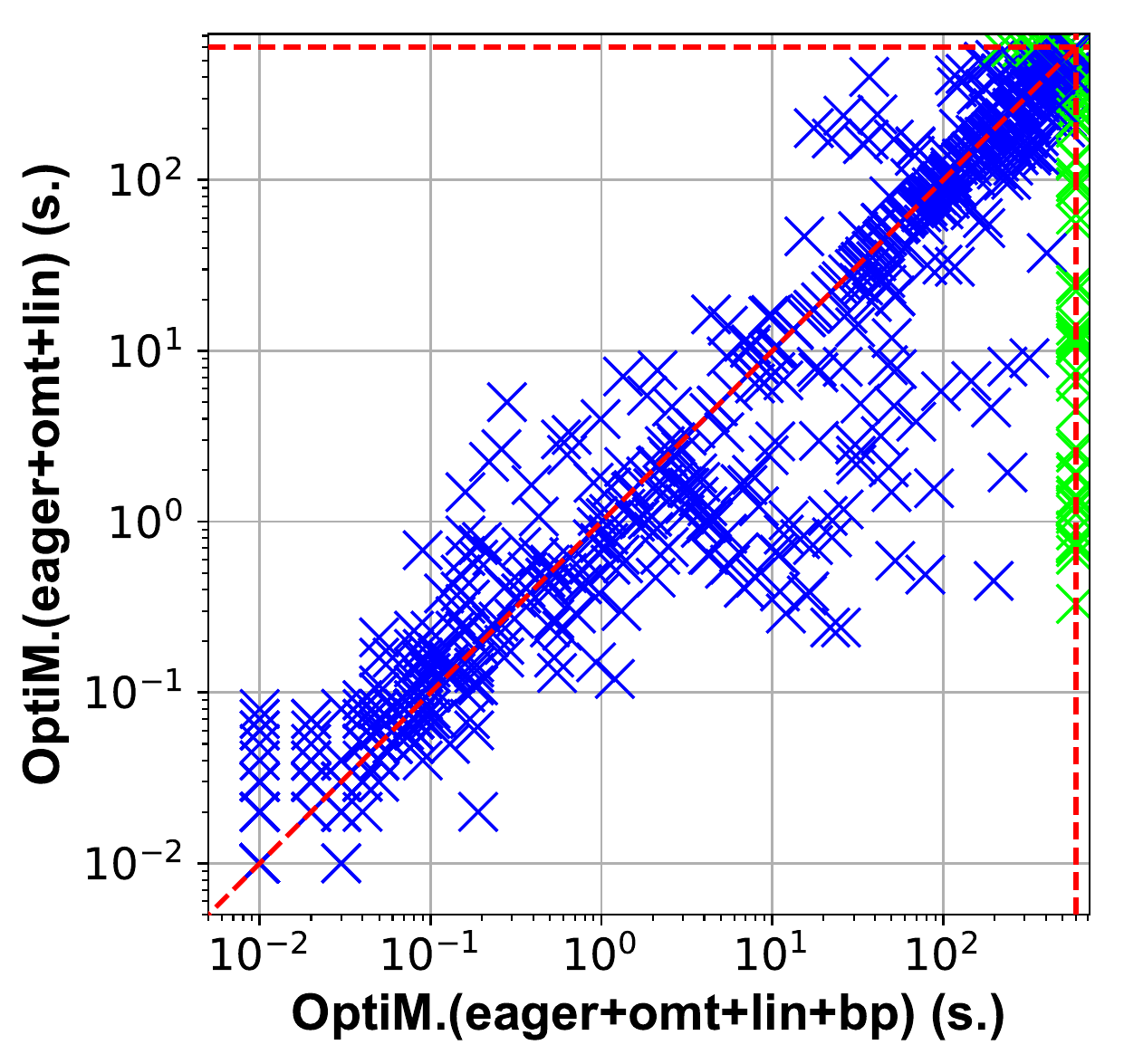}
&
        \includegraphics[scale=0.47]{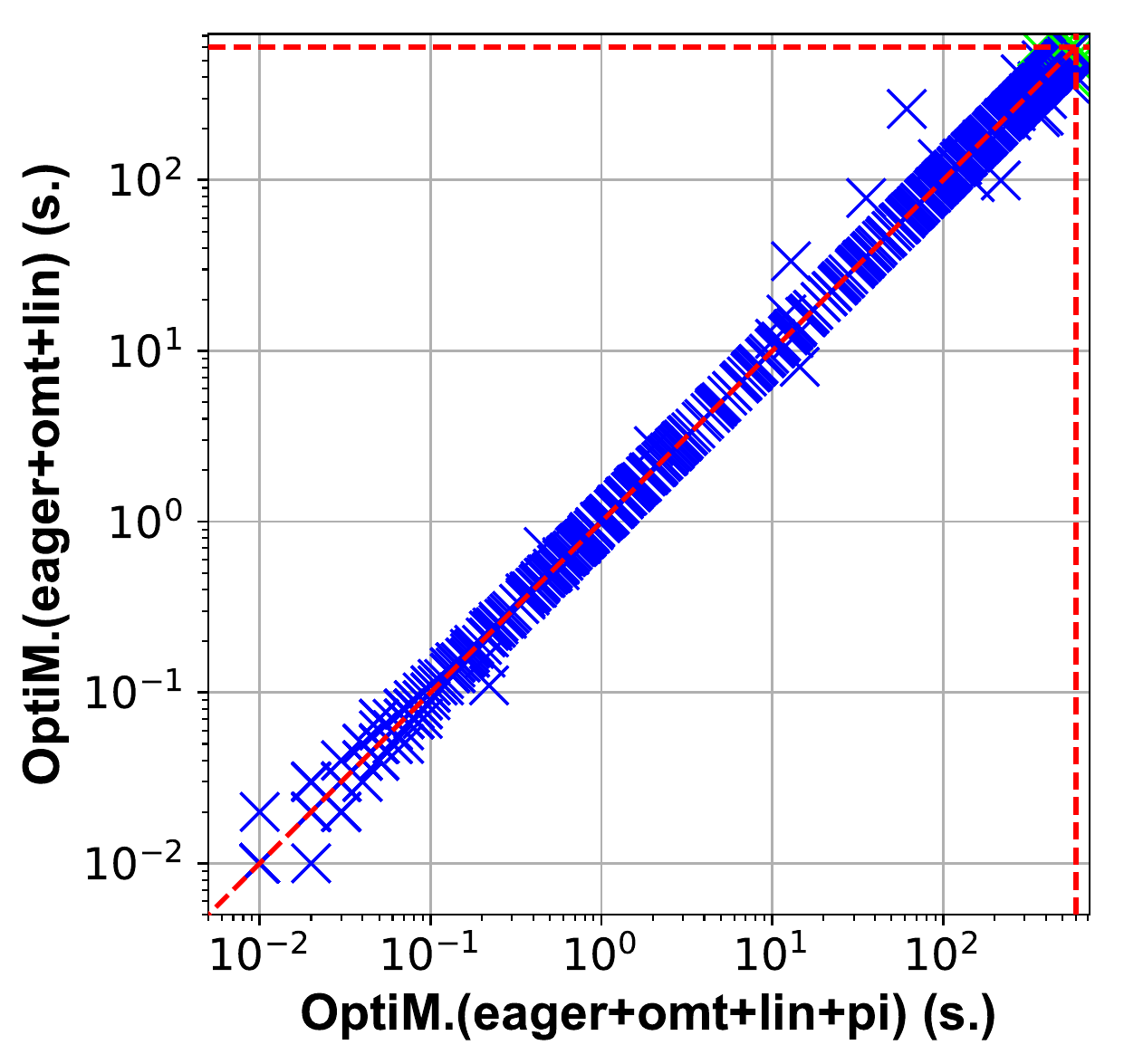}
\\%
        \includegraphics[scale=0.47]{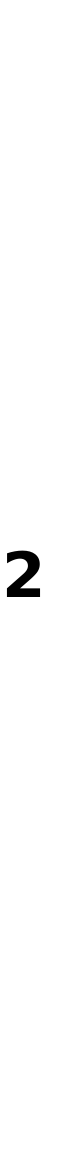}
&
        \includegraphics[scale=0.47]{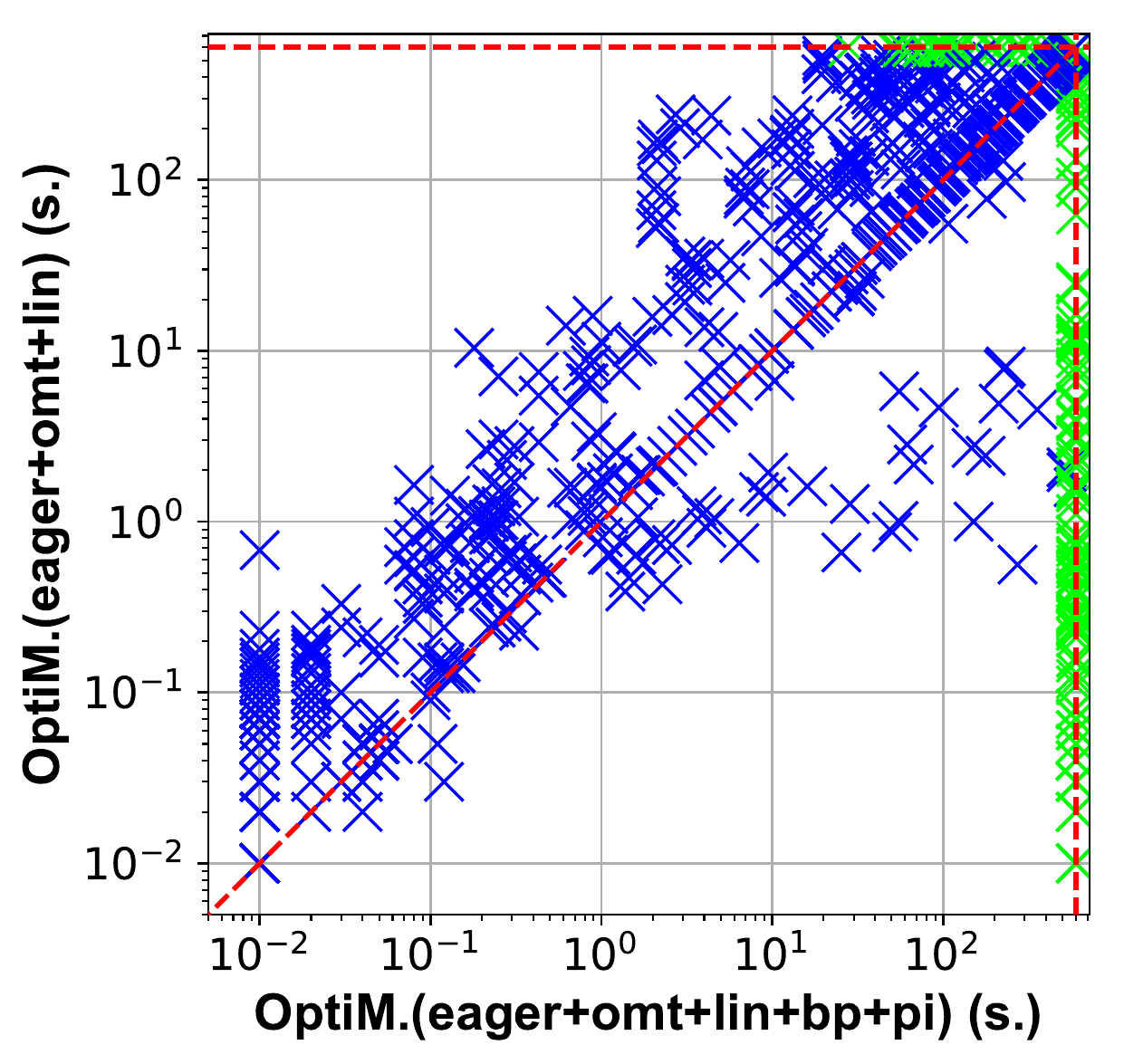}
&
        \includegraphics[scale=0.47]{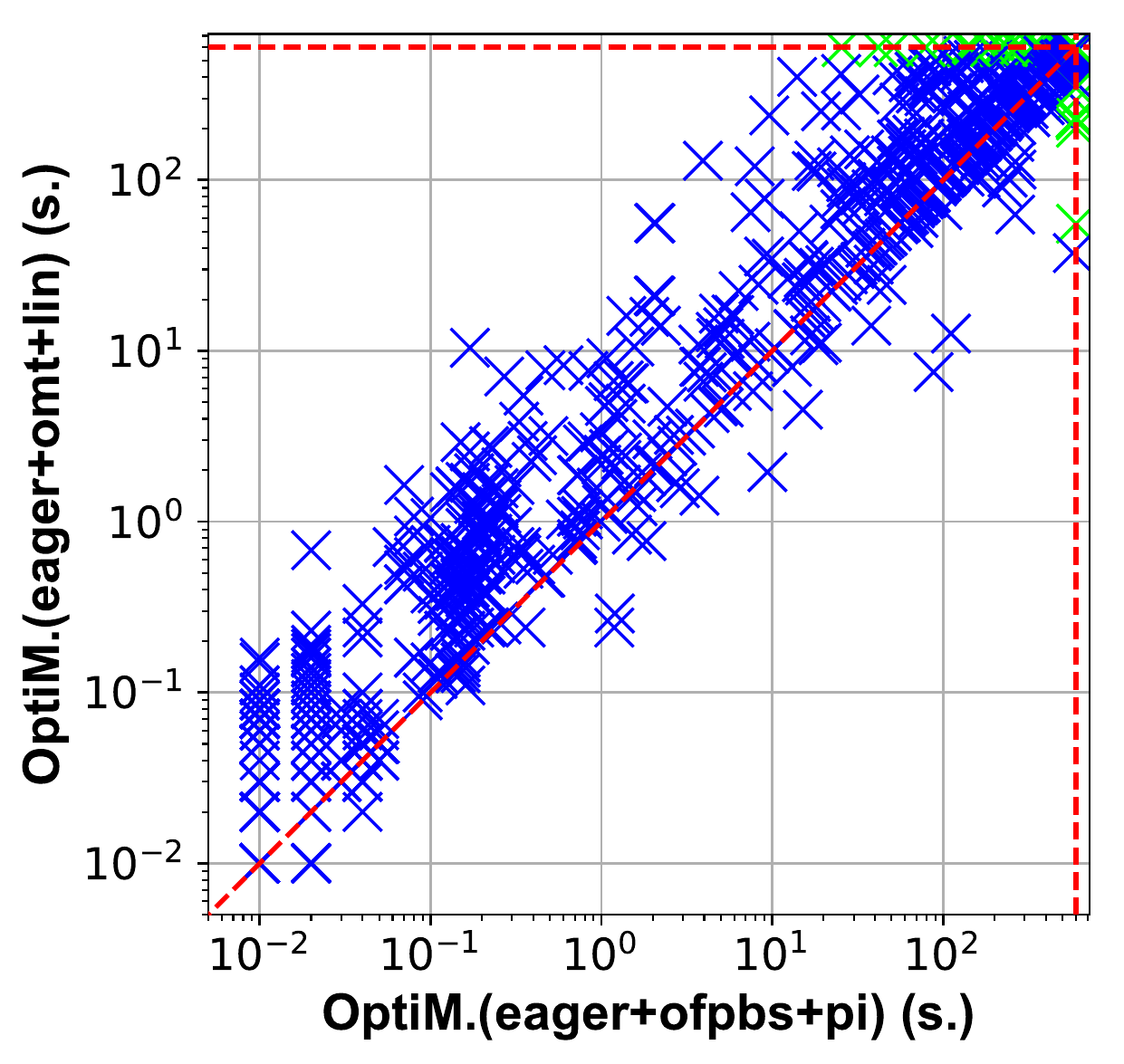}
\\%
        \includegraphics[scale=0.47]{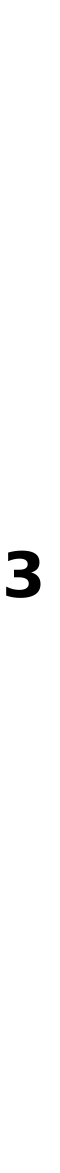}
&
        \includegraphics[scale=0.47]{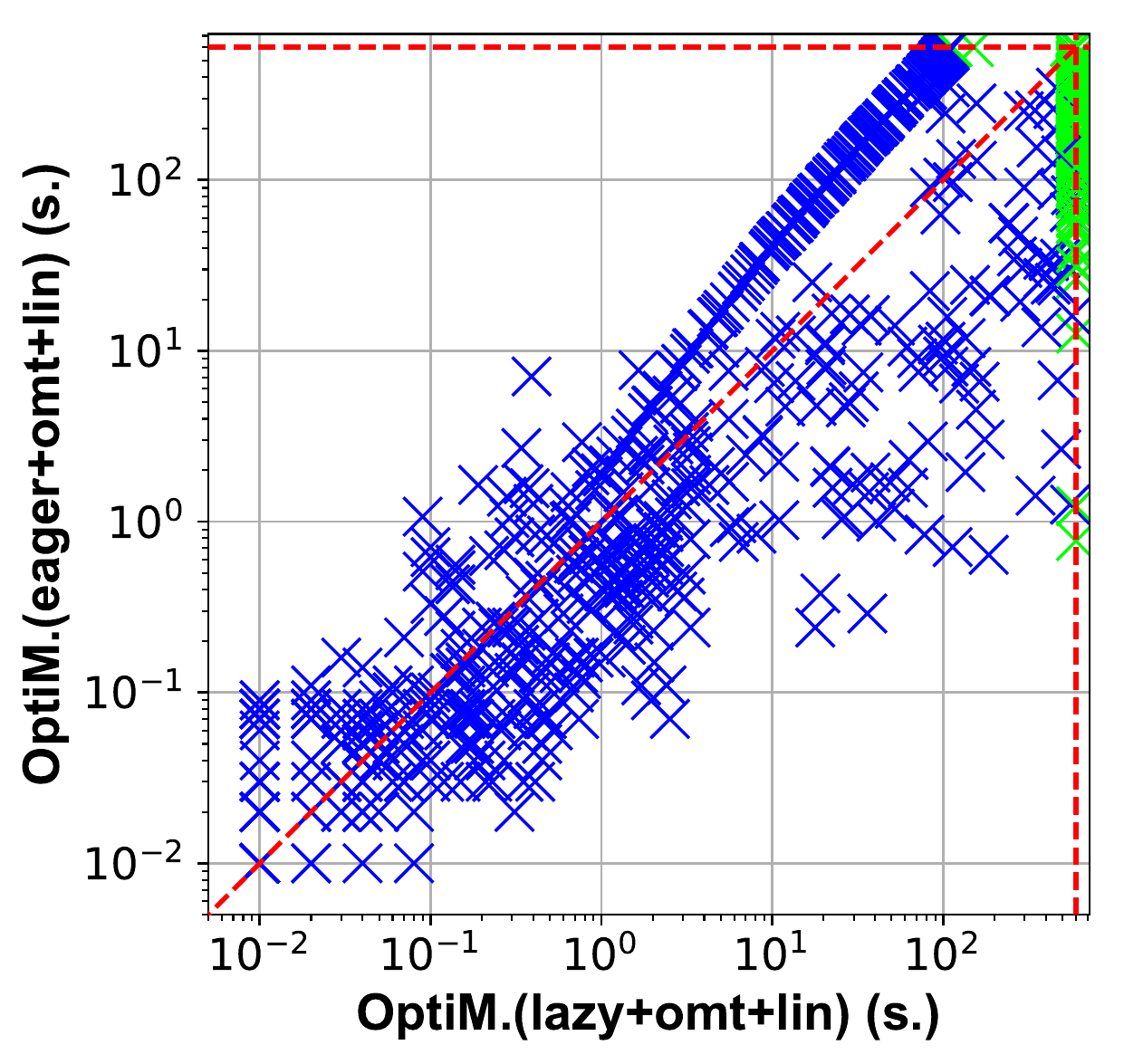}
&
        \includegraphics[scale=0.47]{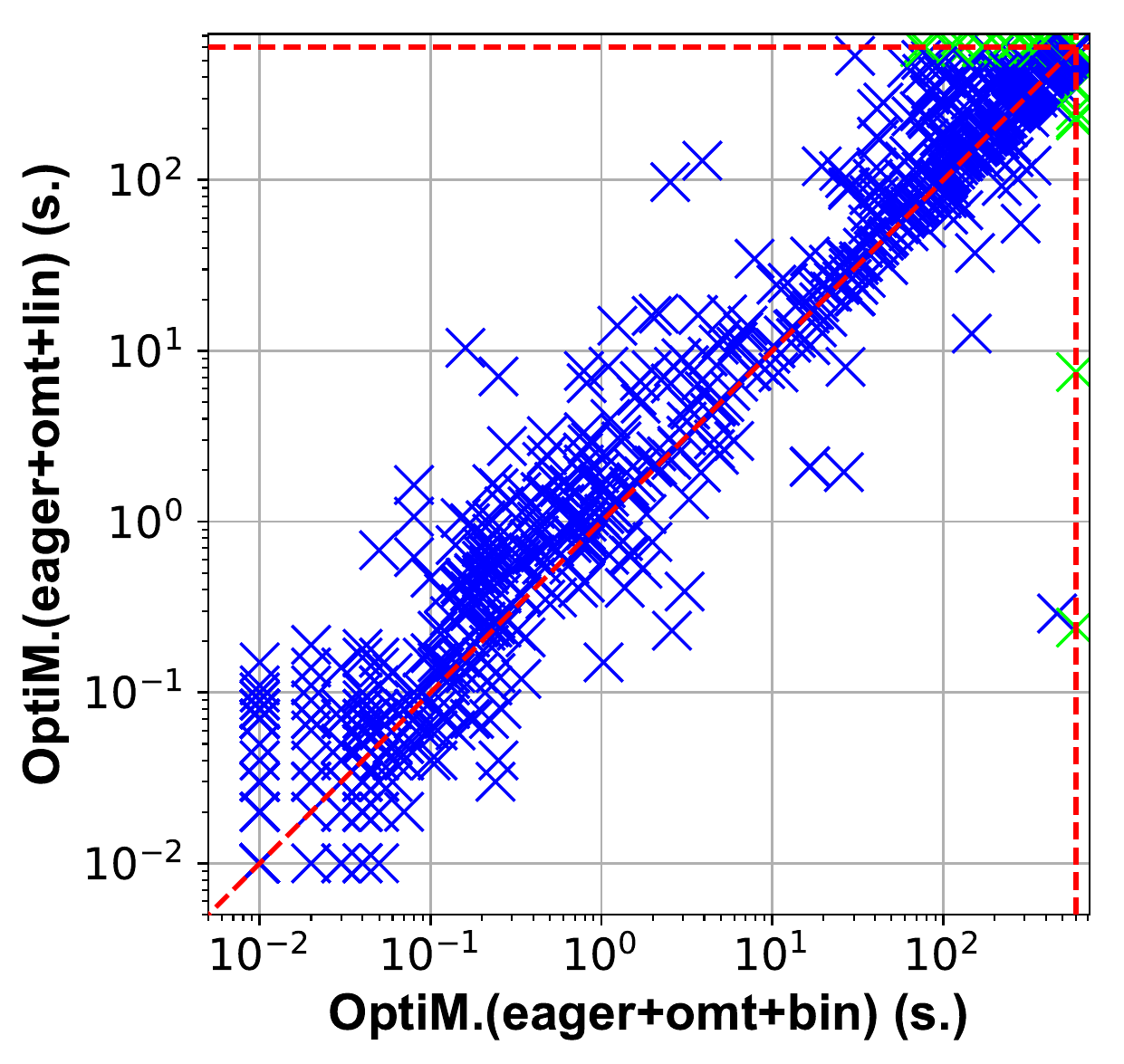}
\\%
    \end{tabular}
\caption[Comparison on \omtfp{} formulas with linear-search (Scatter Plots)]{
\label{fig:fp-sp-lin}
Pairwise comparisons on \omtfp{} formulas using \omt-based{} linear-search
and other configurations.
(\blue{Blue} points denote satisfiable benchmarks, \green{green} denotes a timeout.)
}
\end{figure}
\end{IGNOREINSHORT}


\begin{IGNOREINSHORT}
\begin{figure}[tbp]
    \centering
    \footnotesize
    \begin{tabular}{ccc}%
&\includegraphics[scale=0.47]{figs/A.pdf}&\includegraphics[scale=0.47]{figs/B.pdf}
\\%
        \includegraphics[scale=0.47]{figs/1.pdf}
&
        \includegraphics[scale=0.47]{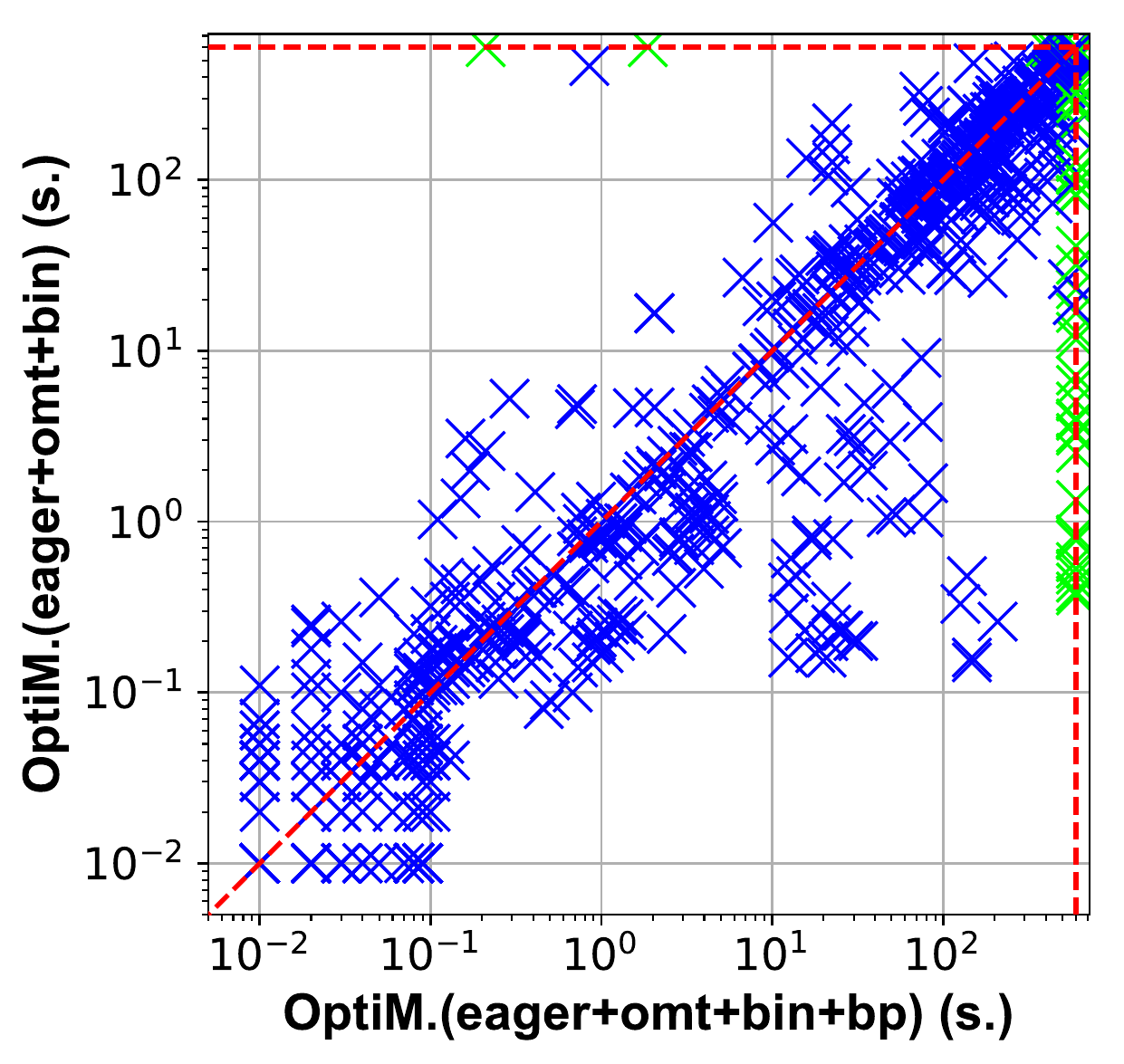}
&
        \includegraphics[scale=0.47]{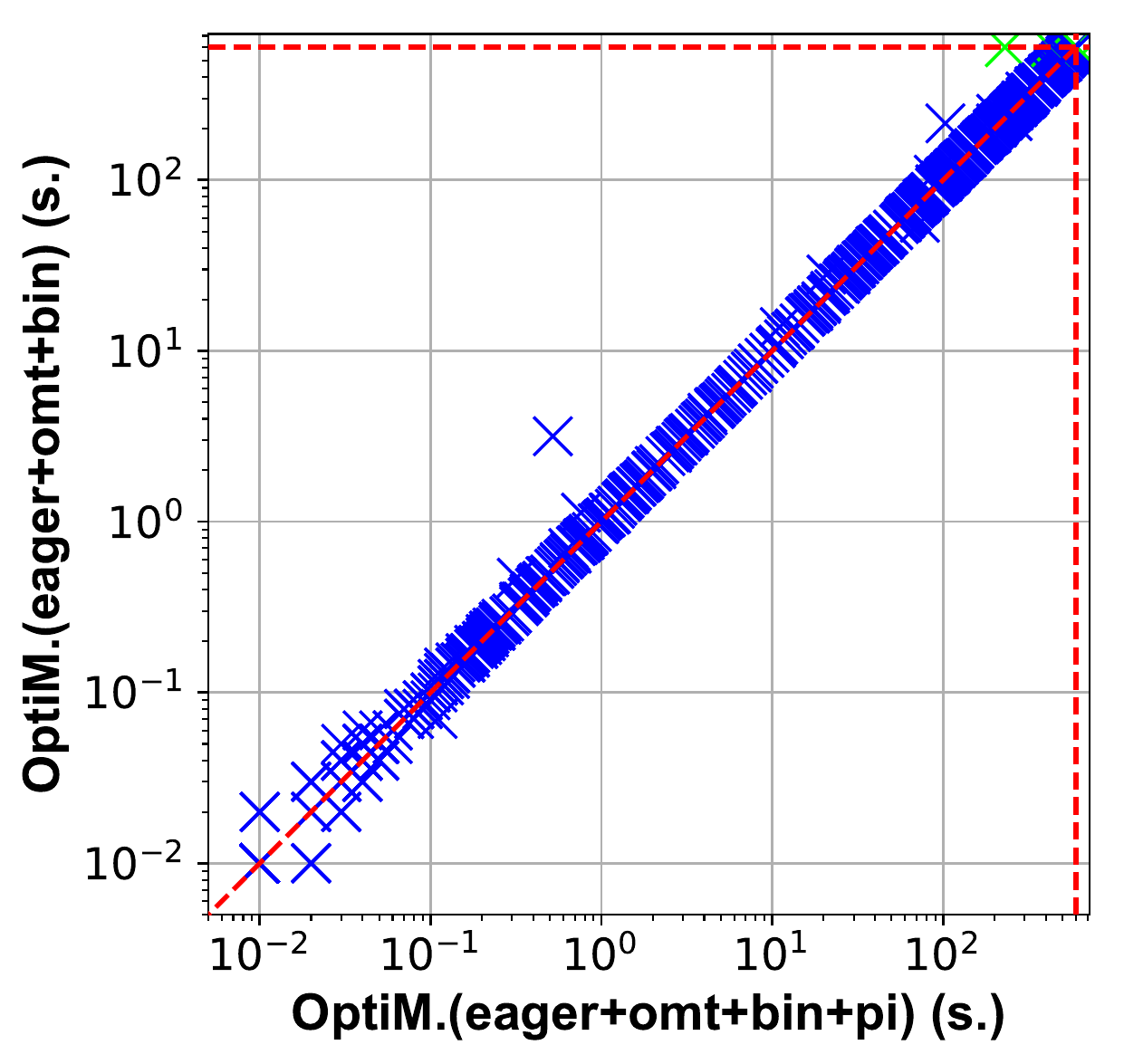}
\\%
        \includegraphics[scale=0.47]{figs/2.pdf}
&
        \includegraphics[scale=0.47]{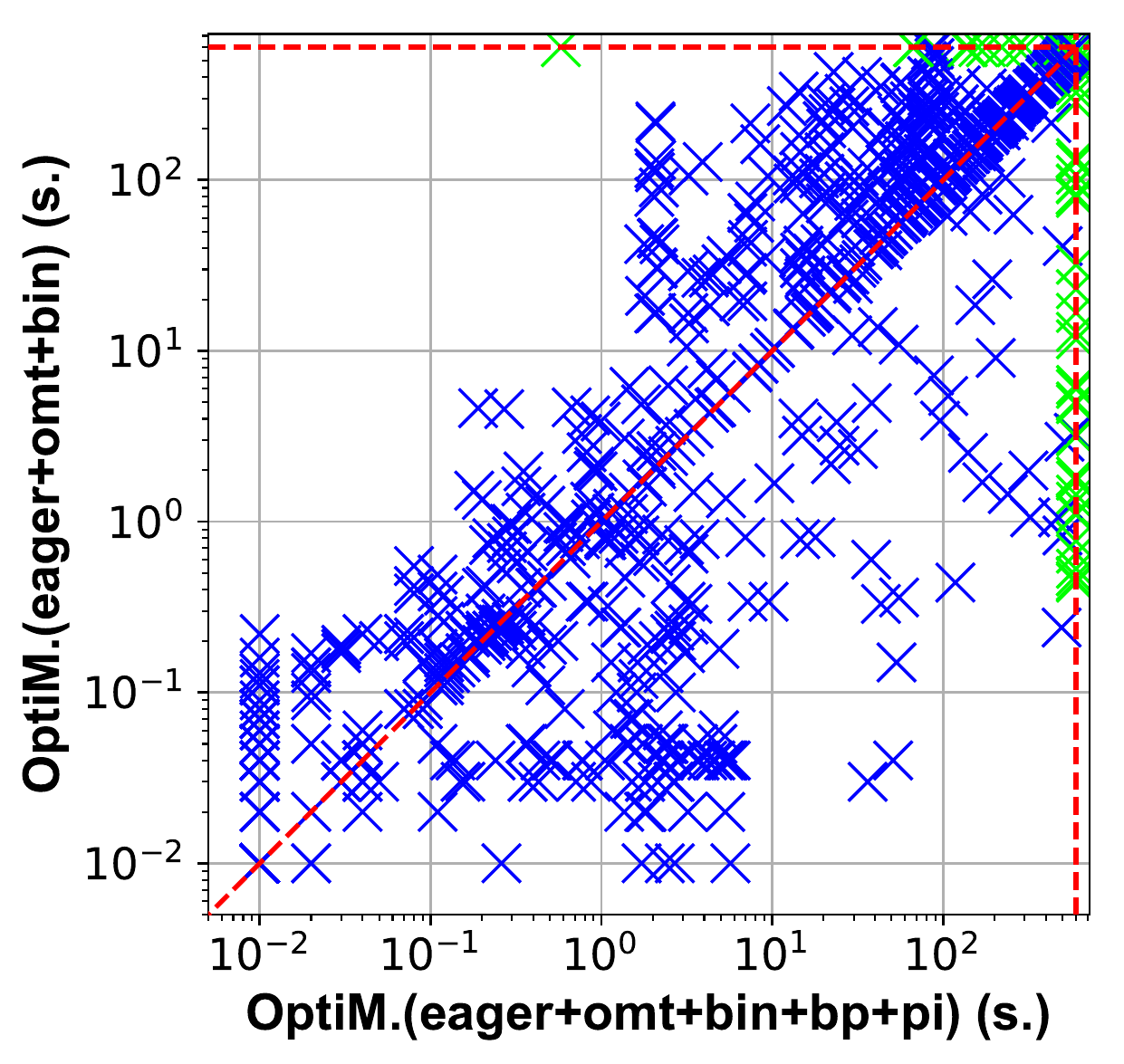}
&
        \includegraphics[scale=0.47]{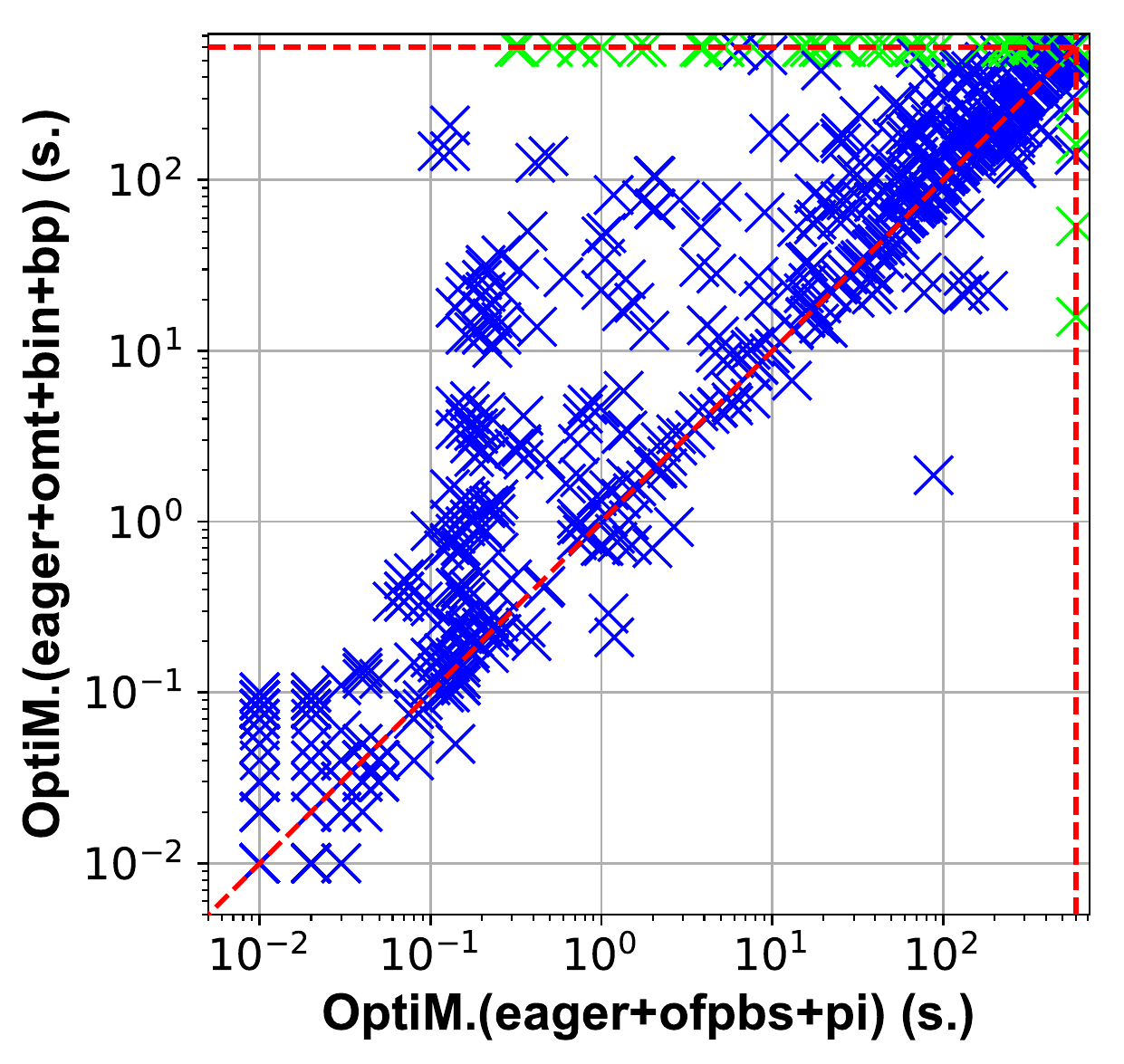}
\\%
        \includegraphics[scale=0.47]{figs/3.pdf}
&
        \includegraphics[scale=0.47]{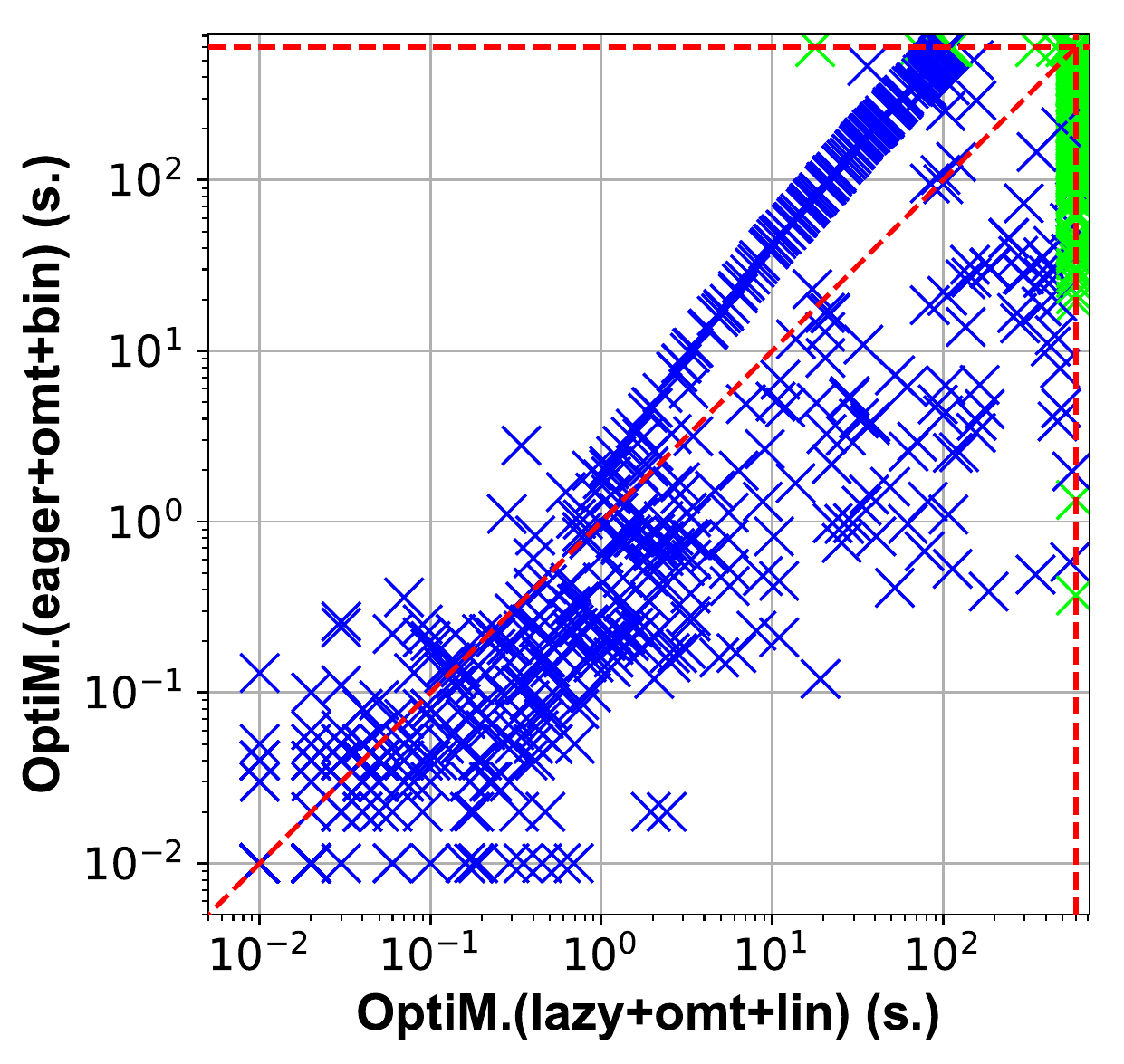}
&
        \includegraphics[scale=0.47]{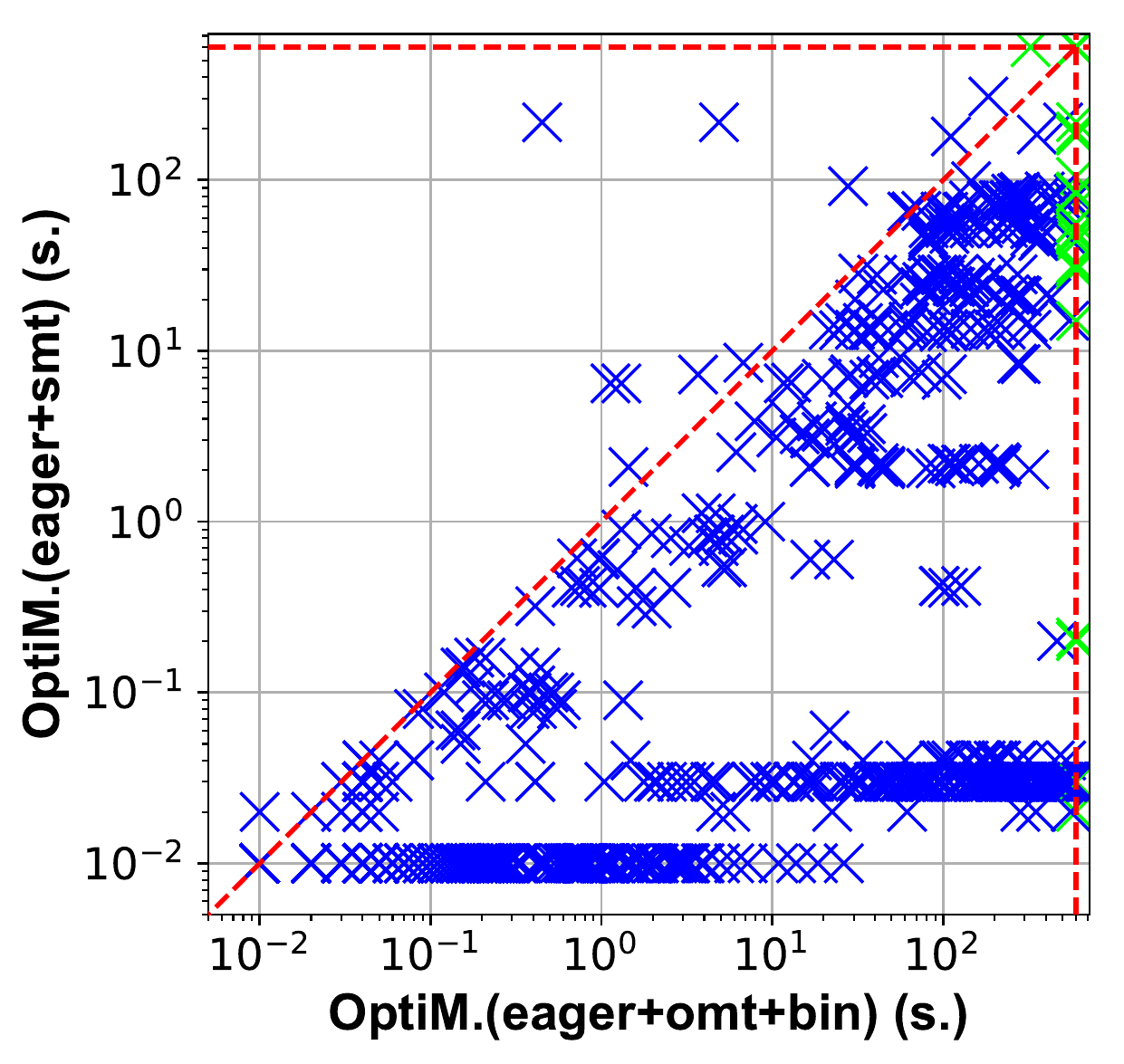}
\\%
    \end{tabular}
\caption[Comparison on \omtfp{} formulas with binary-search (Scatter Plots)]{
\label{fig:fp-sp-bin}
Pairwise comparisons on \omtfp{} formulas using \omt-based{} binary-search
and other configurations.
(\blue{Blue} points denote satisfiable benchmarks, \green{green} denotes a timeout.)
}
\end{figure}
\end{IGNOREINSHORT}


\begin{IGNOREINSHORT}
\begin{figure}[tbp]
    \centering
    \footnotesize
    \begin{tabular}{ccc}%
&\includegraphics[scale=0.47]{figs/A.pdf}&\includegraphics[scale=0.47]{figs/B.pdf}
\\%
        \includegraphics[scale=0.47]{figs/1.pdf}
&
        \includegraphics[scale=0.47]{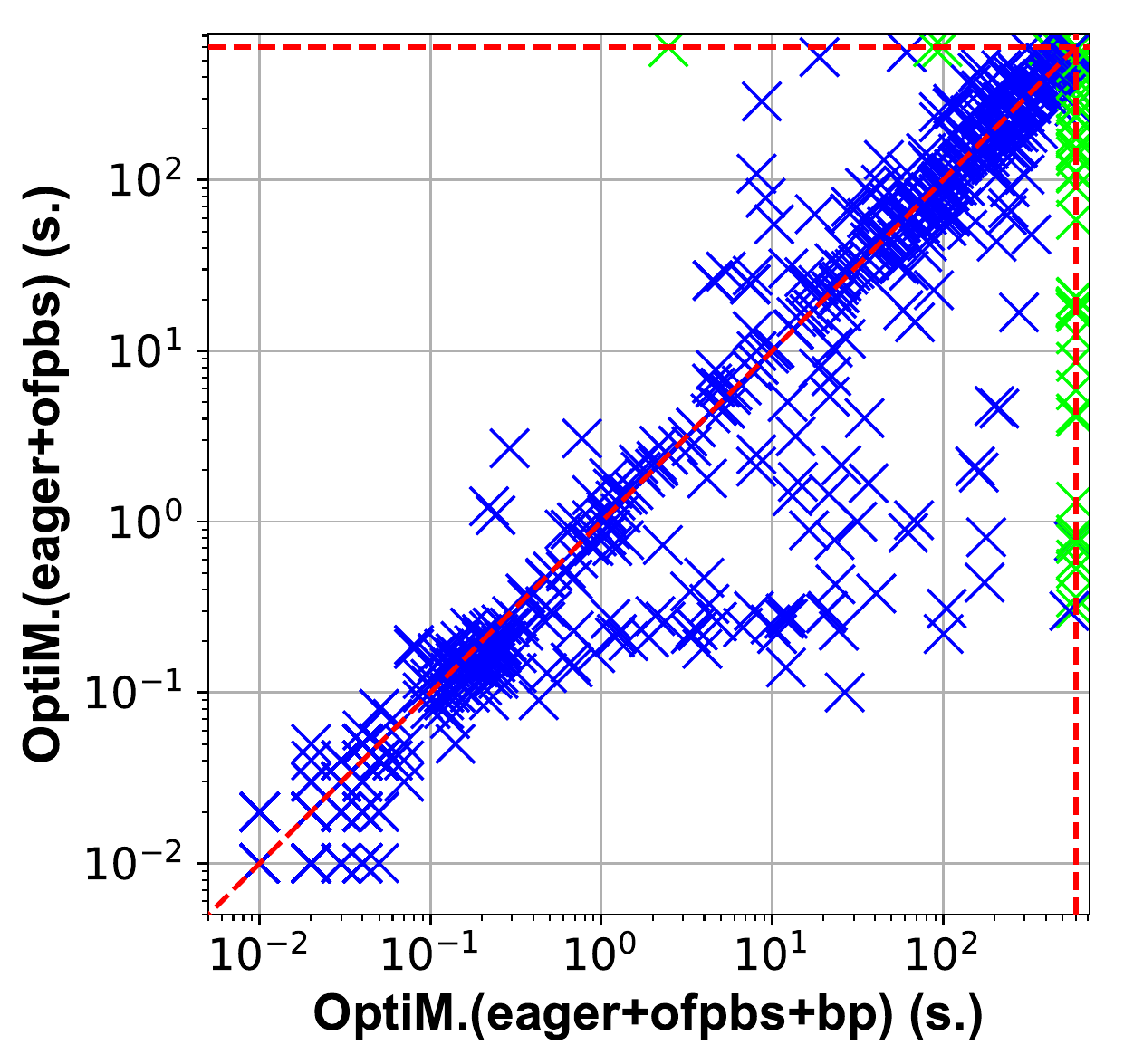}
&
        \includegraphics[scale=0.47]{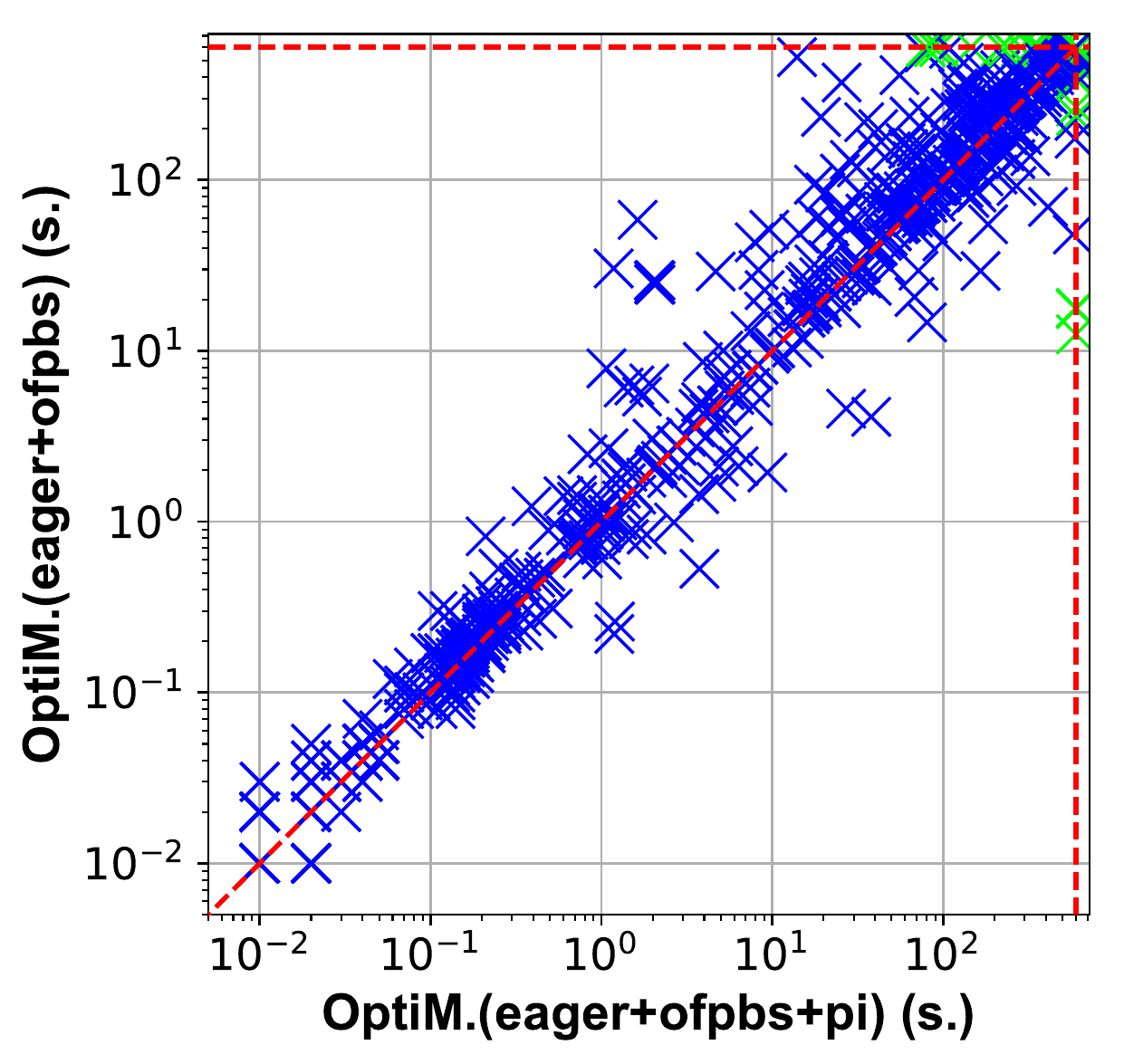}
\\%
        \includegraphics[scale=0.47]{figs/2.pdf}
&
        \includegraphics[scale=0.47]{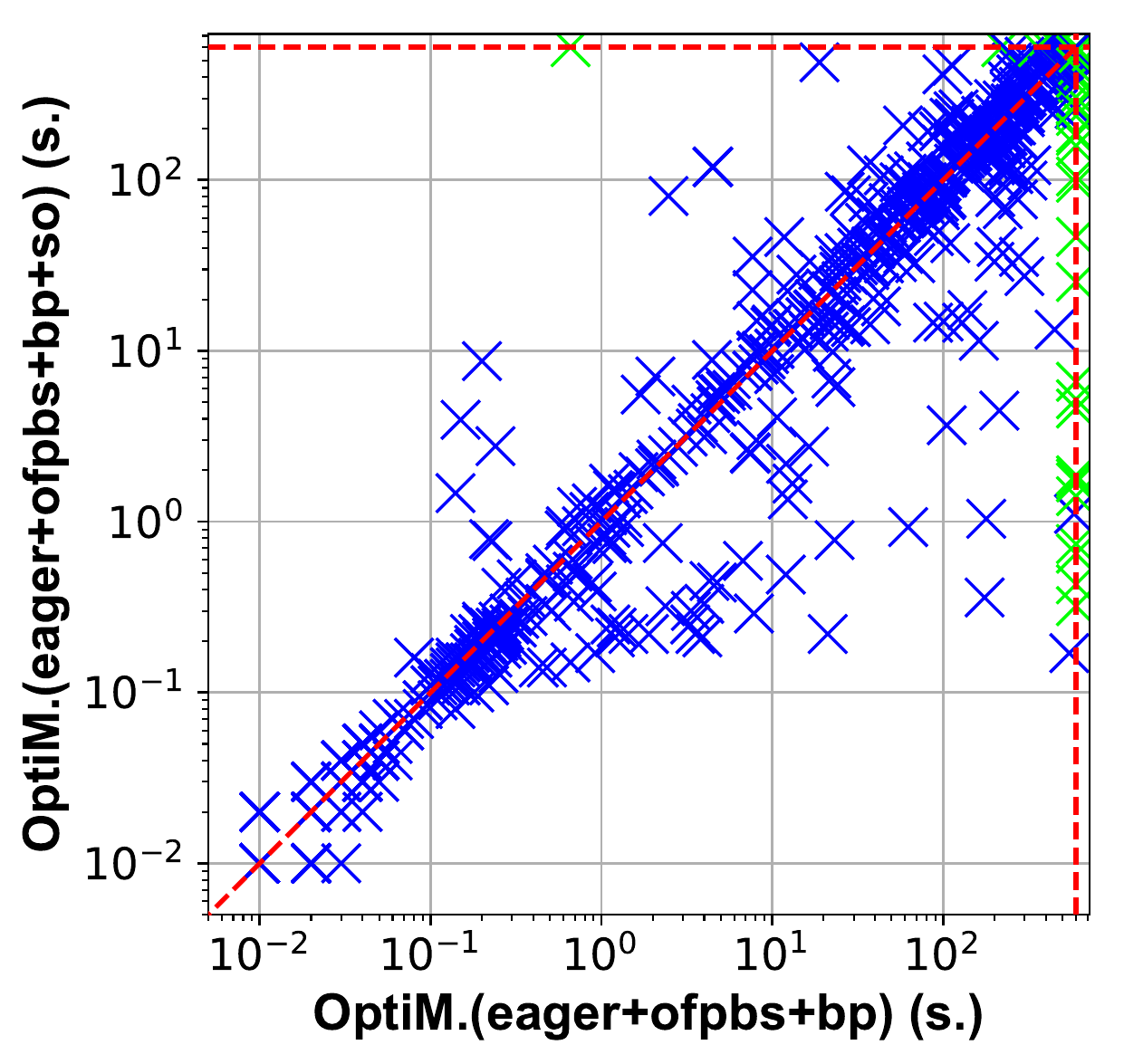}
&
        \includegraphics[scale=0.47]{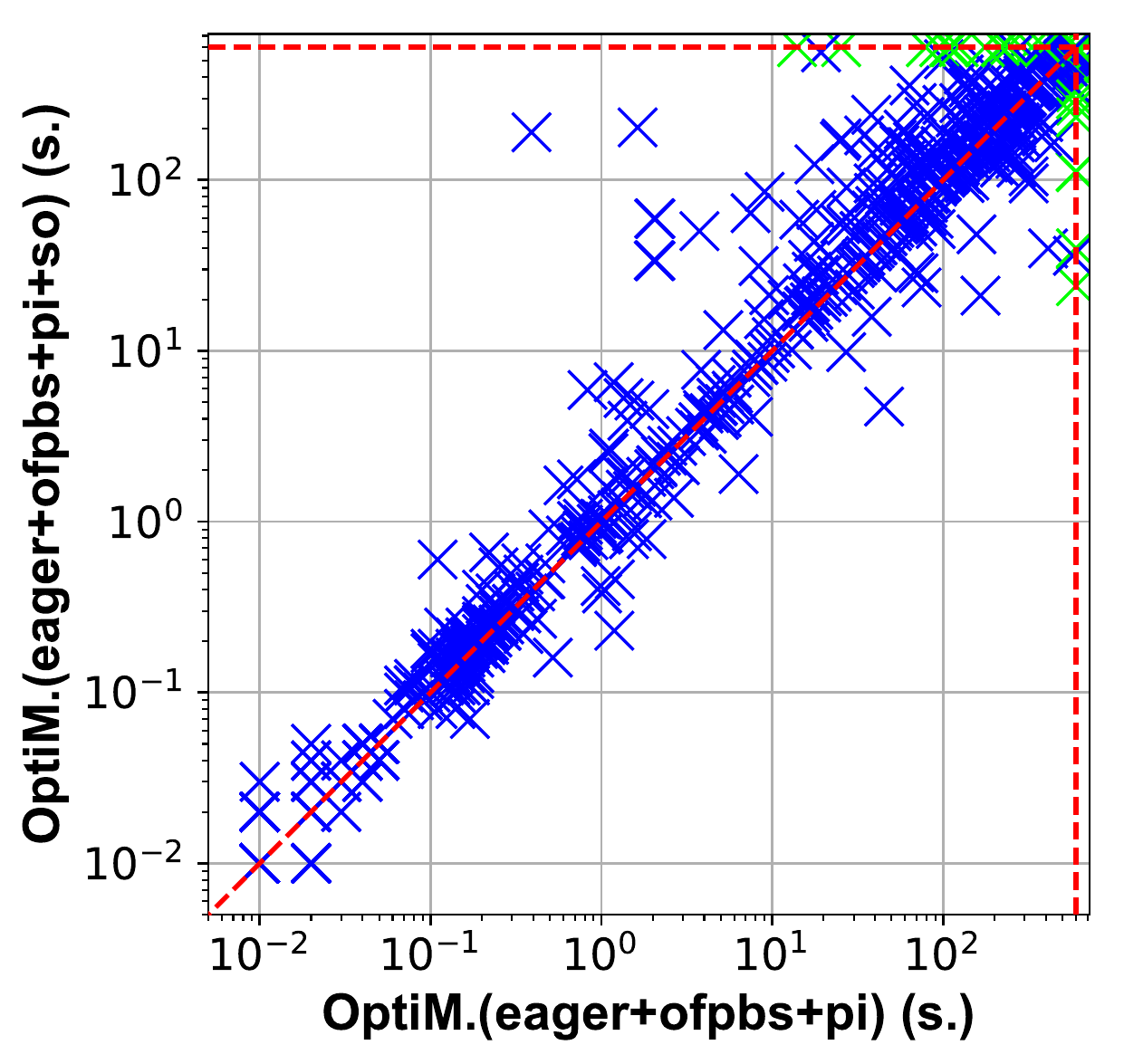}
\\%
        \includegraphics[scale=0.47]{figs/3.pdf}
&
        \includegraphics[scale=0.47]{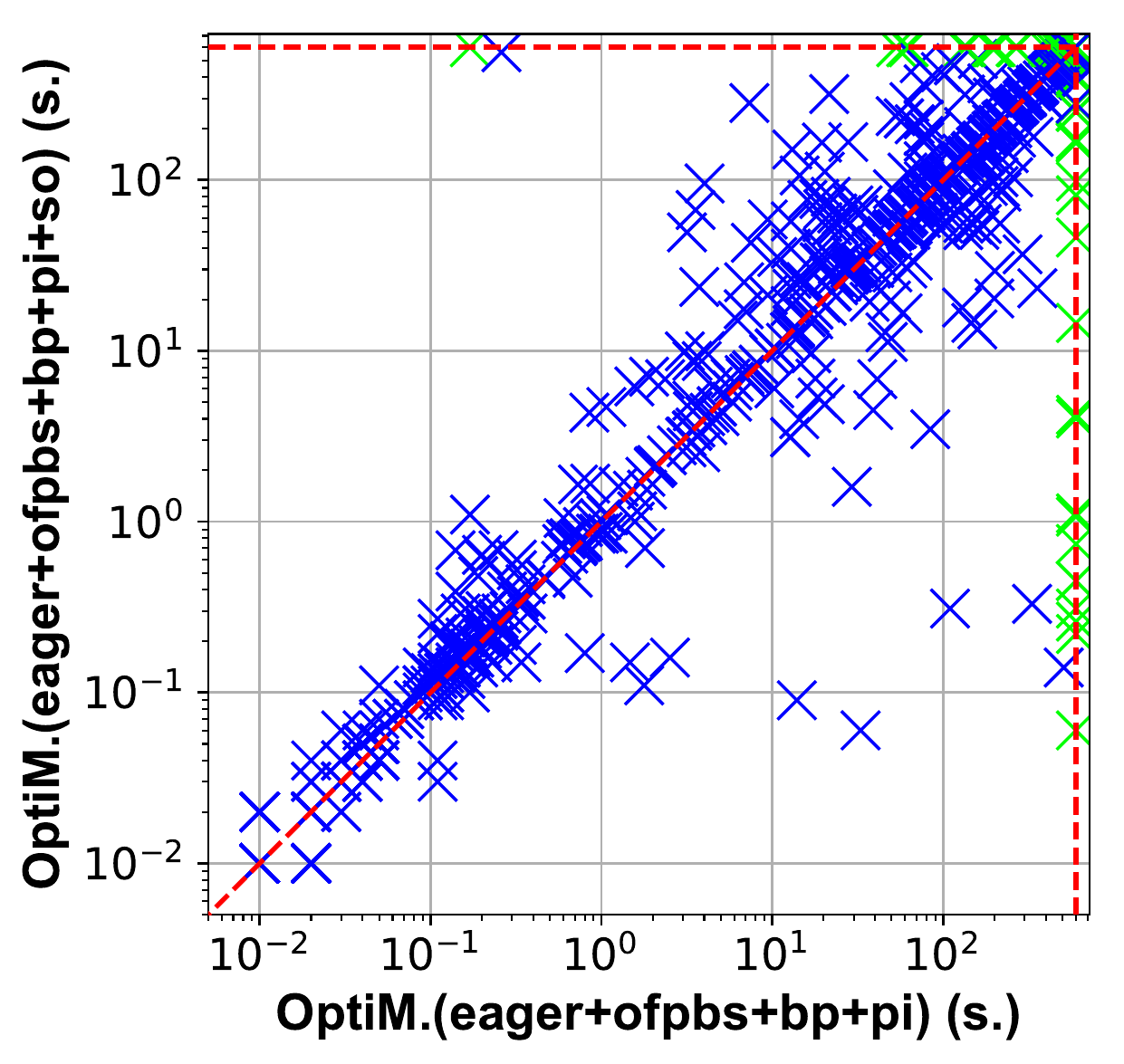}
&
        \includegraphics[scale=0.47]{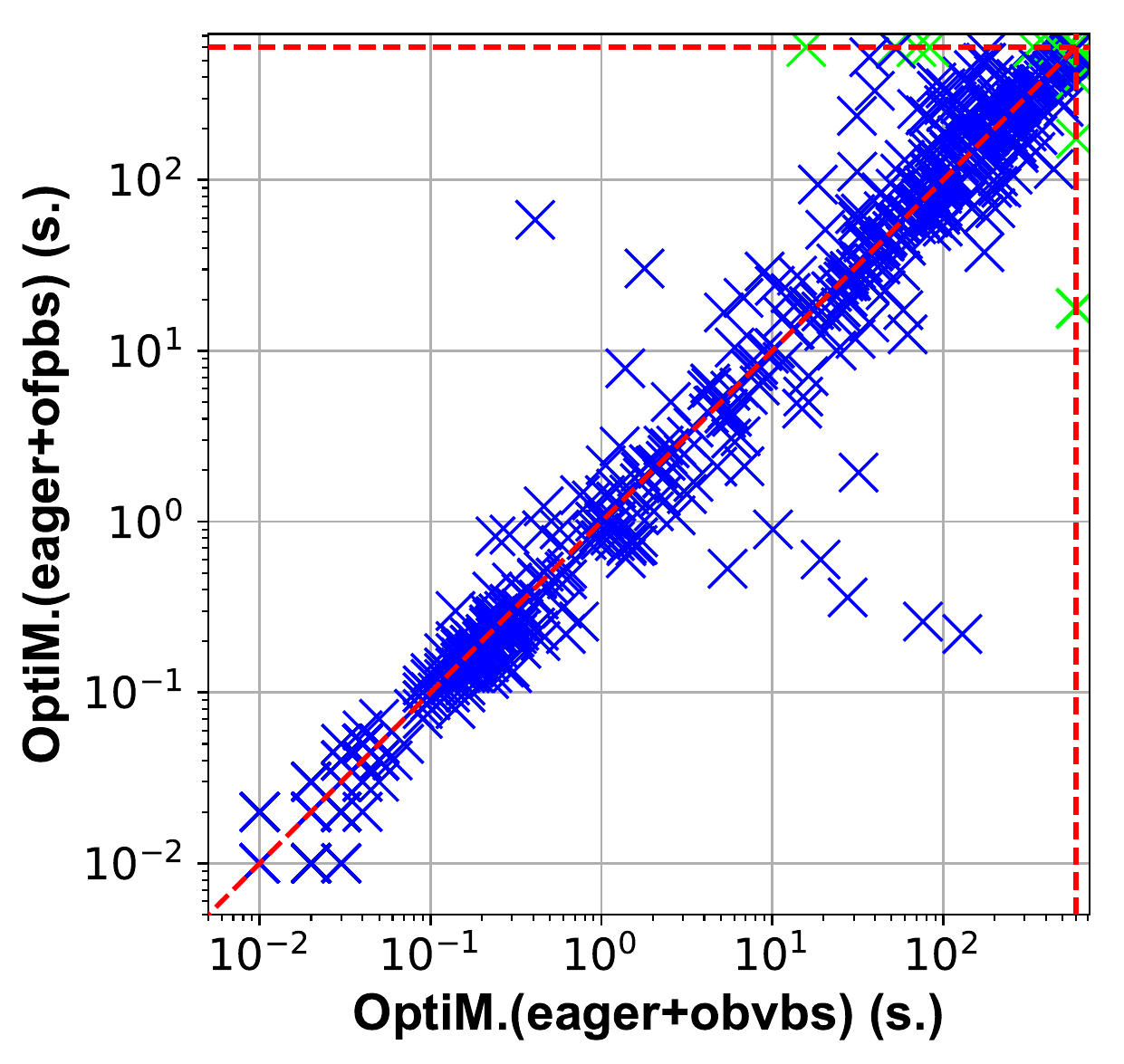}
\\%
    \end{tabular}%
\caption[Comparison on \omtfp{} formulas using \ofpbs{} (Scatter Plots)]{
\label{fig:fp-sp-ofpbs}
Pairwise comparisons on \omtfp{} formulas using the \ofpbs{} engine
and other configurations.
(\blue{Blue} points denote satisfiable benchmarks, \green{green} denotes a timeout.)}
\end{figure}%
\end{IGNOREINSHORT}


The results of this experiment are listed in Table \ref{tab:fp}.
\begin{IGNOREINSHORT}
Figure~\ref{fig:fp-cp} depicts the loc-scale cactus plot of the same
data, for a visual comparison among the different configurations.
In addition, Figures \ref{fig:fp-sp-lin}, \ref{fig:fp-sp-bin} and \ref{fig:fp-sp-ofpbs}
show a selection of relevant pairwise comparisons among various \optimathsat configurations.
Figure \ref{fig:fp-sp-lin} focuses on variants of the \omt-based{} linear-search approach,
Figure \ref{fig:fp-sp-bin} depicts variants of the \omt-based{} binary-search approach,
whereas Figure \ref{fig:fp-sp-ofpbs} focuses on the \ofpbs{} engine.
\end{IGNOREINSHORT}


For what concerns \omt-based \emph{linear-search} optimization, we
observe that \optimathsat{} performs the best when no enhancement
is enabled.
In particular, the empirical evidence suggests that enabling \emph{branching
preference} significantly increases the number of timeouts, generally deteriorating
the performance\ignoreinshort{~(plot $1A$ in Fig. \ref{fig:fp-sp-lin})}.
Enabling only \emph{polarity initialization} does not result in an
appreciable change on the running time of the solver\ignoreinshort{~(plot $1B$ in Fig.
\ref{fig:fp-sp-lin})}.
In contrast, enabling both enhancements at the same time\ignoreinshort{~has
a small chance to result in a small improvement of the search time (plot $2A$
in Fig. \ref{fig:fp-sp-lin}), but it} generally worsens the performance and
results in a drastic increase in the number of timeouts (Table~\ref{tab:fp}).
We justify these results as follows.
First, when only \emph{polarity initialization} is used, the phase-saving value that
is being set by \optimathsat{} does not really matter because the optimization
search is dominated by the structure of the formula itself rather than by the
bits of the \fp objective.
Second, when \emph{polarity initialization} is used on top of \emph{branching
preference}, there is an even more drastic decrease in performance due to the fact
that the initial phase-saving value that is statically assigned by the \omt solver
to the bits of the \fp objective cannot be expected to be ``good enough'' for any
situation.
\ignoreinshort{
In fact, as illustrated in example \ref{ex:fp-sa-bad}, the initial
phase-saving can be misleading and force the \omt solver --when running in
\emph{linear-search}-- to explore an exponential number of intermediate satisfiable
solutions.
}


In the case of the \omt-based{} \emph{binary-search} optimization approach,
we observe that it solves more formulas than linear-search and it generally
appears to be faster\ignoreinshort{~(plot $3B$ in Fig.~\ref{fig:fp-sp-lin})}.
Overall, \emph{polarity initialization} does not seem to be beneficial,
whereas enabling \emph{branching preference} increases the number of
formulas solved within the timeout.
This behavior is different from the linear-search approach, and we conjecture
that it is due to the fact that, with the \omt-based binary-search approach,
branching over the bits of the objective function can reveal in advance any
(partial) assignment to the bits of the objective function that it is
inconsistent wrt. the pivoting cuts learned by the optimization engine.


Using the {\em lazy} \fp engine results in fewer formulas being solved,
although a significant number of these benchmarks is solved faster than
with any other configuration\ignoreinshort{~(over $90$ instances, for both configurations)}.%


The {\sc \optimathsat{}(eager+\obvbs{})} configuration is able to solve $1013$
formulas within the timeout, showing that  \omtfp{} can be reduced to \omtbv{}
effectively, and that --on the given benchmark-set-- the performance of this
approach are comparable with the best \omtfp{} configurations being tested.


Overall, the best performance is obtained by using the \ofpbs{} engine, with up
to $1019$ benchmark-set instances being solved in correspondence to the 
{\sc \optimathsat{}(eager+\ofpbs{}+pi)} configuration.
\ignoreinshort{
In plot $2B$ of Figures \ref{fig:fp-sp-lin} and \ref{fig:fp-sp-bin}, we show the
pairwise comparison of the best \ofpbs{} configuration with the best \omt-based run.
}
Similarly to the case of \omt-based optimization with linear-search, we observe
that enabling \emph{branching preference} generally makes the performance
worse\ignoreinshort{~(plot $1A$ in Fig. \ref{fig:fp-sp-ofpbs})}.
Instead, when \emph{polarity initialization} is used we observe a general
performance improvement that does not only result in an increase in the number
of formulas being solved within the timeout, but also a noticeable reduction
of the solving time as a whole.
This is in contrast with the case of \omt-based optimization, and it can be
explained by the fact that \ofpbs{} uses an internal heuristic function to
dynamically determine and update the most appropriate phase-saving value for
the bits of the objective function.
An equally important role is played by the \emph{safe bits restriction}, that
limits the effects of \emph{branching preference} and \emph{polarity initialization}
to only certain bits of the {\it dynamic attractor}.
\ignoreinshort{
As illustrated by the plots in the second and third rows of Figure \ref{fig:fp-sp-ofpbs}
and by the data in Table \ref{tab:fp}, t}\ignoreinlong{T}his feature is particularly
effective when used in combination with \emph{branching preference}.


\begin{IGNOREINSHORT}
The results of \optimathsat over the \smt-only version of the benchmark-set
are reported in Table~\ref{tab:fp} and in the scatter-plot $3B$
in Fig.~\ref{fig:fp-sp-bin}, and show that for a large number of instances
the \omt problem is considerably harder than its \smt-only version
There are a few exceptions to this rule, that we ascribe to the fact
that the removal of the objective function alters the internal stack
of formulas, and this can have unpredictable consequences on the behavior
of various internal heuristics that depend on it. A solution can
be found in a shorter amount of time when the sequence of (heuristic)
choices is compatible with its assignment and it requires little
back-tracking effort.
\end{IGNOREINSHORT}


\section{Conclusions and Future Work}
\label{sec:concl}
We have presented for the first time \omt procedures (for signed
Bit-Vectors and) Floating-Point numbers, based on the novel notions of
attractor, dynamic attractor and attractor trajectory, which we have
implemented in \optimathsat and tested on modified problems from
SMT-LIB.

Ongoing research involves implementing our \ofpbs procedure on top of
the ACDCL \smtfp procedure ---which is not immediate to do efficiently because
the latter approach does not allow directly accessing and setting the single bits of
the objective (since \bv and \fp are not signature-disjoint).
Future research involves experimenting the new \omt
procedure directly on problems coming from bit-precise SW and HW
verification, produced, e.g.,  by the NuXmv model checker \cite{nuxmv_url}.


\begin{IGNOREINLONG}
\newpage
\end{IGNOREINLONG}
\FloatBarrier
\pagenumbering{roman}
\bibliographystyle{abbrv}
\bibliography{main_long}


\end{document}